\documentclass[numberedappendix]{emulateapj}
\usepackage{amsmath}
\usepackage{multirow}
\usepackage{array}
\newcommand{\fthin}{$f_{\rm{DS}}$}
\newcommand{\nthin}{$n_{\rm{DS}}$}
\newcommand{\fstar}{$f_\star$}
\newcommand{\mstar}{$M_\star$}
\newcommand{\fstarMh}{\fstar($M_{\rm{H}}$)}
\newcommand{\mhalo}{$M_{\rm{H}}$}
\newcommand{\msun}{M$_\sun$}
\newcommand{\lcdm}{$\Lambda$CDM}
\newcommand{\esn}{$\epsilon_{\rm{SN}}$}
\newcommand{\esneq}{\epsilon_{\rm{SN}}}
\newcommand{\arh}{$\alpha_{\rm{RH}}$}
\newcommand{\arheq}{\alpha_{\rm{RH}}}
\newcommand{\tst}{$\tau_\star$}
\newcommand{\vej}{$V_{\rm{eject}}$}

\newcommand{\mh}{$M_{\rm{H}}$}
\newcommand{\gammapl}{$\gamma_{\rm{PL}}$}
\newcommand{\alphapl}{$\alpha_{\rm{PL}}$}
\newcommand{\taupl}{$\tau_{\rm{PL}}$}
\newcommand{\htwo}{H$_2$}
\newcommand{\hi}{H \textsc{i}}

\newcommand{\chirein}{$\chi_{\rm{ReIn}}$}

\def\lesssim{\lower.5ex\hbox{$\; \buildrel < \over \sim \;$}}
\def\gtrsim{\lower.5ex\hbox{$\; \buildrel > \over \sim \;$}}

\shorttitle{High-z overproduction of stars in modeled dwarf galaxies}
\shortauthors{White et al.}

\begin{document}

\title{A parametric study of possible solutions to the high-redshift overproduction of stars in modeled dwarf galaxies}
\author{Catherine E. White\altaffilmark{1}, Rachel S. Somerville\altaffilmark{2}, and Henry C. Ferguson\altaffilmark{3}}

\affil{\altaffilmark{1} Department of Physics and Astronomy, Johns Hopkins University, 3400 N. Charles Street, Baltimore, MD 21218, USA}

\affil{\altaffilmark{2}  Department of Physics \& Astronomy, Rutgers University, 136 Frelinghuysen Road, Piscataway, NJ 08854, USA}

\affil{\altaffilmark{3}   Space Telescope Science Institute, 3700 San Martin Drive,
Baltimore, MD 21218, USA}

\begin{abstract}
Both numerical hydrodynamic and semi-analytic cosmological models of
galaxy formation struggle to match observed star formation histories
of galaxies in low mass halos (M$_{\rm{H}} \lesssim 10^{11}$\ \msun),
predicting more star formation at high redshift and less star
formation at low redshift than observed.  The fundamental problem is
that galaxies' gas accretion and star formation rates are too closely
coupled in the models: the accretion rate largely drives the star
formation rate.  Observations point to gas accretion rates that
outpace star formation at high redshift, resulting in a buildup of gas
and a delay in star formation until lower redshifts.  We present three
empirical adjustments of standard recipes in a semi-analytic model
motivated by three physical scenarios that could cause this
decoupling: 1) the mass-loading factors of outflows driven by stellar
feedback may have a steeper dependence on halo mass at earlier times,
2) the efficiency of star formation may be lower in low mass halos at
high redshift, and 3) gas may not be able to accrete efficiently onto
the disk in low mass halos at high redshift. These new recipes, once
tuned, better reproduce the evolution of \fstar$\equiv M_\star/M
_{\rm{H}}$~as a function of halo mass as derived from abundance
matching over redshifts $z=0$~to 3, though they have different effects
on cold gas fractions, star formation rates, and metallicities.
Changes to gas accretion and stellar-driven winds are promising, while
direct modification of the star formation timescale requires drastic
measures that are not physically well-motivated.

\end{abstract}

%====================================================================
%====================================================================

\section{Introduction}
%Look at dark matter!  It's so cool!
The formation of dark matter halos is fairly well understood: a
universe dominated by dark energy and cold dark matter
(\lcdm\ cosmology) matches many large-scale observations
\citep{primack:03} and makes clear predictions about the evolution of
dark matter.  N-body simulations run with \lcdm\ physics and
observed initial conditions allow us to model the formation and
interactions of dark halos and have been run on both cosmological
scales \citep[e.g.][]{springel:05} and galactic scales
\citep[e.g.][]{diemand:07, stadel:09}.

%Baryons are hard
The formation of the baryonic component of galaxies is less well
understood.  Despite the success of \lcdm\ in describing the formation
and evolution of dark matter halos, attempts to produce realistic
galaxies within the \lcdm\ framework have had only limited
success. When modeling dark matter alone, the inability to model large
and small scales simultaneously results in finite resolution, but does
not affect the large scale behavior.  With baryonic matter, the
inability to explicitly follow processes on small scales requires
models to make assumptions about small scale physics such as star
formation and feedback from massive stars and supernovae.  These
processes are very important for the overall behavior and properties
of galaxies. Hydrodynamic simulations numerically model the baryonic
matter to the resolution limit, then use recipes for sub-grid physics
to approximate the smaller scale processes.  Semi-analytic models
instead use a set of analytic recipes to approximate the overall
properties and evolution of entire galaxies. Both types of model can
achieve fairly good agreement with many observations of nearby
galaxies, including the local stellar mass function (SMF) and
luminosity function (LF), cold gas fractions, and the mass-metallicity
relation (MZR) \citep[e.g.][]{bower:06, somerville:08, guo:11,
  dave_of:11, vogelsberger:14}.  This is in part possible by tuning
sub-grid recipes and their free parameters.

%Description of the problem we're looking at
One of the difficulties in galaxy formation models of both types is
reproducing the properties of galaxies in low mass halos (virial mass
$M_{\rm{H}}\lesssim10^{11}$~\msun).  Simulated low mass galaxies tend
to form stars too early and too efficiently, producing a population of
low mass galaxies at redshift $z=0$\ with redder colors, lower star
formation rates, and older stellar population ages than are observed
\citep{fontanot:09}.  Star formation histories for these galaxies peak
too early, resulting in an excess of low mass galaxies ($10^9$~\msun
$\lesssim M_\star \lesssim 10^{10}$~\msun) at
$z>0$\ \citep{weinmann:12}. In addition, observed star forming
galaxies tend to have decreasing specific star formation rates (sSFR;
$\dot{M}_{\star}/M_{\star}$) with increasing mass, a trend that is not
reproduced in the models: the models produce constant sSFRs
over a large range of stellar mass or even show higher sSFRs at higher
stellar mass.  These are all symptoms of the same fundamental problem:
in the models, gas accretion closely follows dark matter accretion and
star formation traces gas accretion.  The net result is that the star
formation history mirrors the dark matter accretion history, which in
\lcdm\ is nearly self-similar for different halo masses.  Something
must break the self-similarity between accretion rate and star
formation rate for models to reproduce low mass galaxies' observed
star formation histories
\citep{conroy:09,behroozi:10,moster:13,behroozi:13}.\footnote{Although
  here we focus on low mass galaxies, it is worth noting briefly that
  high mass galaxies show the same sort of ``breaking'' of the
  self-similar scaling governed by the halo mass-accretion histories
  but in the opposite sense in time: massive galaxies' stellar masses
  apparently grow more slowly than their halos at late times. This
  trend has been more successfully reproduced by models that implement
  feedback from Active Galactic Nuclei (AGN).}
  
Nearly all current models of galaxy formation set within the
\lcdm\ framework rely on qualitatively similar recipes for the crucial
sub-grid processes. First, powerful outflows driven by massive stars
and supernovae (hereafter ``stellar-driven winds''), are assumed to
efficiently heat and eject cold gas from the interstellar medium (ISM)
of galaxies. In order to match the observed slope of the $z\sim
0$\ SMF as well as the observed MZR, outflows must be more
  efficient in lower mass galaxies, ejecting more gas per unit star
  formation.  Matching high gas fractions at redshift $z=0$\ requires
  lower efficiency star formation in low mass galaxies, which is
usually achieved by imposing a minimum gas surface density threshold
below which star formation does not occur.  A second
consequence of the adopted sub-grid recipes is that star formation in
low mass halos is strongly self-regulated, meaning that
  modifying the star formation and stellar feedback in the models can
  have a smaller impact than anticipated on many observables: lower
  efficiency star formation leads to less stellar feedback, leading to
  more cold gas available and therefore more star formation
\citep{dave_fo:11, haas:13}.

%Previous attempts to fix
Several previous modelers have attempted to address these problems. For instance, \citet{krumholz:12}\ implemented a metallicity-dependent star formation efficiency in an analytic toy model. In their model, stars can form only in molecular gas (\htwo) and the formation efficiency of molecular gas depends on metallicity.  They suggested that this would delay star formation in low mass galaxies since it would take time for sufficient metals to build up to provide efficient formation of \htwo\ and stars. However, they did not actually show that their model quantitatively reproduces galaxy stellar mass functions at low and high redshift. In another toy model, \citet{bouche:10} cut off accretion for halo masses $M_{\rm{H}}<10^{11}$\ \msun\ and successfully match the slope of the star forming sequence and the Tully-Fisher relation.  However, the accretion floor means halos with $M_{\rm{H}}<10^{11}$\ \msun\ will have no stars at all, which is inconsistent with observations of nearby dwarf galaxies. 

%What H13 does
\citet[H13]{henriques:13} attempted to solve the problems with the redshift evolution of the SMF and LF by altering the timescale for re-accretion of previously ejected gas. They found that changing the ejecta reincorporation timescale and retuning parameters controlling star formation and gas handling could match observed B- and K-band luminosity functions and the SMF over redshifts $0\leq z \leq 3$\ while no retuning of their standard model could. In their altered model, ejecta reincorporation timescales are inversely proportional to halo mass but independent of redshift.  In their standard model, the timescale is inversely proportional to halo dynamical time, which is a fairly strong function of redshift but independent of halo mass. In addition, they showed that their model predictions agree with the \mstar-\mhalo\ relation derived from abundance matching by \citet{moster:13} over the same redshift range and produced higher sSFRs and younger ages for galaxies with $M_* \simeq 10^{9}-10^{9.5}$ \msun, in better agreement with observations.  Although H13 point out that the form of the reincorporation timescale that they adopt is similar to that found in some hydrodynamic simulations \citep[e.g.][]{oppenheimer:10}, recent hydrodynamic simulations that implement similar recipes for stellar winds still overproduce low mass galaxies at intermediate redshifts \citep{weinmann:12, torrey:13}. 

In general, any solution to the problems in modeled dwarf galaxies must suppress star formation preferentially at higher redshift and lower halo masses.  In this paper, we explore three different physical scenarios that seem promising for solving the problems with low mass galaxies.  We alter 1) the scaling of the mass-loading factor for stellar-driven winds, defined as the outflow rate divided by the star formation rate, 2) the timescale for turning cold gas into stars, or 3) the timescale for gas to accrete into dark matter halos. In the present work, we restrict ourselves to modifications of only one of these recipes at a time. Although our scenarios are physically motivated, we adopt a flexible and empirical approach with the aim of identifying general properties of the necessary scalings, which may provide clues to more physically motivated forms of solution.

%Can I compress the next three paragraphs into one?  Yes, because I'm awesome like that.
As discussed above, models already require high mass-loading factors for low mass galaxies in order to reproduce SMFs.  Most models define mass-loading factors that depend only on the halo circular velocity $V_{\rm{circ}}$, but it is likely that in reality the mass-loading factor depends on other galaxy properties as well.  These other properties, such as metallicity, gas density, and pressure, could introduce an effective redshift dependence in the mass-loading factor expressed in terms of $V_{\rm{circ}}$.  To reproduce dwarf galaxy properties, reheating must be preferentially more efficient at low masses and high redshifts, a scenario we call ``preferential reheating.''  If we instead consider the problem in terms of star formation itself, star formation efficiencies in low mass halos must be lower than in high mass halos.  Models usually implement this as a gas surface density threshold, another quantity that is most likely dependent on galaxy properties not taken into account in the model.  A changing star formation efficiency can be viewed as either a redshift and/or halo mass dependent fraction of a galaxy's gas available for star formation or as a changing star formation timescale.  In either case, adjusting the star formation efficiency could allow low mass galaxies to have their star formation suppressed until more gas accumulates onto the disk, delaying the galaxy's star formation. We will refer to this scenario as ``direct suppression.''  Lastly, small halos may have trouble accreting gas in the first place. The photo-ionizing background suppresses gas accretion in the very smallest halos at high redshift \citep[and refs.~therein]{efstathiou:92, quinn:96, somerville:02, somerville:08}, but there could be some ``pre-heating'' mechanism that acts on halos up to $\sim10^{10}$\ \msun, preventing accretion of gas \citep[e.g.][]{lu:14}.  The gas would instead remain in a ``parking lot,'' waiting to be accreted at later times.  Star formation in low mass galaxies would then be delayed until gas is released from the parking lot at lower redshifts. We will refer to this as the parking lot scenario.

%So, that's all cool.  What are we doing in this paper?
We implement these three model variants, preferential reheating, direct suppression, and parking lot, within the Santa Cruz semi-analytic model \citep{somerville:08,somerville:12}, and explore the implications for a set of complementary observations. These include the fraction of stellar mass to dark matter mass as a function of halo mass, the comoving number density of low mass galaxies as a function of redshift, and scaling relations between the galaxy stellar mass and the cold gas fraction, sSFR, and metallicity.

The baseline semi-analytic model is described in \textsection \ref{sec:sam_desc}~and we discuss the predictions of the fiducial model in \textsection \ref{sec:fiducial}.  In \textsection \ref{sec:param_space}, we describe an exploration of parameter space in the existing model in order to gain insight into how various model ingredients affect the observables.  A more detailed presentation of these results is presented in Appendix \ref{sec:param_appendix}.  In \textsection \ref{sec:ad_hoc}, we present the results of our preferential reheating, direct suppression, and parking lot models. We discuss the implications of our results and conclude in \textsection \ref{sec:conclusions}. Appendix~\ref{sec:appendix_h13} explores possible reasons for differences between our results and those presented by H13.

%====================================================================
%====================================================================
\section{Summary of the model}
\label{sec:sam_desc}
In this paper, we use the baseline model described in
\citet[S08]{somerville:08} and \citet[S12]{somerville:12}. We adopt a
Chabrier stellar initial mass function (IMF) and WMAP5 cosmological
parameters: $\Omega_0=0.28$, $\Omega_\Lambda=0.72$, $h_{100}=0.70$,
and $f_{\rm{baryon}}=0.1658$\ \citep{wmap5}.  We shut off AGN
feedback, both radio mode and quasar mode, in all of the 
simulations presented.  As implemented in our
models, the AGN feedback does not noticeably affect galaxies with halo
masses M$_{\rm{H}} \lesssim 10^{11.75}$\ and omitting AGN feedback
isolates any effects from our adjusted recipes on high mass halos.

The merging histories of dark matter halos are constructed based on
the Extended Press-Schechter (EPS) formalism following the method
described in \citet{somerville_kolatt:99} and \citet{somerville:08}
We follow the merger history of a particular halo back to a minimum progenitor mass of 0.01 times the final mass of the halo. Whenever dark matter halos merge, the central galaxy of the
largest progenitor halo becomes the new central galaxy and all the
other galaxies become ``satellites.''  Satellite galaxies may
eventually merge with the central galaxy due to dynamical friction and
these merger timescales are estimated using a variant of the
Chandrasekhar formula from \citet{boylan-kolchin:08}. Tidal stripping
and destruction of the satellites is included as described in
\citet{somerville:08}.

%----------------------------------------------------------------------------------------
\subsection{Gas handling}
\label{sec:gas_stuff}

The SAM tracks four ``boxes'' of gas: a cold disk representing
  the ISM, a hot halo representing the intra-group or
  -cluster medium (ICM), an ejected gas reservoir holding gas
  that has been ejected from the galaxy and pristine gas prevented
  from accreting from the IGM by the photo-ionizing background, and a
reservoir containing intergalactic medium (IGM) gas that has never
been inside a resolved halo.  Gas is assigned to reservoirs
  and moves between them as follows.  Gas in the ejected reservoir is
allowed to accrete into the hot halo, and gas in the hot halo may cool
and fall onto the cold disk.  Stellar feedback reheats gas in the cold
disk, moving it to the hot halo or ejecting it to the ejected
reservoir.  New gas is added to the hot halo through pristine gas
accretion and stripping of gas from satellites as they fall into the
main halo.

On creation, halos are assigned a certain mass of gas.  Before
  the reionization of the universe, each halo is assigned its
  universal baryon fraction's worth of gas, but after reionization the
  collapse of gas into low-mass halos is suppressed by the
  photo-ionizing background (``squelching''). In the published models
  of S08 and S12, the fraction of baryons that can collapse into halos
  of a given mass after reionization is modeled using the fitting
  functions provided by \citet{gnedin:00} and \citet{kravtsov:04}.
  Some more recent studies indicate that the characteristic mass below
  which squelching strongly prevents accretion, called the filtering
  mass, predicted by \citet{gnedin:00} may be too large
  \citep[e.g.][]{okamoto:08}. According to this more recent work, the
  halo mass at which halos have their baryon fractions reduced by a
  factor of two on average due to squelching is only about $M_{\rm{H}}
  \sim 9.3 \times 10^9$\ \msun\ at $z=0$~rather than $M_{\rm{H}} \sim
  10^{10.5}$\ \msun\ used in \citet{gnedin:00} and
  \citet{kravtsov:04}. This halo mass is much lower than the
  lowest-mass host halos considered in this work, and therefore simply
  turning off squelching provides a good approximation to implementing
  this lower filtering mass. We therefore turn squelching off before
  conducting our experiments with other aspects of the model. 

Hot halo gas is assumed to be distributed in an isothermal
sphere at the halo's virial temperature. Halo gas cools
through collisionally excited atomic lines as described in S08, based
on the model originally proposed in \citet{white:91}.  All cooled gas is added to the
cold disk of the central galaxy.  When halos become satellites, they
are stripped of their hot gas and their ejected reservoir and are not
allowed to accrete any more gas.  The stripped gas is added to the
central galaxy's hot gas halo.  Gas in the ejected gas reservoir is
allowed to re-accrete into the hot gas halo at a rate given by:
\begin{equation}
\dot{M}_{\rm{ReIn}} =\chi_{\rm{ReIn}}  \left(\frac{M_{\rm{eject}}}  {t_{\rm{dyn}}}\right)
\label{eq:re-infall}
\end{equation}
Here $\dot{M}_{\rm{ReIn}}$\ is the rate at which gas falls into the
hot halo from the ejected gas reservoir, $M_{\rm{eject}}$\ is the mass
in the ejected gas reservoir, $t_{\rm{dyn}}$\ is the central halo's dynamical
timescale, and \chirein\ is the efficiency parameter with a default
value of \chirein$=0.1$. When halos merge, the ejected reservoirs from
all but the largest progenitor are added to the hot gas reservoir of
the new host halo.

%----------------------------------------------------------------------------------------

\subsection{Star formation}
\label{sec:star_formation}
Stars form in a ``normal'' (disk) mode and in merger-driven starbursts.  Heavy elements are generated with a fixed yield per stellar mass formed and recycled instantaneously.  Details on the collisional starburst treatment may be found in S08.  These do not affect the work in this paper significantly since the star formation density due to bursts is about an order of magnitude below that due to normal star formation over most of the age of the universe (see Fig.~14 in S08). The model assumes that both the cold gas and stars in the disk are distributed with radial exponential profiles with separate scale lengths related by a factor, $r_{\rm{gas}}=\chi_{\rm{gas}}r_\star$, with the fiducial value $\chi_{\rm{gas}}=1.7$. The scale length of the gas disk is calculated using angular momentum conservation arguments \citep{mo_mao_white:98, somerville_barden:08} based on the halo spin parameter.

Normal star formation in the fiducial model follows a Kennicutt-Schmidt relation \citep{kennicutt:98}:
\begin{equation}
\dot{\Sigma}_{\rm{SFR}}=\frac{A_{\rm{K}}}{\tau_\star}\Sigma_{\rm{gas}}^{N_{\rm{K}}}
\label{eq:kennicutt}
\end{equation}
Here, \tst\ is a dimensionless free parameter with a fiducial value of $\tau_\star=1.5$~and $\Sigma_{\rm{gas}}$\ is the surface density of the cold gas disk.  The values of $A_{\rm{K}}$\ and $N_{\rm{K}}$\ are set by observations to be $A_{\rm{K}}=0.167$\ \msun\ yr$^{-1}$kpc$^{-2}$\ and $N_{\rm{K}}=1.4$.  At each time step, the model applies the Kennicutt law to all cold gas with surface density greater than a fixed critical value, $\Sigma_{\rm{crit}}=6$\ \msun\ pc$^{-2}$.  

In some cases, to interpret the effects of the modified recipes more easily, we use a constant efficiency star formation recipe instead of the Kennicutt law.  For this, we let the star formation timescale be a free parameter and set
\begin{equation}
\dot{M}_\star=\frac{M_{\rm{cold}}}{\tau_{\rm{CE}}}
\label{eq:constsfr}
\end{equation}
We take the default value of the timescale to be $\tau_{\rm{CE}}=10^9$~yr. 

%----------------------------------------------------------------------------------------

\subsection{Stellar feedback}
\label{sec:snfb}
Massive stars and supernovae reheat some of the cold gas following star formation and deposit it in either the hot gas halo or in the ejected gas reservoir. The fraction of the reheated gas which is ejected is determined by the halo virial velocity with the parameter $V_{\rm{eject}}$\ setting the transition from mostly ejected at lower $V_{\rm{vir}}$~to mostly retained in the hot halo at higher $V_{\rm{vir}}$.  In the fiducial model, $V_{\rm{eject}}=130$ km/s. The total mass of gas reheated is given by:
\begin{equation}
\dot{M}_{\rm{RH}}=\esneq \left(\frac{V_{\rm{circ}}}{200~\text{km/s}}\right)^{-\arheq} \dot{M}_\star
\label{eq:snfb}
\end{equation}
where $V_{\rm{circ}}$\ is the circular velocity of the disk, defined as the maximum rotation velocity of the dark matter halo, and \arh\ and \esn\ are dimensionless free parameters.  In the fiducial model, \arh=2.2 and \esn=1.5.

%----------------------------------------------------------------------------------------

\subsection{Main free parameters}
\label{sec:fid_main_free}
The parameters most relevant to the properties of low mass galaxies are the re-infall rate normalization \chirein, the star formation normalization \tst\ and critical surface density $\Sigma_{\rm{crit}}$, and the stellar feedback parameters, power \arh, normalization \esn, and ejection/retention transition velocity $V_{\rm{eject}}$.  A summary of how these parameters enter into the recipes can be found in Sec.~\ref{sec:gas_stuff} for the re-infall parameter, Sec.~\ref{sec:star_formation}~for star formation parameters, and Sec.~\ref{sec:snfb}~for the stellar feedback parameters.  Complete descriptions can be found in S08.

Some of the fiducial model's parameters correspond to values that can be derived from observations or numerical simulations and are set to those values.  Others are not directly measurable and are adjusted so that the simulated galaxy population matches certain sets of observations.  The stellar feedback parameters were tuned to match the low mass end of the SMF.  The fiducial values are \arh=2.2, \esn=1.5, and $V_{\rm{eject}}=130$\ km/s.  The value of \chirein\ is degenerate with the wind mass-loading parameters, so the fiducial model adopts \chirein$=0.1$, the minimal value that allows the model to fit both cluster baryon fractions and the mass function of $z=0$\ low mass galaxies.  For normal star formation, the values of $N_{\rm{K}}=1.4$\ and $A_{\rm{K}}=0.167$\ \msun~yr$^{-1}$kpc$^{-2}$~in the star formation law (Eqn.~\ref{eq:kennicutt}) are taken from observations \citep{kennicutt:98}.  The fiducial value of \tst\ is set to be 1.5 to match observed cold gas fractions.  The value $\Sigma_{\rm{crit}}=6$\ \msun\ pc$^{-2}$ is consistent with direct observations and reproduces the observed turn-over in the relationship between SFR density and total gas density \citep{bigiel:08}, as well as reproducing gas fractions in low mass galaxies.

%====================================================================
%====================================================================

\section{Properties of the fiducial model}
\label{sec:fiducial}
The fiducial model is tuned by hand to match a subset of
$z=0$\ observations, the SMF in particular, and does fairly well at
matching a larger set of $z=0$\ observations
\citep{somerville:08,somerville:12}. However, as already discussed and
shown in \citet{fontanot:09} and \citet{lu_candels:14}, it suffers
from the usual set of dwarf galaxy problems.  This can be
seen in observables such as the SMF, cold gas fractions, sSFR, and
metallicities. In this section, we show the predictions of our
fiducial model for the set of properties which are most
enlightening.  These are $f_* \equiv M_\star/M_{\rm{H}}$
(Fig.~\ref{fig:fiducial_fstar}), the stellar mass function (SMF)
(Fig.~\ref{fig:fiducial_smf}), the cold gas fraction in disks
(Fig.~\ref{fig:fiducial_cgas}), the specific star formation rate
(sSFR), $\dot{M}_\star/M_\star$ (Fig.~\ref{fig:fiducial_ssfr}), and
the ISM metallicity (Fig.~\ref{fig:fiducial_zgas}).  These properties
are shown with the scatter about the median: the ``$\pm 1
\sigma$~region" which contains 68\% of the galaxies in the model.  The
reader should keep in mind that all of the simulations run for this
paper have AGN feedback switched off, including the fiducial model.
This means the high mass galaxies will not necessarily match
observations.  In addition, as we discuss above, the
implementation of squelching in the fiducial model is most likely too
aggressive so we also show the fiducial model without squelching in
these figures.

%----------------------------------------------------------------------------------------
\subsection{Stellar mass function and \fstar}
\label{sec:fid_fstar_smf}

\begin{figure*}
\centering
\includegraphics[width=\textwidth]{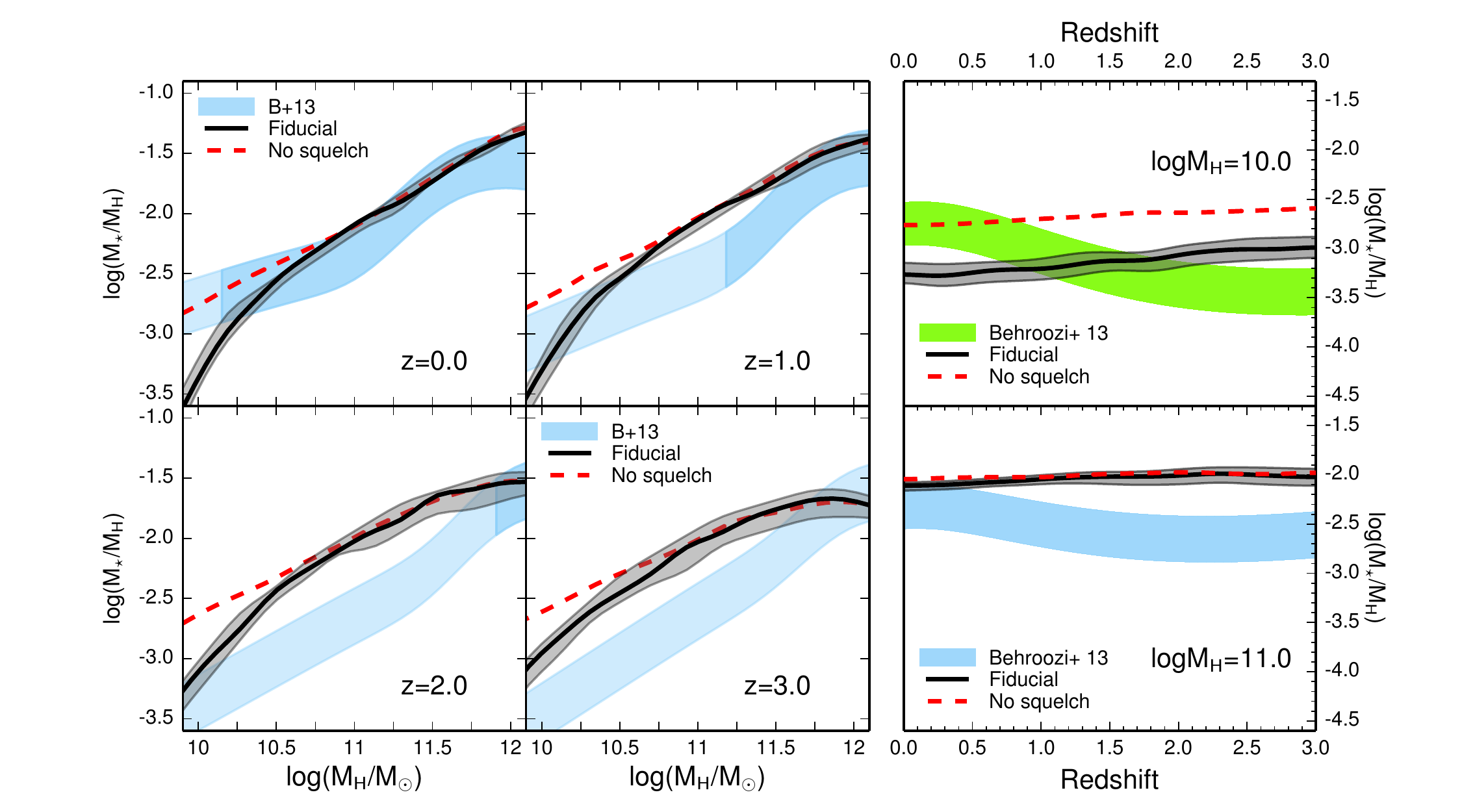}
\caption{Ratio of stellar mass to halo mass (\fstar) in the fiducial model.  In all panels, the fiducial model is shown in black with the $\pm1\sigma$~region shaded gray. The red dashed line shows the median of the fiducial model with photo-ionization squelching switched off (see text).  Left panel: \fstar\ as a function of halo mass shown for four redshifts.  Empirical constraints on \fstar\ from \citet{behroozi:13}~are shown by their $\pm1\sigma$~region shaded blue where there are observational constraints and light blue where the relation is extrapolated.  Right panel: \fstar\ as a function of redshift for halo mass $M_{\rm{H}}=10^{10}$\ \msun\ in the top panel and halo mass $M_{\rm{H}}=10^{11}$\ \msun\ in the lower panel.  \citet{behroozi:13}~results are shown as shaded regions indicating $\pm1\sigma$\ region.
\label{fig:fiducial_fstar}}
\end{figure*}

\begin{figure*}
\centering
\includegraphics[width=\textwidth]{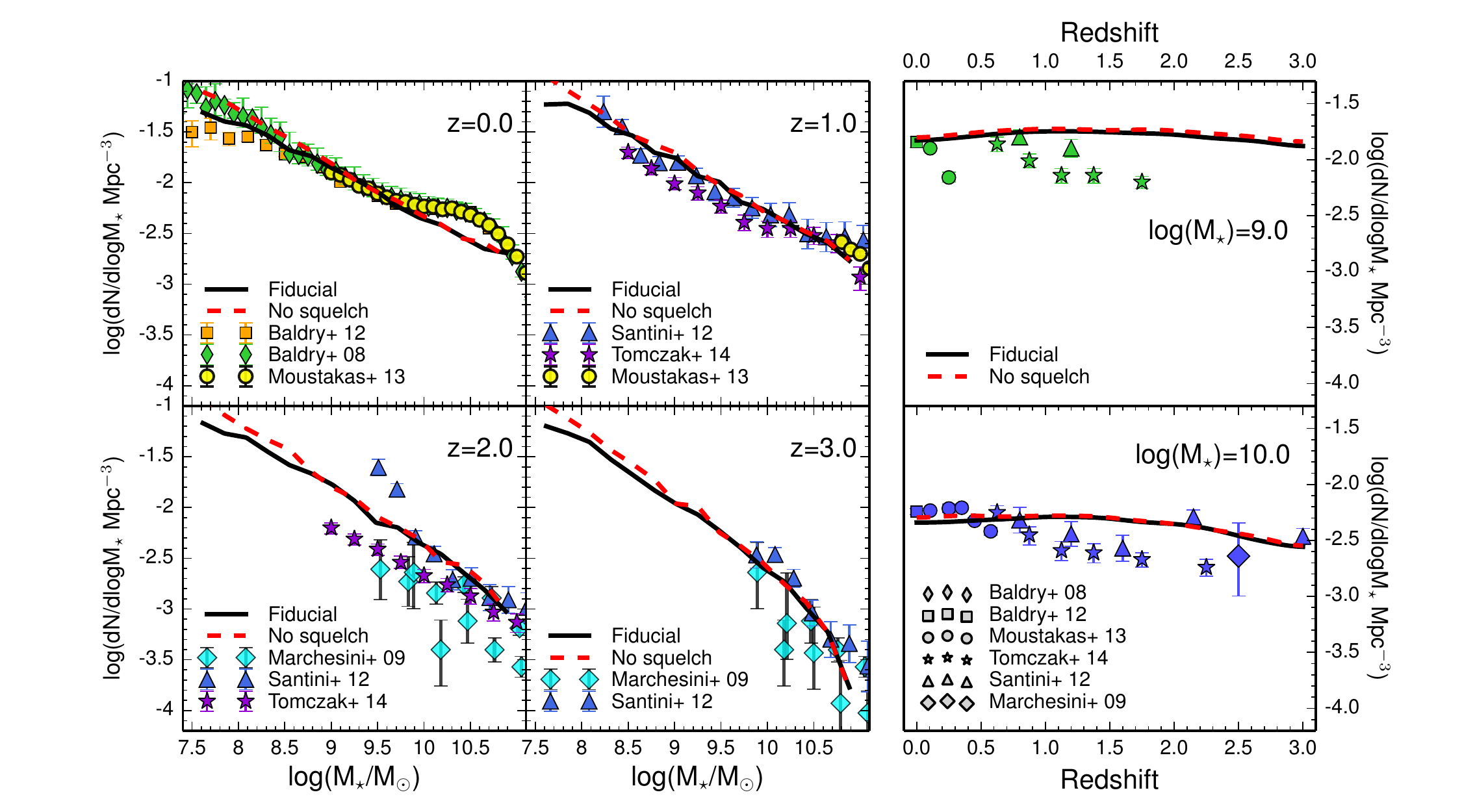}
\caption{Stellar mass function for the fiducial model. In all panels, the fiducial model is shown in black and the red dashed line shows the fiducial model with photo-ionization squelching switched off (see text).  Left panel: stellar mass functions for four redshifts.  Colored points show observations from \citet{baldry:08} as green diamonds, \citet{baldry:12} as orange squares, \citet{moustakas:13} as yellow circles, \citet{santini:12} as dark blue triangles, \citet{tomczak:14} as purple stars, and \citet{marchesini:09} as wide cyan diamonds.  Right panel: the number densities as a function of redshift for galaxies with \mstar$=10^9$\ \msun\ in the top panel and \mstar$=10^{10}$\ \msun\ in the lower panel.  Data are shown as points with the same shape as in the left panel.
\label{fig:fiducial_smf}}
\end{figure*}

\begin{figure*}
\centering
\includegraphics[width=\textwidth]{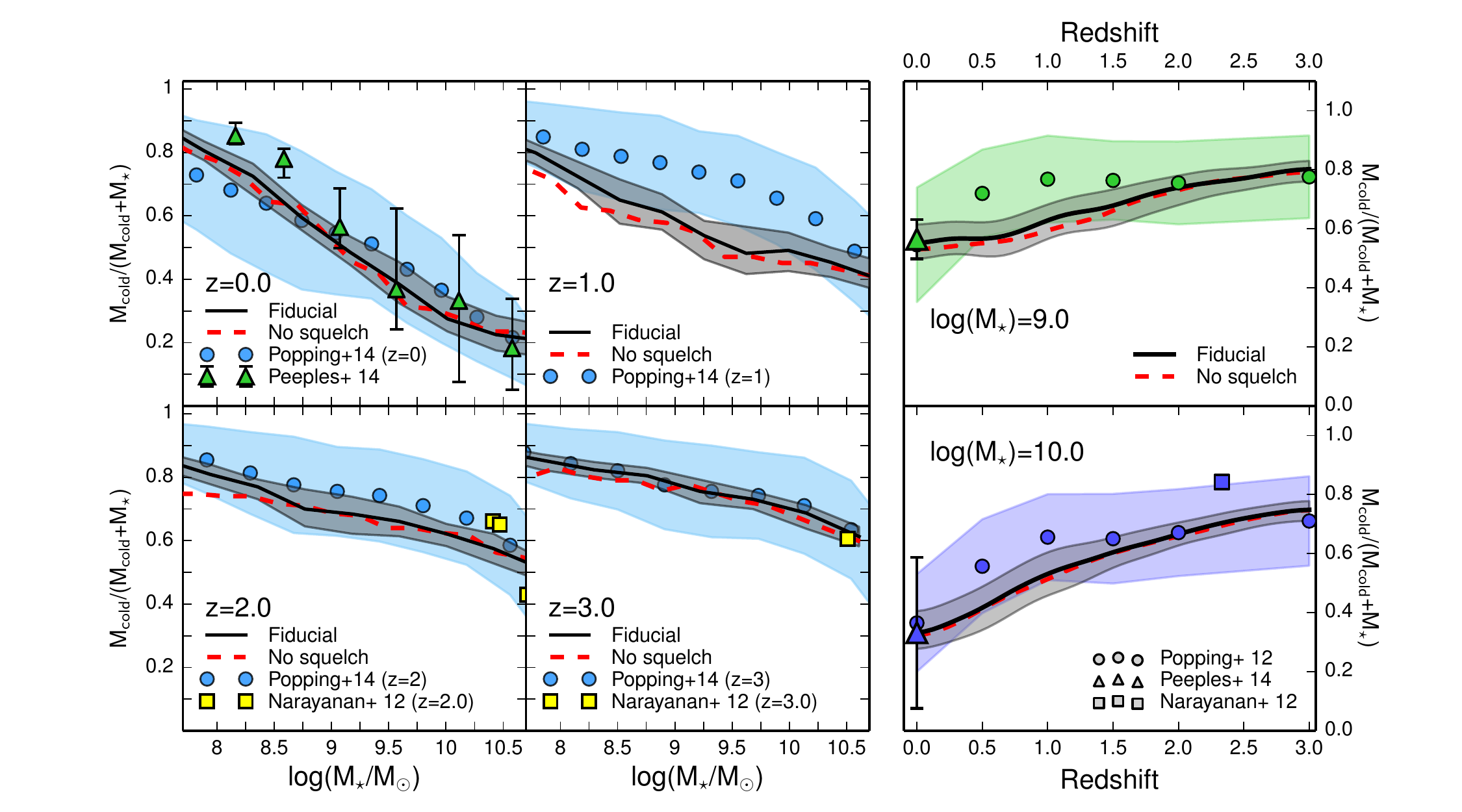}
\caption{Cold gas fractions for the fiducial model. In all panels, the fiducial model is shown in black with the $\pm1\sigma$~region shaded gray. The red dashed line shows the median of the fiducial model with photo-ionization squelching switched off (see text).  We show only galaxies with nonzero gas fractions and bulge to total ratios B/T\textless0.4.  Left panel: the cold gas fraction shown as a function of stellar mass at four redshifts.   The \citet{peeples:14}~points, shown as green triangles in the $z=0$\ panel, are averages of a collection of data sets in stellar mass bins.  We also show the indirect cold gas fraction estimates from the \citet{popping:14gasfrac} empirical model as blue circles with the $\pm1\sigma$~region shaded light blue and direct estimates of molecular gas fraction from \citet{narayanan:12} as yellow squares. Note that the \citet{narayanan:12} points are individual galaxies rather than binned results.  See Sec.~\ref{sec:fid_coldgas} for a more complete discussion of cold gas observations.  Right panel: gas fractions as a function of redshift for galaxies with stellar mass \mstar$=10^9$\ \msun\ in the top panel and \mstar$=10^{10}$\ \msun\ in the lower panel.  Data are shown as points with the same shape as in the left panel and the errors on the \citet{popping:14gasfrac} gas fractions shown as colored shaded regions.
\label{fig:fiducial_cgas}}
\end{figure*}

The fiducial model is tuned to approximately match the
$z\sim0$\ observations of the SMF.  However, to match the ``kink'' in
the SMF precisely at low masses ($M_\star \lesssim
10^{9.5}$\ \msun), we would need to adopt a more complicated scaling
for the mass-loading factor than our single power-law
\citep{lu_candels:14}. Our fiducial model lies within the error bars
on the SMF observations of \citet{baldry:08} for $M_\star \gtrsim
10^{8}$\ \msun.  We also compare to the observed SMFs from
\citet{baldry:08, baldry:12, santini:12, moustakas:13, tomczak:14,
  marchesini:09}, and to \citet{behroozi:13}, who calculate the SMF as
a function of redshift with their subhalo abundance matching, fit to a
large compilation of observed SMFs
(Fig.~\ref{fig:fiducial_smf}). We also compare to the
  \citet{behroozi:13}~\fstarMh\ relation, which we expect to match
  about as well as we match the SMF since \fstarMh\ is derived from
  the SMF (Fig.~\ref{fig:fiducial_fstar}).

The fiducial model fits the observed \fstar\ and SMF well
over the range of masses of interest, $10^{10} M_\odot \lesssim
M_{\rm{H}} \lesssim 10^{11.5} M_\odot$~at $z=0$.  The high mass end of
the \fstar\ relation shown in Fig.~\ref{fig:fiducial_fstar}~is high at
$z=0$\ due to the absence of AGN feedback in these simulations. Note
that the \emph{deficit} of galaxies in the SMF at stellar masses
$M_\star \gtrsim 10^{10}$\ \msun~is also due to the absence of AGN
feedback -- some of these galaxies should arise from more massive
halos, due to the ``turnover'' in \fstar\ at larger halo masses (see
e.g.~S08).  The model's sharp decrease in \fstar\ at the lowest masses
(below $M_{\rm{H}} \sim 10^{10.5}$\ \msun) is due to photo-ionization
squelching of low mass halos as described in \ref{sec:gas_stuff}.

At higher redshift, both \fstar\ and the SMF show an excess of stellar mass, increasingly so towards $z\sim 2$.   Constraints on \fstar\ for low mass galaxies show that for a fixed halo mass, the stellar mass increases over time, whereas the SAM predicts that stellar mass at a fixed halo mass decreases: the dark matter halos of modeled low mass galaxies grow slightly faster than the stellar mass.  Additionally, the slope of the SAM's \fstarMh\ relation becomes steeper from high redshift to low, indicating that at high redshift, the overall efficiency of star formation is not suppressed enough in the lower mass halos relative to the higher mass halos.  The model's SMF shows an excess of all low mass galaxies towards higher redshift because the SAM galaxies at a particular stellar mass have lower halo mass due to their too-high overall star formation efficiency, and therefore reside in more abundant halos than the observed galaxies.

\subsection{Cold gas fractions}
\label{sec:fid_coldgas}

At low redshift, direct estimates of cold gas content can be obtained from 21 cm emission which traces \hi, and CO emission which traces \htwo. We compare with the compilation of \citet{peeples:14}, which includes \hi\ and \htwo. At high redshift $z\gtrsim 0.2$, direct estimates of \htwo\ content from CO observations are available for only a small number of relatively massive galaxies. Therefore, we also compare to the \citet{popping:14gasfrac} work, which uses an empirical model to estimate gas fractions.  The \citet{popping:14gasfrac} work uses a subhalo abundance matching procedure to determine typical star formation rates as a function of halo mass and redshift.  They invert these typical star formation rates with an empirical molecular hydrogen-based star formation law to find the gas mass in \hi\ and \htwo\ assuming gas distributions and using a pressure-based recipe dictating the ratio of molecular to atomic hydrogen.  We also plot estimates of the \htwo\ fraction for individual galaxies from \citet{narayanan:12}, which are obtained from a re-analysis of CO measurements with a more sophisticated model for the conversion of CO emission to \htwo\ mass than the standard single CO to total gas ratio.

Cold gas fractions in the fiducial model match well with observations at $z=0$\ (Fig.~\ref{fig:fiducial_cgas}), unsurprisingly as the model was tuned to match these data. However, at intermediate redshifts $0.5<z<1.5$, gas fractions are too low for galaxies with $M_\star \lesssim 10^{10} M_\odot$, as also noted in \citet{popping:14}.  This may be another symptom of the fact that these galaxies are forming stars too early.  At higher redshifts, $z\gtrsim2$, the fiducial model's cold gas fractions again match up with the predictions from the empirical model.  It is important to note that the galaxies included in the cold gas fraction plots are selected to be late type (bulge stellar mass to total stellar mass ratio of less than 0.4) and have non-zero cold gas mass.

\subsection{Specific star formation rates}
\label{sec:fid_ssfr}
\begin{figure*}
\centering
\includegraphics[width=\textwidth]{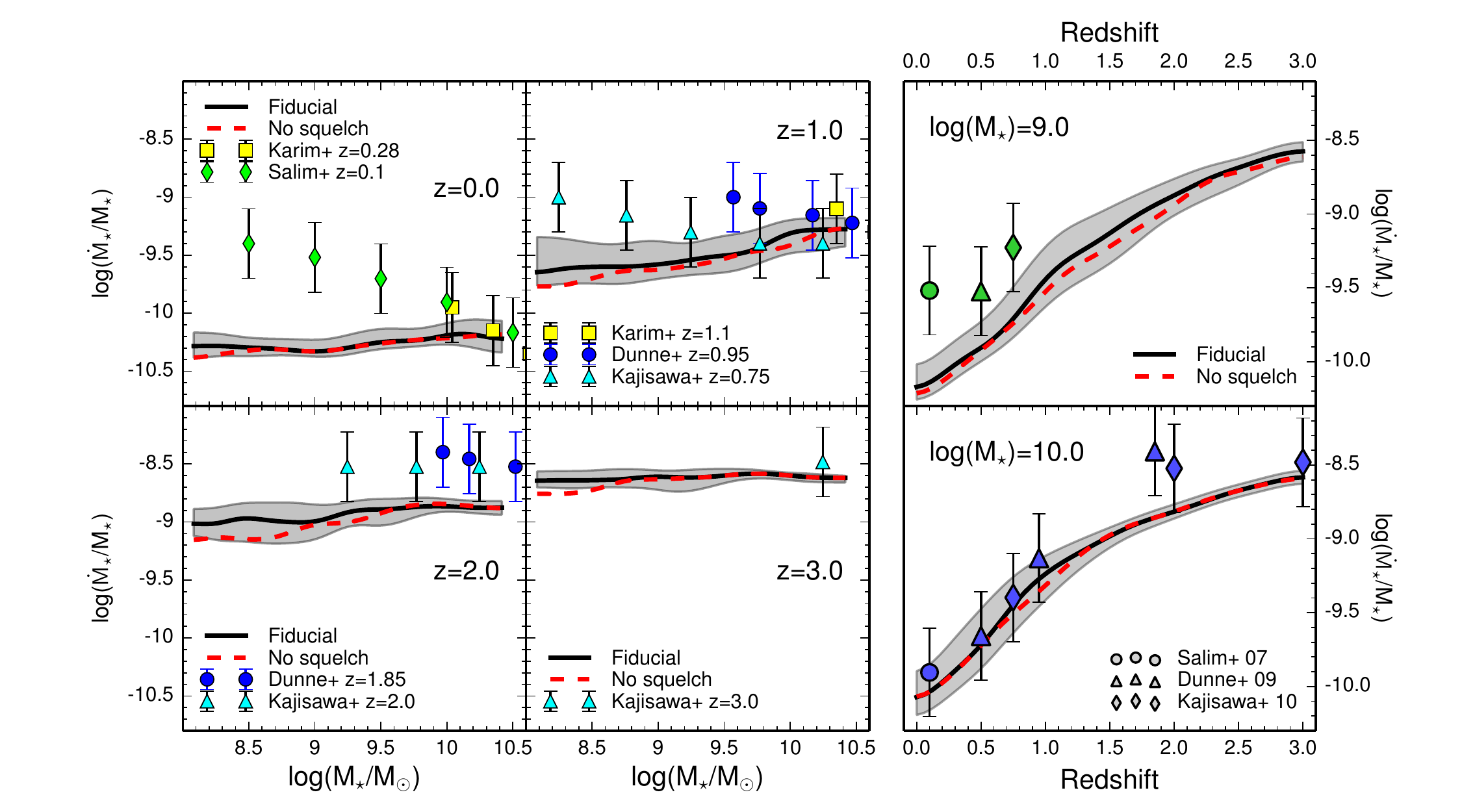}
\caption{Specific star formation rates ($\dot{M}_\star/M_\star$) for the fiducial model.   In all panels, the fiducial model is shown in black with the $\pm1\sigma$~region shaded gray. The red dashed line shows the median of the fiducial model with photo-ionization squelching switched off (see text).  Left panel: specific star formation rates vs.~stellar mass for four redshifts.  Points show data from the sources given in the legends.  Note that the errors on the sSFR data are not from the original works but from \citet{behroozi:13}, calculated using many data sets to estimate any systematic errors unaccounted for in original sources.  Right panel: sSFR as a function of redshift for \mstar$=10^9$\ \msun\ in the top panel and \mstar$=10^{10}$\ \msun\ in the lower panel.  The shapes of the data points denote the source.
\label{fig:fiducial_ssfr}}
\end{figure*}

Fig.~\ref{fig:fiducial_ssfr} shows the average sSFR vs.~stellar mass and redshift in the fiducial model.  The ``star forming sequence'' in observations has a slightly negative slope: the lowest mass galaxies have somewhat higher sSFRs than intermediate mass galaxies.  In the fiducial model, sSFRs of low mass galaxies are too low and the SF sequence is flat or even has a positive instead of negative slope. The fiducial model does more or less match the observed increase in sSFR with increasing redshift, but the normalization is too low for low mass galaxies from $0 \lesssim z \lesssim 1$.  We compare to observations from \citet{salim:07, dunne:09, kajisawa:10, karim:11}.

\subsection{Metallicity}
\label{sec:fid_zgas}
\begin{figure}
\centering
\includegraphics[width=\columnwidth]{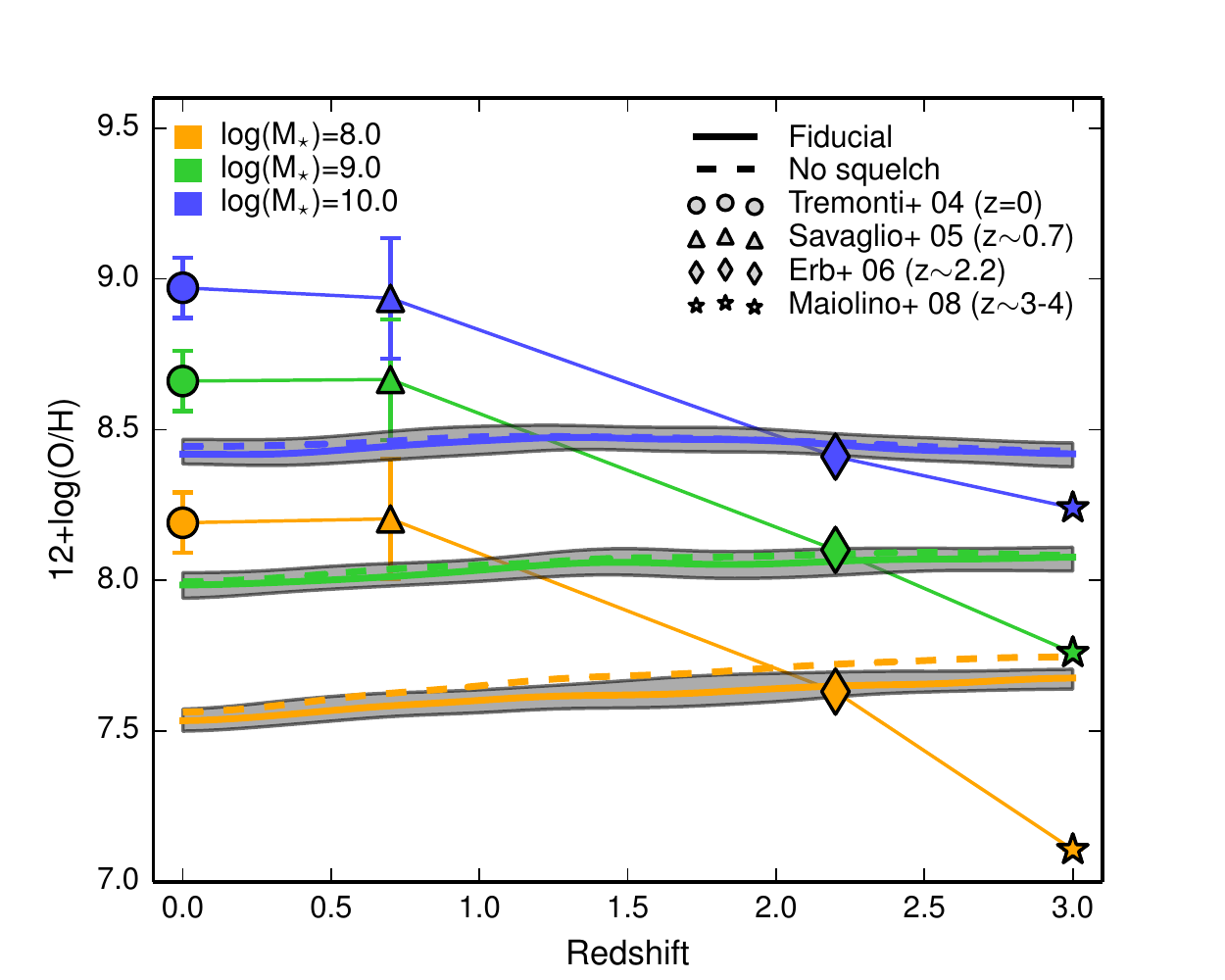}
\caption{Gas phase metallicities for galaxies in the fiducial model
  for selected stellar masses as a function of redshift.  Only
  galaxies with gas fractions greater than 0.2 are included.
  Metallicities are color coded according to the stellar mass bin they
  represent.  Fiducial model curves are shown as thick lines with shaded gray
  regions giving $\pm1\sigma$\ and observations are shown as points
  whose shapes indicate which data set they represent. Dashed lines
  show the fiducial model with squelching switched off.
\label{fig:fiducial_zgas}}
\end{figure}

Fig.~\ref{fig:fiducial_zgas} shows the gas-phase metallicity as a function of redshift for three different stellar mass bins.  The metallicity estimates at $z\sim0$\ are from \citet{tremonti:04}, obtained using photo-ionization and stellar population evolution models fit to SDSS spectroscopy. At $z\sim1$, we use the estimates of \citet{savaglio:05}, derived from Gemini Deep Deep Survey spectra with the $R_{23}$\  method. At $z\sim 2$, we use the results of \citet{erb:06}, from H$\alpha$\ and N\textsc{II} in spectra of star-forming galaxies, and at $z\sim 3$, the results of \citet{maiolino:08}, using strong line diagnostics such as H$\beta$, O\textsc{II}, O\textsc{III}, and Ne\textsc{III}.  We note that the absolute normalization of the metallicity from different indicators is highly uncertain, which may impact the redshift evolution implied by the observations shown here \citep{kewley:08}. We also note that the chemical yield in our model ($y= 1.5$\ in solar units) was chosen to match the \emph{stellar} metallicity of Milky Way mass galaxies at $z\sim 0$\ from observations \citep{gallazzi:05}. The gas-phase mass-metallicity relation predicted by our models is a steeper function of stellar mass than the observed relations, and is a pure power-law in stellar mass, unlike observations which turn over at high mass. The fiducial model also builds up metals very early, predicting almost no evolution in the metallicity of gas in galaxies at fixed stellar mass, or even a slight decrease.  This disagrees with the trend implied by the observations taken at face value, which suggest an increase of more than an order of magnitude in metallicity at fixed stellar mass since $z\sim 3$.

%====================================================================
%====================================================================

\section{Exploration of existing parameter space}
\label{sec:param_space}
Before changing the model recipes, we examine how the fiducial model
responds to parameter variations by running the model with all
  parameters set to the fiducial values, except for one quantity which
  is set to a fixed value significantly higher or lower than the
  fiducial value.  Here we look at the effects of changing
four parameters: the stellar feedback parameters \arh\ and \esn, the
star formation efficiency parameter \tst, and the re-infall efficiency
\chirein.  The other set of parameters we examined, \vej,
$\chi_{\rm{gas}}$, and $\Sigma_{\rm{crit}}$, do not have very much
impact on the quantities of interest.  A more quantitative
illustration of the effects of changing these parameters is presented
in Appendix \ref{sec:param_appendix}.

Adjusting the stellar feedback parameters strongly affects the ratio of stellar mass to halo mass, \fstar.  The amount of gas reheated by stellar feedback is determined by Eqn.~\ref{eq:snfb} and the values of \esn\ and \arh.  The value of \esn\ controls the normalization of the stellar feedback relation.  Increasing \esn~decreases the total mass of stars formed independent of halo mass.  This is seen most clearly in the first row, second column of  Fig.~\ref{fig:all_grid}.  The cold gas fraction, $M_{\rm{cold}}/(M_{\rm{cold}}+M_\star)$, remains more or less the same at $z=0$~as \esn\ increases because both the cold gas and stellar masses decrease.  Specific star formation rates are also largely independent of \esn\ because stellar masses and star formation rates react similarly to changes in \esn.  These trends are illustrated in the second columns of Figs.~\ref{fig:all_grid}\ and \ref{fig:vs_z_grid}.

The value of \arh\ determines how much more strongly low mass halos are affected by stellar feedback than higher mass halos. Making the value of \arh\ larger leads to a steeper dependence of mass-loading on halo circular velocity, suppressing star formation in low mass halos more strongly relative to high mass halos.  This results in a steeper dependence of \fstar\ on halo mass and a reduced comoving number density of low mass galaxies relative to high mass galaxies.  This is most evident in the redshift $z=0$\ \fstarMh\ relation as shown in the first row, first column of Fig.~\ref{fig:all_grid}.  The slope of the sSFR-$M_\star$\ relation depends weakly on \arh, with larger values of \arh\ leading to lower values of sSFR in low mass galaxies, and therefore to flatter or more positive slopes in sSFR-$M_\star$.  Cold gas fractions are also impacted, with larger values of \arh\ producing higher gas fractions in low mass galaxies because with higher reheating rates, less gas can turn into stars. The value of \arh\ also affects the slope of the MZR, and to a lesser extent, its evolution; larger values of \arh\ lead to a steeper MZR and a slightly larger decline in gas-phase metallicity at fixed stellar mass with cosmic time. These relations are shown in the first columns of Figs.~\ref{fig:all_grid}\ and \ref{fig:vs_z_grid}.

In contrast, changing the star formation timescale \tst\ has almost no impact on the \fstarMh\ relation or SMF at $z\lesssim 3$\ as seen in the third columns of Figs.~\ref{fig:all_grid}\ and \ref{fig:vs_z_grid}. Changing \tst\ mainly impacts gas fractions, particularly for high mass galaxies.  This happens because as star formation efficiency decreases, gas mass builds up to compensate. Note that \tst\ multiplies a timescale so higher \tst\ means lower star formation efficiencies and higher gas fractions. Lower SF efficiency (higher \tst) therefore leads to a flatter relation between gas fraction and stellar mass since the gas supply in low mass galaxies is more modulated by the stellar feedback. For the most part, the slope of the gas fraction with redshift at fixed mass does not depend strongly on \tst, nor does the slope of the sSFR-$M_\star$\ relation or the redshift evolution of the sSFR at fixed mass.  

Modulating the re-infall timescale by varying \chirein\ mainly changes low redshift galaxy properties as shown in the right-most column of Fig.~\ref{fig:vs_z_grid}.  This is because re-infall timescales are long, so most re-accretion occurs at late times.  In the fiducial model, the re-infall timescale is constant with halo mass and increases with time.  Higher \chirein, meaning more efficient re-infall, increases the total stellar mass of all but the lowest mass galaxies due to the halo mass-independent increased availability of gas from re-infall as shown in the right-most column of Fig.~\ref{fig:all_grid}.  The lowest mass galaxies are unaffected by changes in \chirein\ primarily because of squelching.  Squelched galaxies have much or all of the in-falling IGM diverted to the ejected reservoir, rather than it falling into the hot halo as in high mass galaxies, significantly decreasing the accessibility of this gas relative to high mass halos no matter the \chirein.  Contributing to this effect, low mass halos eject a large fraction of their stellar-driven winds and winds have a high mass-loading factor.  Even with efficient re-infall, gas re-accreted onto low mass halos spends little time in the disk before being re-ejected, rendering the re-infall timescale largely irrelevant to squelched galaxies.  If squelching is turned off, this is no longer the case and high \chirein\ leads to higher stellar masses in low mass halos.

%====================================================================
%====================================================================

\section{Results with Modified Recipes}
\label{sec:ad_hoc}
The insight gained by varying the parameters in our current recipes
informs how we should alter our recipes.  For example, we have learned
that the slope of the mass-loading dependence on halo mass appears to
have the most leverage on \fstar, while modifying \fstar\ by changing
the star formation efficiency will require drastic measures because of
the strongly self-regulated nature of star formation in the
models. Note that the purpose of our experiments is to gain a
qualitative understanding of which physical scenarios are most
promising for solving all facets of the problem, as well as to gain
insights into the requirements for a solution.  As such, we do not
attempt to obtain precise fits to the observations.  In addition, we
find that models without squelching do a better job of reproducing the
normalization and evolution of low mass galaxies' \fstarMh, so we do
not include squelching in the models presented below.  This is not
unreasonable given the more recent work on squelching, which
  suggests that the filtering mass is much lower than the mass of the
  smallest host halos considered here (see
Sec.\ \ref{sec:gas_stuff}).  We also continue to omit AGN
  feedback.  Altered models run with AGN feedback included show the
  same behavior at low masses as the models shown but obscure any
  effects of the alterations on high mass galaxies.

\subsection{Preferential reheating: changing stellar feedback scalings}
\label{sec:pref_rh}
\label{sec:discussion_pref_rh}

\begin{figure}
\centering
\includegraphics[width=\columnwidth]{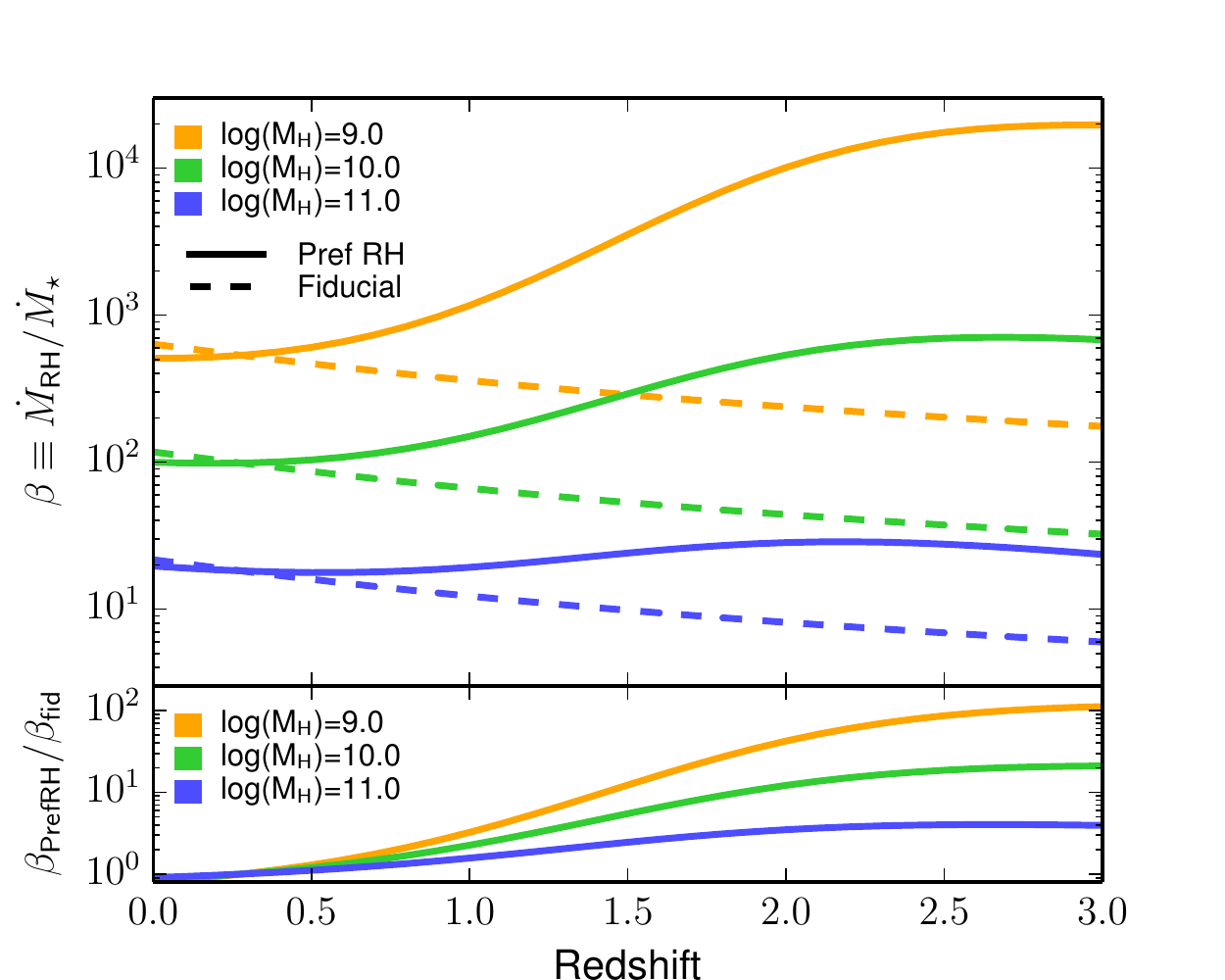}
\caption{A comparison of the fiducial mass-loading factor
  $\beta_{\rm{fid}} \equiv \dot{M}_{\rm{RH}}/\dot{M}_\star$~to the
  preferential reheating mass-loading factor $\beta_{\rm{pref RH}}$.
  The top panel shows the values of $\beta$~for three halo masses, solid
  lines for preferential reheating and dotted for fiducial.  The
  bottom panel shows the ratio $\beta_{\rm{pref
      RH}}/\beta_{\rm{fid}}$~for the same three halo masses.
\label{fig:alphavar}}
\end{figure}

\begin{figure*}
\centering
\includegraphics[width=\textwidth]{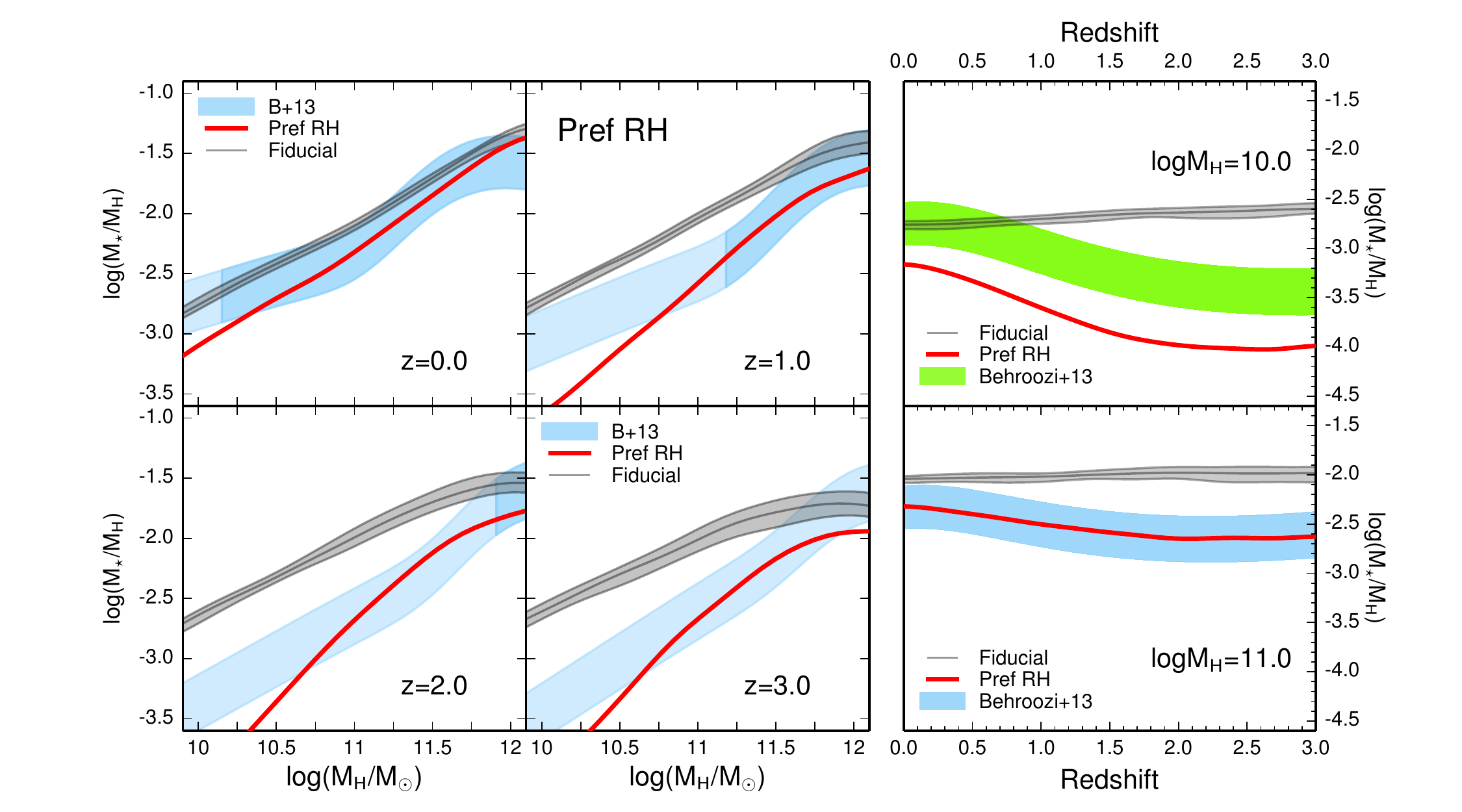}
\caption{Ratio of stellar mass to halo mass, \fstar, for the preferential reheating model.  In all panels, the preferential reheating model is shown in red and the median and $\pm1\sigma$\ region of the fiducial model are shown in gray.  Left panel: the \fstarMh\ relation for four redshifts.  Right panel: \fstar(z) for halo mass $M_{\rm{H}}=10^{10}$\ \msun\ in the top panel and halo mass $M_{\rm{H}}=10^{11}$\ \msun\ in the lower panel.  Constraints from \citet{behroozi:13} (colored shaded regions) are as described in Fig.~\ref{fig:fiducial_fstar}.
\label{fig:steep_fstar}}
\end{figure*}

\begin{figure*}
\centering
\includegraphics[width=\textwidth]{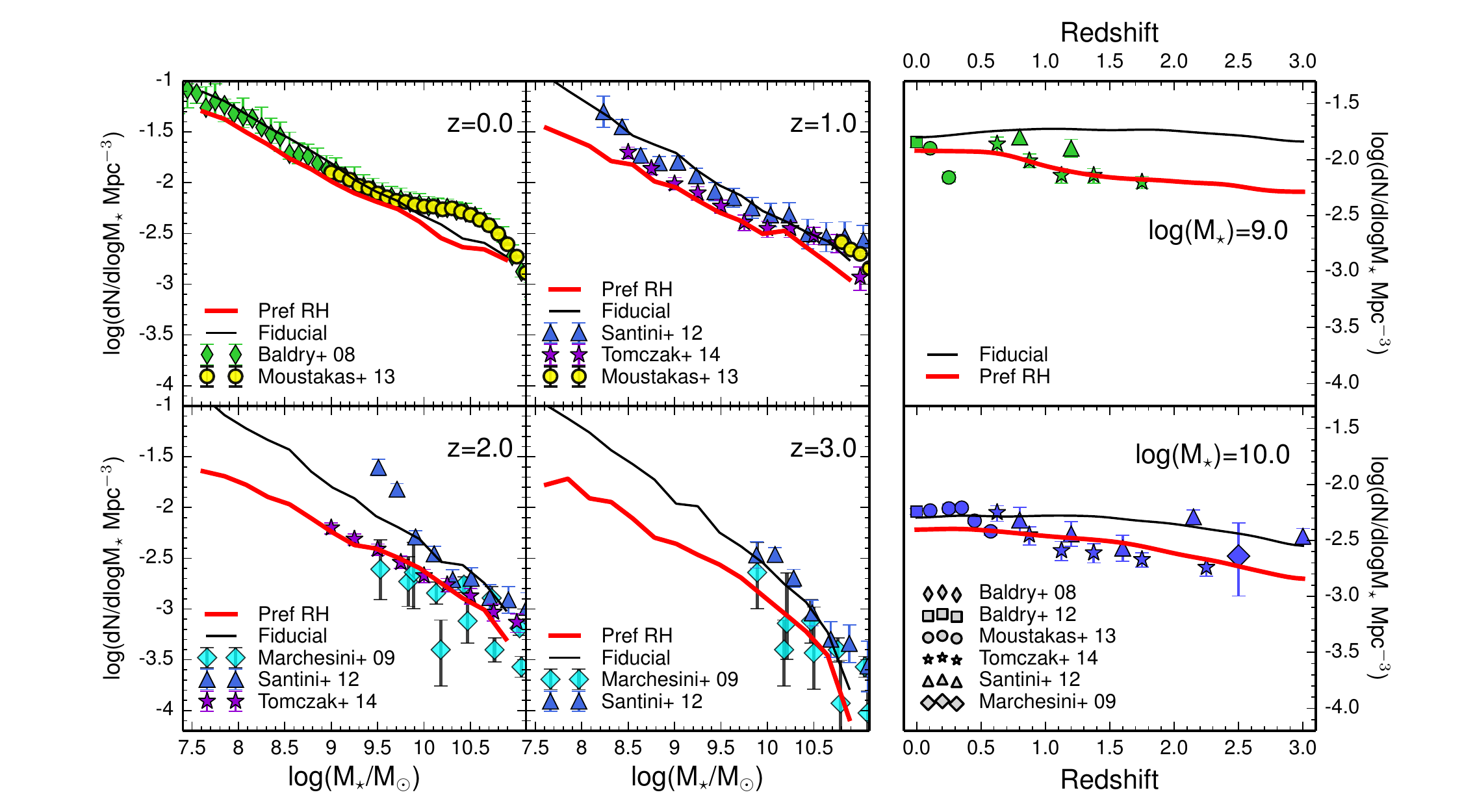}
\caption{Stellar mass function for the preferential reheating model. In all panels, the preferential reheating model is shown in red and the fiducial model is shown in black.  Data are as in Fig.~\ref{fig:fiducial_smf}.  Left panel: stellar mass functions for four redshifts. Right panel: number densities as a function of redshift for galaxies with \mstar$=10^9$\ \msun\ in the top panel and \mstar$=10^{10}$\ \msun\ in the lower panel.
\label{fig:steep_smf}}
\end{figure*}

\begin{figure*}
\centering
\includegraphics[width=\textwidth]{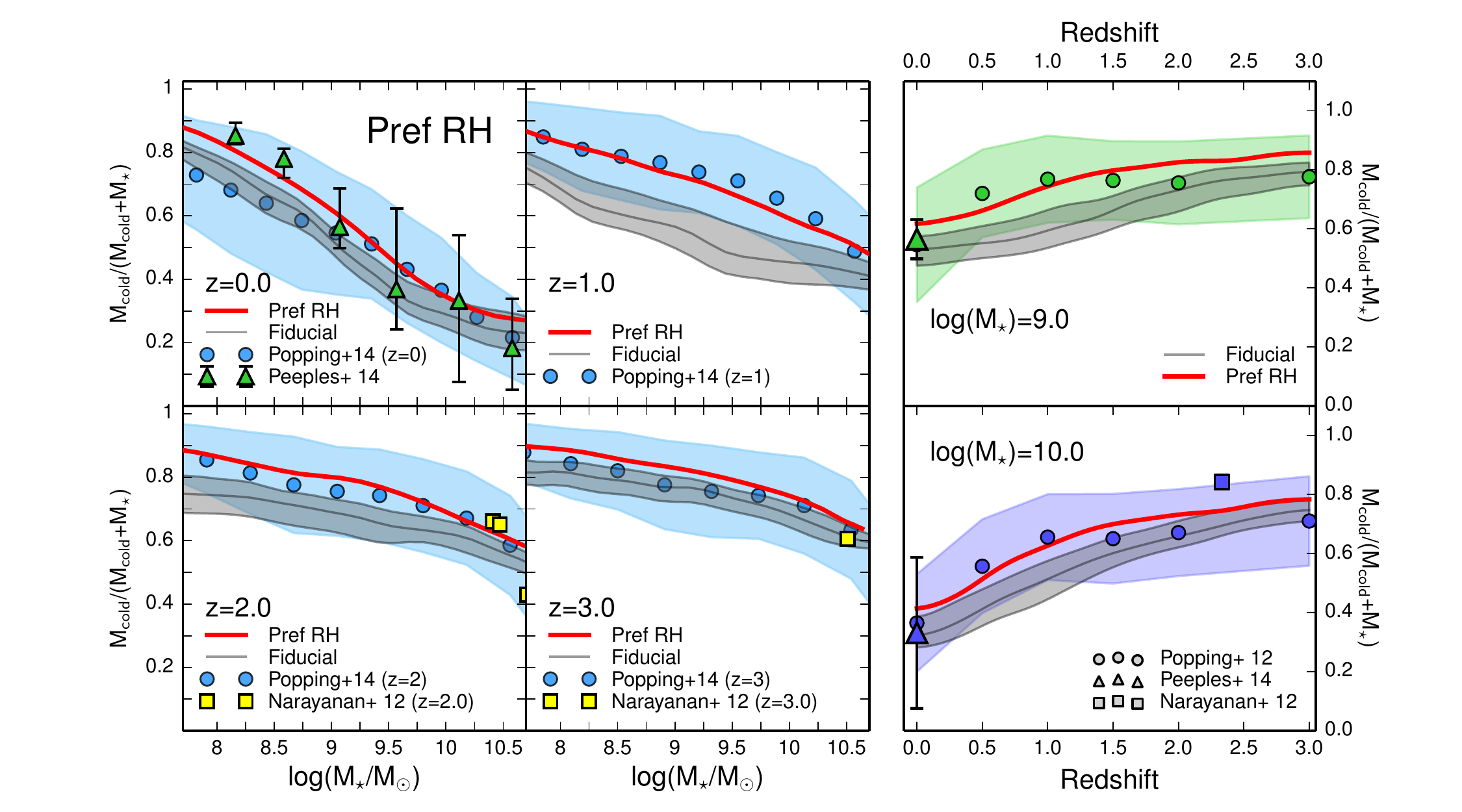}
\caption{Cold gas fractions for the preferential reheating model. In all panels, the preferential reheating model is shown in red and the median and  $\pm1\sigma$~region of the fiducial model are shown in gray.  Data are as in Fig.~\ref{fig:fiducial_cgas}.  Left panel: cold gas fraction as a function of stellar mass for four redshifts.  Right panel: cold gas fraction as a function of redshift for galaxies with stellar mass \mstar$=10^9$\ \msun\ in the top panel and \mstar$=10^{10}$\ \msun\ in the lower panel.  In all panels, only galaxies with nonzero gas fraction and bulge to total ratio B/T\textless0.4 are included.
\label{fig:steep_cgas}}
\end{figure*}

\begin{figure*}
\centering
\includegraphics[width=\textwidth]{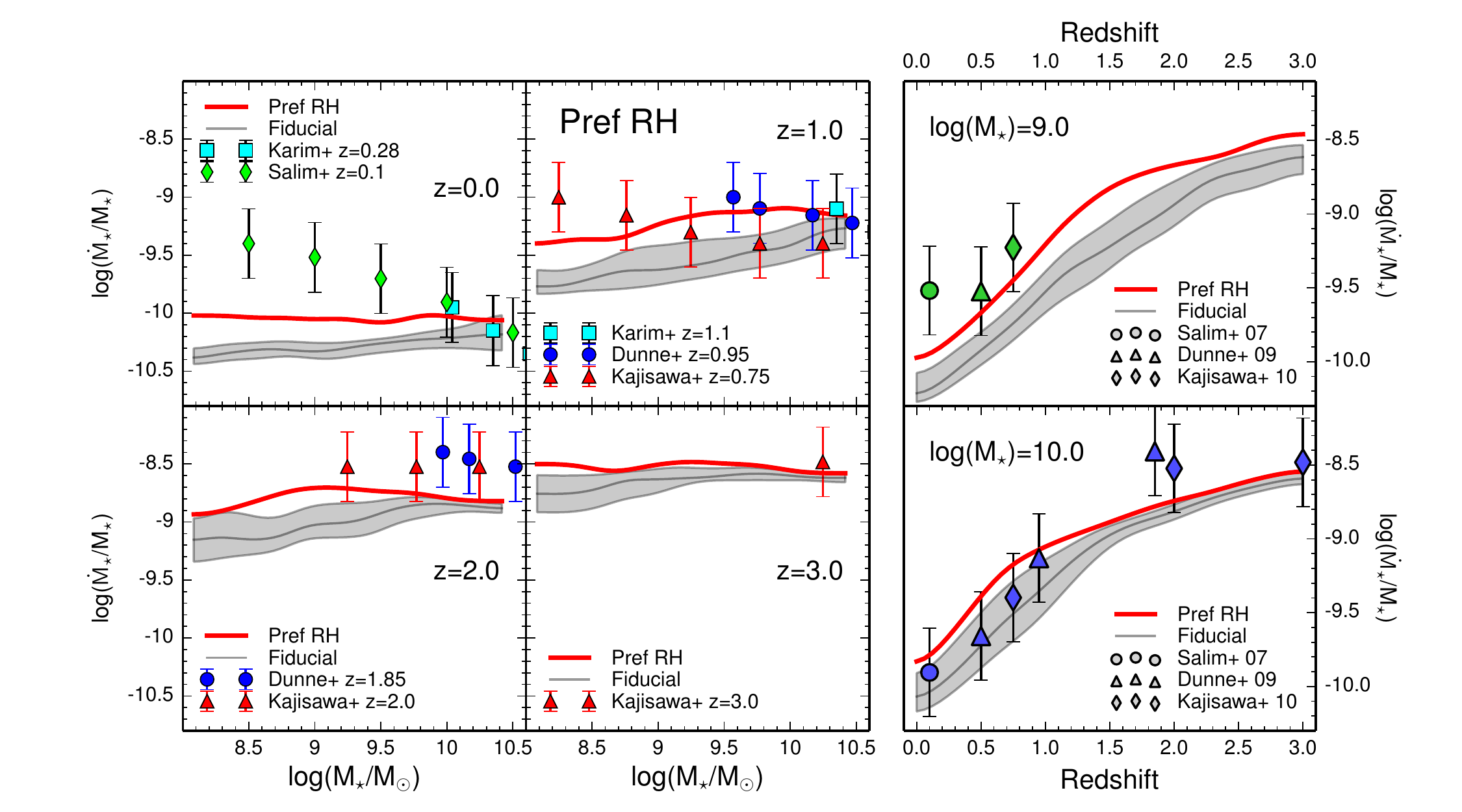}
\caption{Specific star formation rates ($\dot{M}_\star/M_\star$) for  the preferential reheating model.  In all panels, the preferential reheating model is shown in red and the median and $\pm1\sigma$~region of the fiducial model are shown in gray.  Data are as in Fig.~\ref{fig:fiducial_ssfr}.  Left panel: specific star formation rates as a function of stellar mass for four redshifts.  Right panel: specific star formation rates as a function of redshift for \mstar$=10^9$\ \msun\ in the top panel and \mstar$=10^{10}$\ \msun\ in the lower panel. 
\label{fig:steep_ssfr}}
\end{figure*}

\begin{figure}
\centering
\includegraphics[width=\columnwidth]{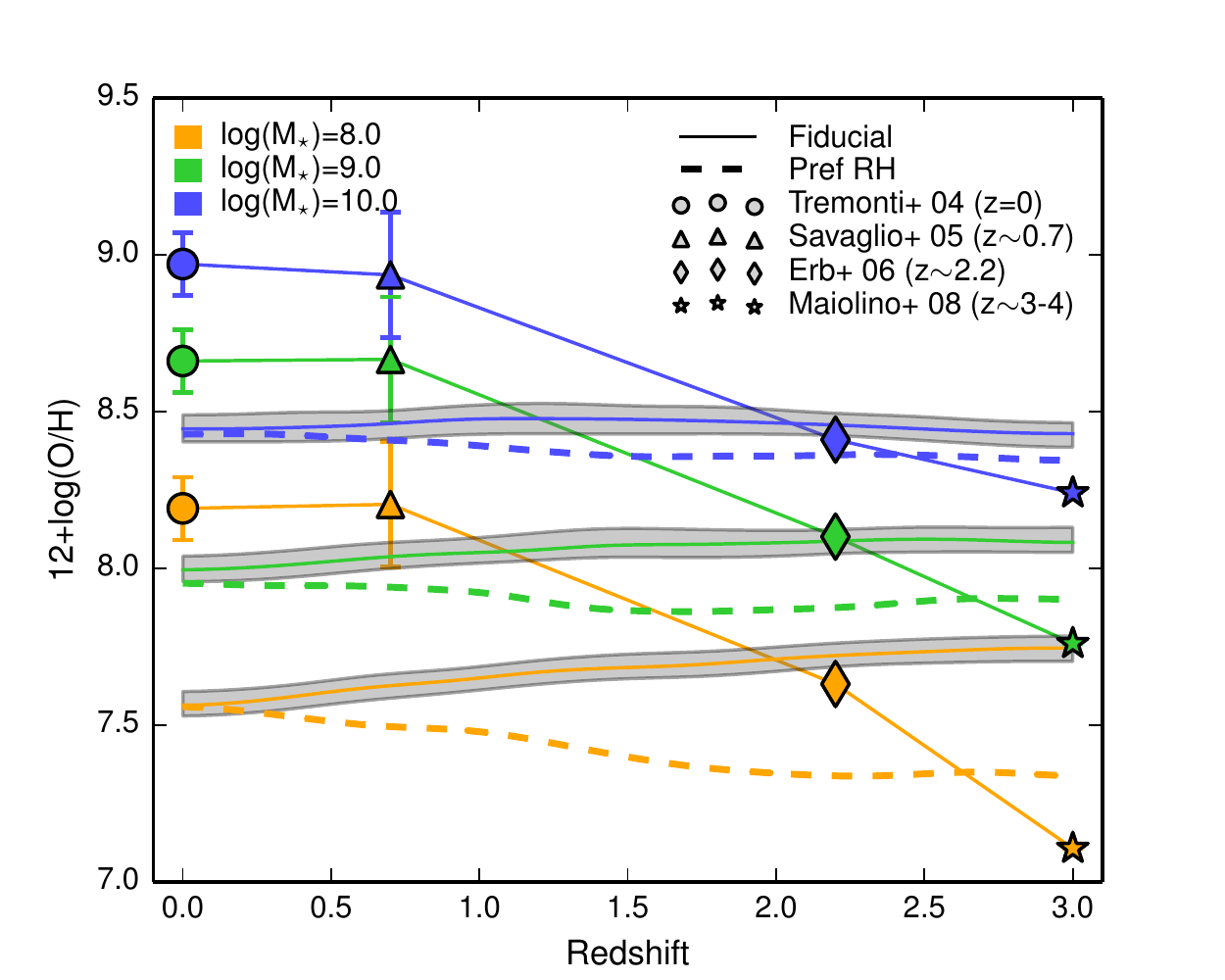}
\caption{Gas phase metallicities for galaxies in the preferential reheating model for selected stellar masses as a function of redshift.  Metallicities are color coded according to the stellar mass bin they represent. The thin solid lines and shaded gray regions show the fiducial model's median and $\pm1\sigma$~region and the thick dashed lines show the preferential reheating model.  Only galaxies with gas fractions greater than 0.2 are plotted and observations are shown as points whose shapes indicate which data set they represent.
\label{fig:pref_rh_zgas}}
\end{figure}

% alpha_rh governs the low mass fstar slope.  how?
As we have seen, the value of \arh\ in the stellar feedback recipe
(Eqn.~\ref{eq:snfb}) controls the slope of the low mass end of
\fstar. The efficacy of stellar feedback, or mass-loading factor
$\beta \equiv \dot{M}_{\rm{RH}}/\dot{M}_\star$, depends on the halo's
maximum circular velocity and thus the halo's mass.\footnote{The
  conversion from halo mass to circular velocity is redshift
  dependent.  At high redshift, the same circular velocity corresponds
  to a lower mass halo.  For instance, $V_{\rm{c}}\sim 160$ km/s
  corresponds to a $\sim 10^{12}$\ \msun\ halo at $z=0$\ or a $\sim
  10^{11}$\ \msun\ halo at $z=3$. \citep{somerville:99}} We expect
$\beta$\ to be mass dependent because ejecting cold gas from a galaxy
is much easier in low mass halos than in high mass halos: a parcel of
reheated gas has a much shallower potential well to climb out of in a
low mass galaxy than it would in a high mass one.  Because gas is
reheated more efficiently in small halos, less star formation is
required to reheat or eject the cold gas.  The higher the power
\arh~of $1/\rm{V}_{\rm{circ}}$~in Eqn.~\ref{eq:snfb}, the more drastic
this difference between low- and high-mass galaxies becomes and the
steeper the low-mass end of the \fstarMh\ relation becomes.
The original form of the supernova reheating recipe
  (Eqn.~\ref{eq:snfb}), $\dot{M}_{\rm{RH}} \propto
  V_{\rm{circ}}^{-\arheq} \dot{M}_\star$, originates in simple energy
  or momentum conservation arguments.  If each supernova produces
  energy $E_{\rm{SN}}$~and there are $N$~supernovae per solar mass
  formed, the energy available to reheat gas will be some fraction of
  the total energy: $E_{\rm{available}} \propto E_{\rm{SN}} N \Delta
  M_\star$.  If all reheated gas is brought to exactly escape
  velocity, the amount of gas that can be reheated is determined by
  $\Delta M_{\rm{RH}}v_{\rm{esc}}^2=2 E_{\rm{available}}$, giving
  $\Delta M_{\rm{RH}} \propto \Delta M_\star/v_{\rm{esc}}^2 \propto
  \Delta M_\star/V_{\rm{circ}}^2$.  The same argument made with
  momentum gives $\Delta M_{\rm{RH}} \propto \Delta
  M_\star/V_{\rm{circ}}$.

Our exploration of parameter space showed that no value of \arh\ that
is redshift independent can fit the observed evolution of \fstar.
In order to reproduce the observed trends, the slope of
  \fstarMh\ must be steep and therefore the value of \arh\ must be
  high at high redshift and lower at low redshift.  If \arh\ is left
  high, star formation is over-suppressed at low redshifts (see
  Fig.~\ref{fig:all_grid}~for a summary of the effects of constant
  high \arh).  We tried making \arh\ constant at low redshifts and
rising linearly to high redshift, but this over-suppressed high
redshift galaxies.  This over-suppression suggested that \arh\ could
not increase indefinitely towards high redshifts.  We also tried to
tie \arh\ to ISM metallicity rather than giving it an explicit
redshift dependence since it is plausible that the mass-loading factor
depends on the metallicity of the gas being reheated.  For
  example, it may be that higher metallicity implies faster cooling,
  which would leave less energy to drive winds and tend to make
  supernova reheating less efficient at lower redshifts.
Unfortunately, one of the symptoms of the problem we are trying to
solve is a mass-metallicity relation that evolves too slowly and gives
higher metallicity at high redshift than at low redshift, contrary to
observations.  The models based on metallicity behaved much like the
fiducial model because the variation in metallicity and therefore
\arh\ was minimal.

In order to fit the observed evolution of \fstar, we find that \arh\ must be very large at high redshift, \arh$\sim$4-5, then decline to around \arh$\sim 2$\ in a fairly narrow region around z$\sim1.5$, then stay approximately constant afterwards.  We parametrize \arh$(z)$\ as a hyperbolic tangent.
\begin{equation}
\arheq (z)=A {\rm{tanh}} \left(B \left( z-z_{\rm{trans}} \right) \right) + C
\label{eq:steepening}
\end{equation}
The values of $A$~and $C$~are determined by the minimum \arh\ ($\alpha_{\rm{min}}$) and the maximum \arh\ ($\alpha_{\rm{max}}$).  The value of $B$~dictates the sharpness of the transition between the high and low \arh; the higher the value of $B$, the more abrupt the transition.  The model we show has $\alpha_{\rm{min}}=2$, $\alpha_{\rm{max}}=4.5$, $z_{\rm{trans}}=1.5$, and $B=1$.  The mass-loading factor for this model is shown for three halo masses as a function of redshift in Fig.~\ref{fig:alphavar}.  This model reproduces several important trends in the evolution of \fstar\ that the fiducial model does not.  The slopes of the low mass end of the \fstarMh\ relation at each redshift are much closer to the predictions from \citet{behroozi:13}~(see Fig.~\ref{fig:steep_fstar}).  The redshift evolution of \fstar\ is also significantly closer to the \citet{behroozi:13} evolution. The M$_{\rm{H}}=10^{11}$\ \msun\ bin reproduces the \citet{behroozi:13} result almost exactly, while the M$_{\rm{H}}=10^{10}$\ \msun\ bin reproduces the shape but not the normalization.  However, the M$_{\rm{H}}=10^{10}$\ \msun\ curve in \citet{behroozi:13} is entirely extrapolated.  This improvement is also seen in the SMF (Fig.~\ref{fig:steep_smf}), where there is no extrapolation.  We also see improvement in the cold gas fractions.  The cold gas fractions from the preferential reheating model are roughly parallel to the fiducial model but somewhat higher (see Fig.~\ref{fig:steep_cgas}).  The net result is that the preferential reheating cold gas fractions follow the estimates from the \citet{popping:14gasfrac} empirical model well at all redshifts, perhaps overestimating them slightly at high redshift.

The other low mass galaxy properties are altered in the right direction, but not by enough to be consistent with the observations.  Specific star formation rates are somewhat increased at $z\sim 0$\ but are still too low and too flat. However, the sSFRs at intermediate redshifts are now marginally consistent with the observations. By redshift $z\sim 3$, the fiducial model and the preferential reheating model have similar sSFRs (Fig.~\ref{fig:steep_ssfr}).  On a positive note, the preferential reheating model predicts that metallicities increase slightly towards the present day at fixed stellar mass, in better qualitative agreement with the observations, but the metallicity evolution is still too weak (Fig.~\ref{fig:pref_rh_zgas}).

Recent ultra-high resolution numerical simulations that
  attempt to explicitly model the most important physical processes
  associated with stellar and supernova feedback suggest that the wind
  mass-loading factor does scale with galaxy $V_{\rm{circ}}$\ in a
  manner that is similar to energy or momentum driven winds, but that
  there is significant scatter in
  $\dot{M}_{\rm{RH}}/\dot{M}_\star$\ at fixed
  $V_{\rm{circ}}$\ \citep{hopkins:12}. They also find that the
  mass-loading scales with other galaxy parameters, such as star
  formation rate and gas surface density, which could introduce an
  effective redshift dependence in the mass-loading factor. In
  addition, increasing attention has been paid recently to other
  possible mechanisms for driving large-scale galactic outflows, such
  as cosmic rays \citep[e.g.][]{hanasz:13}.

%----------------------------------------------------------------------------------------

\subsection{Direct suppression: changing the star formation efficiency}
\label{sec:discussion_direct}
\begin{figure}
\centering
\includegraphics[width=\columnwidth]{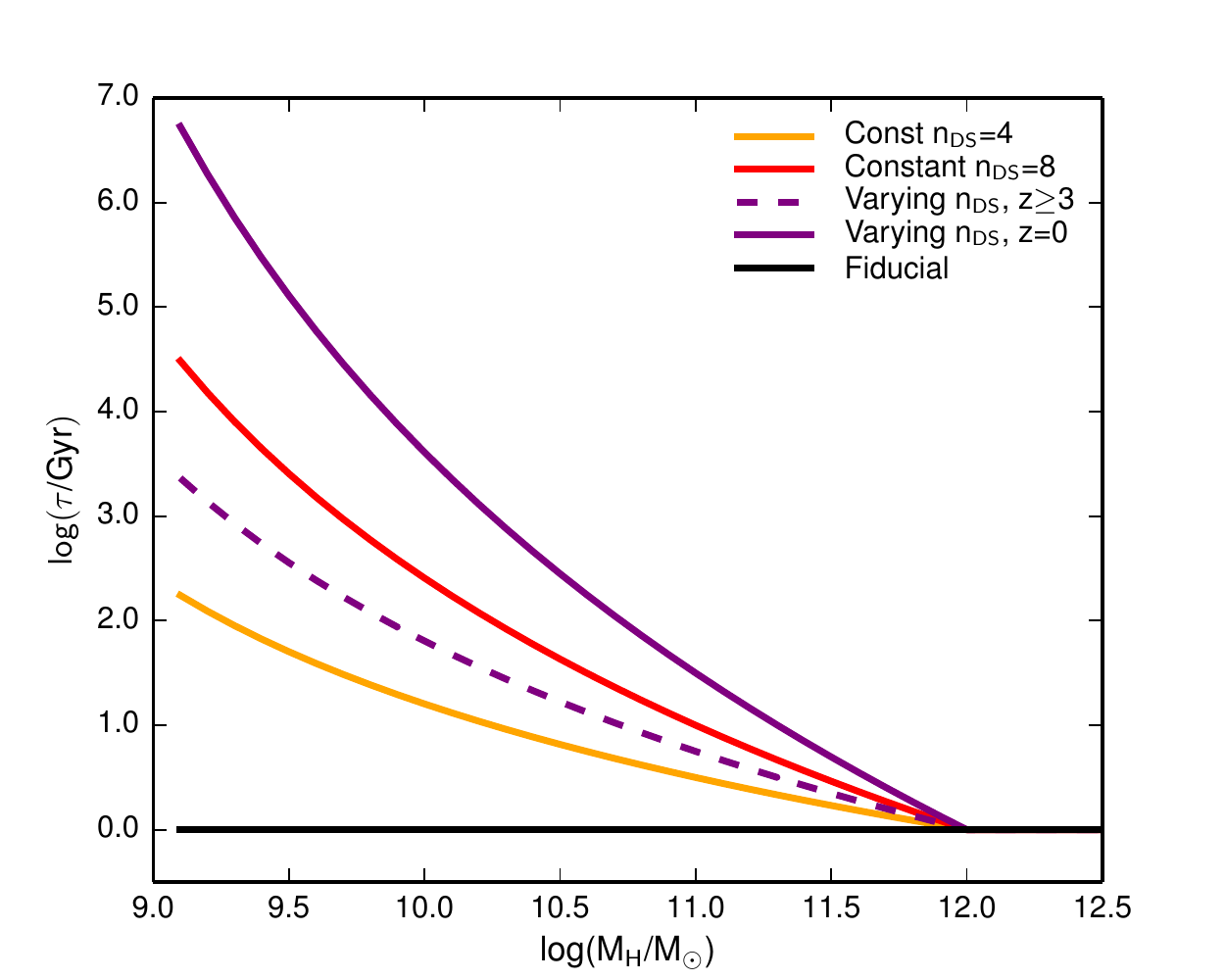}
\caption{Star formation timescale, $\tau_{\rm{DS}}(M_{\rm{H}}, z)
  \equiv \tau_{\rm{CE}}/$\fthin, for three variants of the direct
  suppression model.  The two constant \nthin\ models are shown in
  yellow and red and the varying \nthin\ model is shown in purple with
  the $z=0$\ relation as a solid line and the $z=3$\ relation as a
  dashed line.
\label{fig:thinning}}
\end{figure}

\begin{figure*}
\centering
\includegraphics[width=\textwidth]{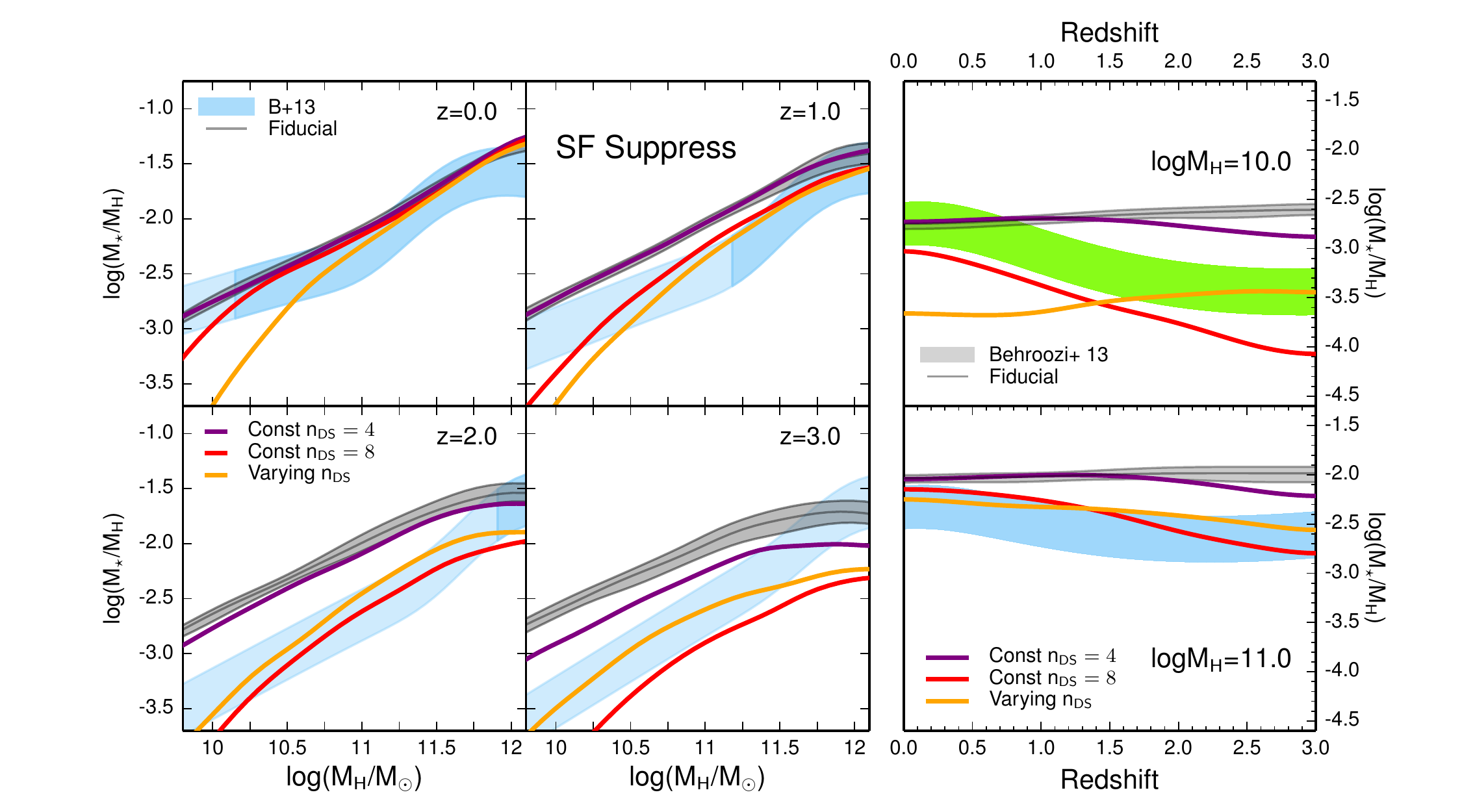}
\caption{Ratio of stellar mass to halo mass, \fstar, for the three direct star formation suppression models.  In all panels, the three direct suppression models are shown in yellow, red, and purple and the median and $\pm1\sigma$\ region of the fiducial model are shown in gray.  Left panel: the \fstarMh\ relation for four redshifts.  Right panel: \fstar(z)\ for halo mass $M_{\rm{H}}=10^{10}$\ \msun\ in the top panel and halo mass $M_{\rm{H}}=10^{11}$\ \msun\ in the lower panel.  Empirical constraints (shaded colored regions) are as described in Fig.~\ref{fig:fiducial_fstar}.
\label{fig:thin_fstar}}
\end{figure*}

\begin{figure*}
\centering
\includegraphics[width=\textwidth]{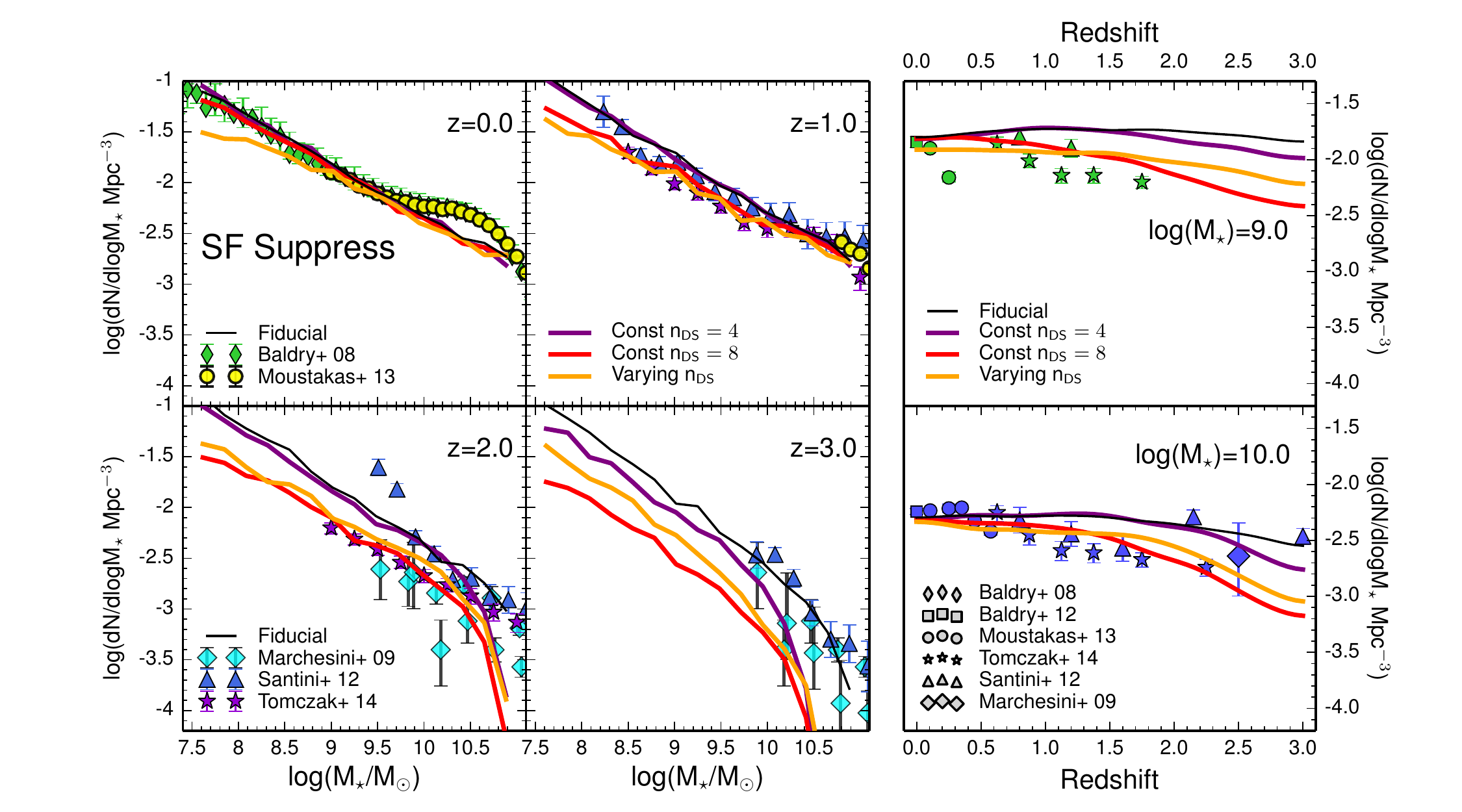}
\caption{Stellar mass function for the direct star formation suppression model. In all panels, the direct suppression models are shown in yellow, red, and purple, and the fiducial model is shown in black.  Data are as in Fig.~\ref{fig:fiducial_smf}.  Left panel: stellar mass functions for four redshifts. Right panel: number densities as a function of redshift for galaxies with \mstar$=10^9$\ \msun\ in the top panel and \mstar$=10^{10}$\ \msun\ in the lower panel.
\label{fig:thin_smf}}
\end{figure*}

\begin{figure*}
\centering
\includegraphics[width=\textwidth]{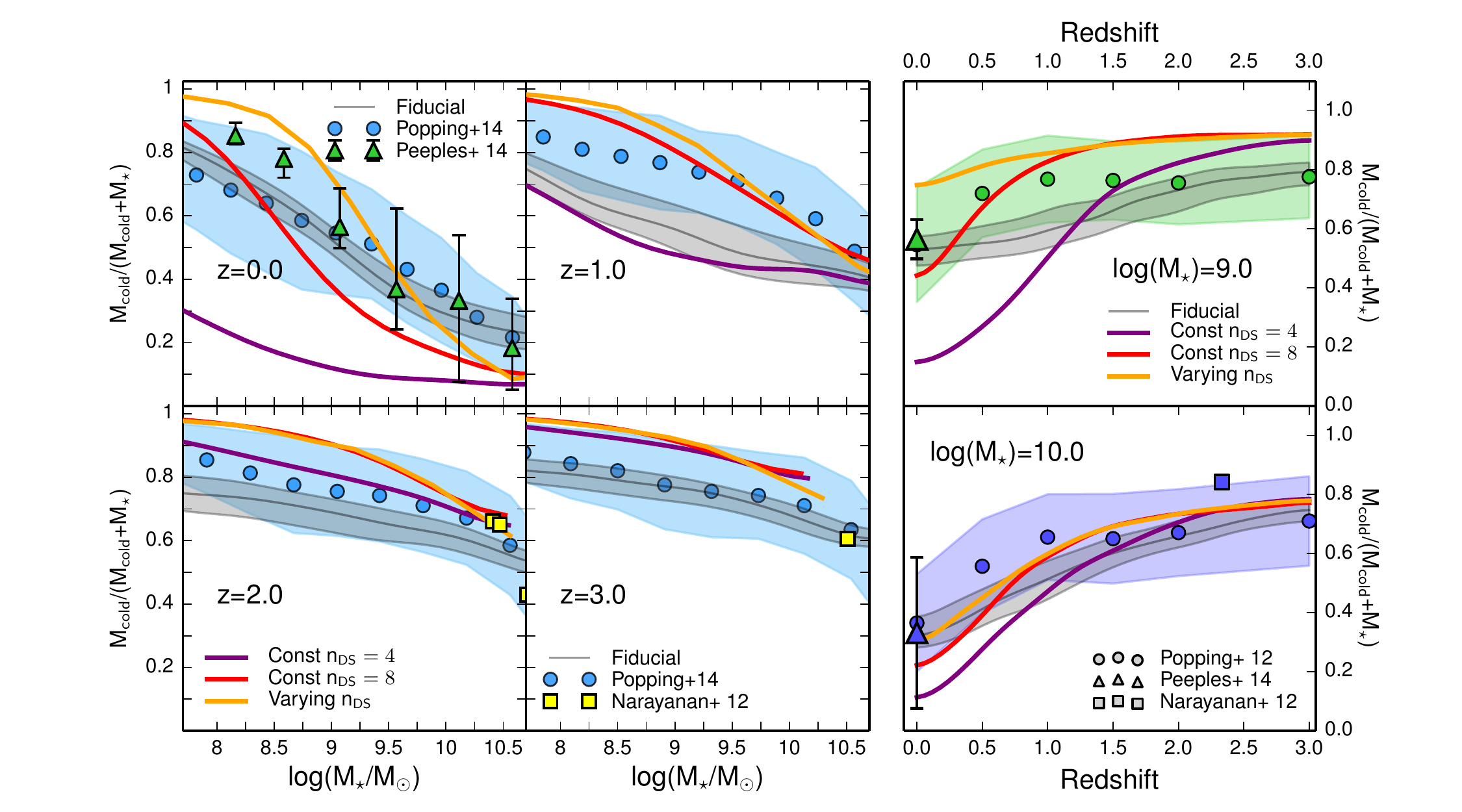}
\caption{Cold gas fractions for the three direct suppression models.  In all panels, the direct suppression models are shown in yellow, red, and purple, and the median and  $\pm1\sigma$~region of the fiducial model are shown in gray.  Data are as in Fig.~\ref{fig:fiducial_cgas}. Left panel: cold gas fractions as a function of stellar mass for four redshifts.  Right panel: cold gas fraction as a function of redshift for galaxies with stellar mass \mstar$=10^9$\ \msun\ in the top panel and \mstar$=10^{10}$\ \msun\ in the lower panel.  Only model galaxies with nonzero gas fraction and bulge to total ratio B/T\textless0.4 are included.
\label{fig:thin_cgas}}
\end{figure*}

\begin{figure*}
\centering
\includegraphics[width=\textwidth]{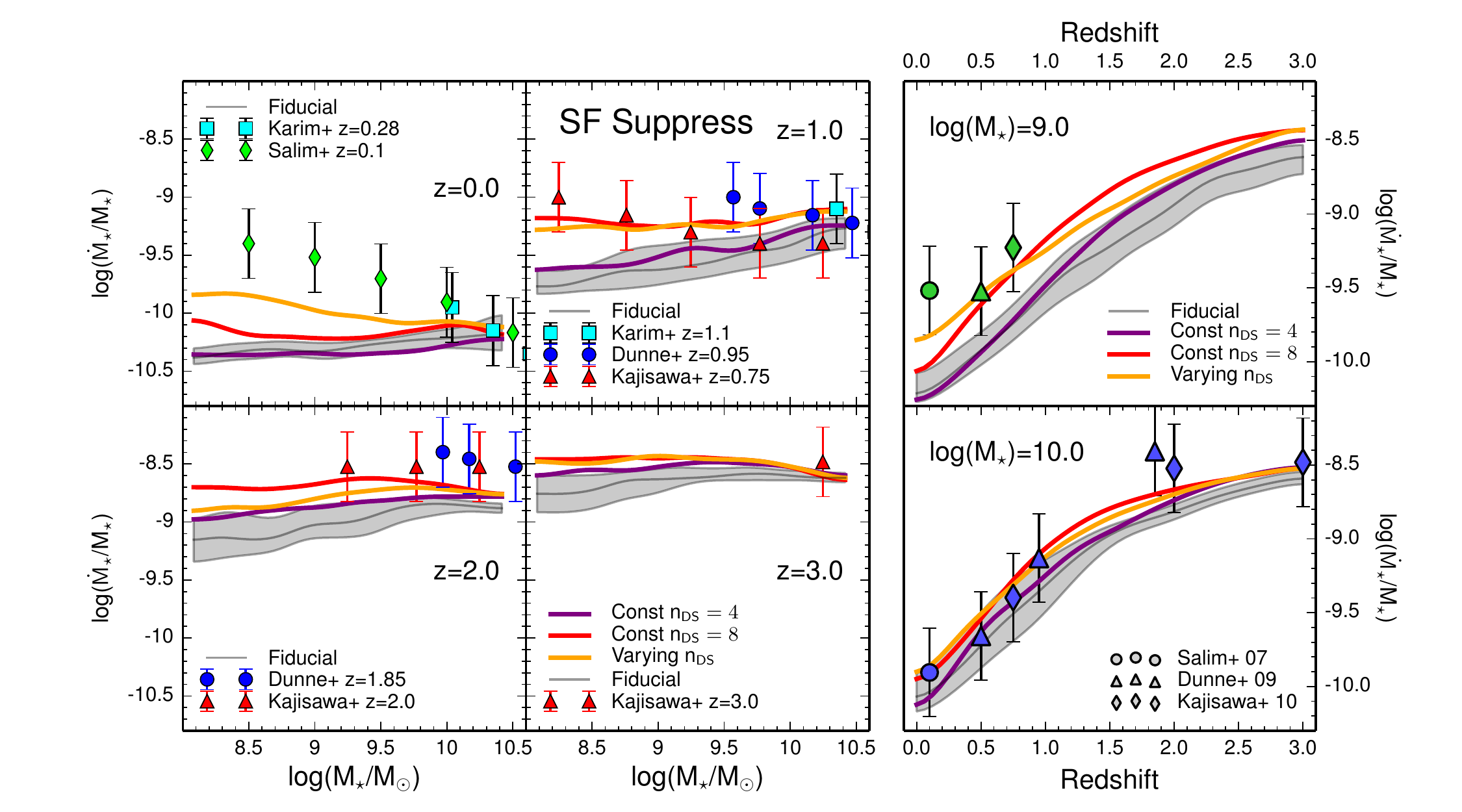}
\caption{Specific star formation rates for the three direct suppression models.  In all panels, the direct suppression models are shown in yellow, red, and purple, and the median and  $\pm1\sigma$~region of the fiducial model are shown in gray.  Data are as in Fig.~\ref{fig:fiducial_ssfr}.   Left panel: sSFRs as a function of stellar mass at four redshifts. Right panel: sSFR as a function of redshift for \mstar$=10^9$\ \msun\ in the top panel and \mstar$=10^{10}$\ \msun\ in the lower panel. 
\label{fig:thin_ssfr}}
\end{figure*}

\begin{figure}
\centering
\includegraphics[width=\columnwidth]{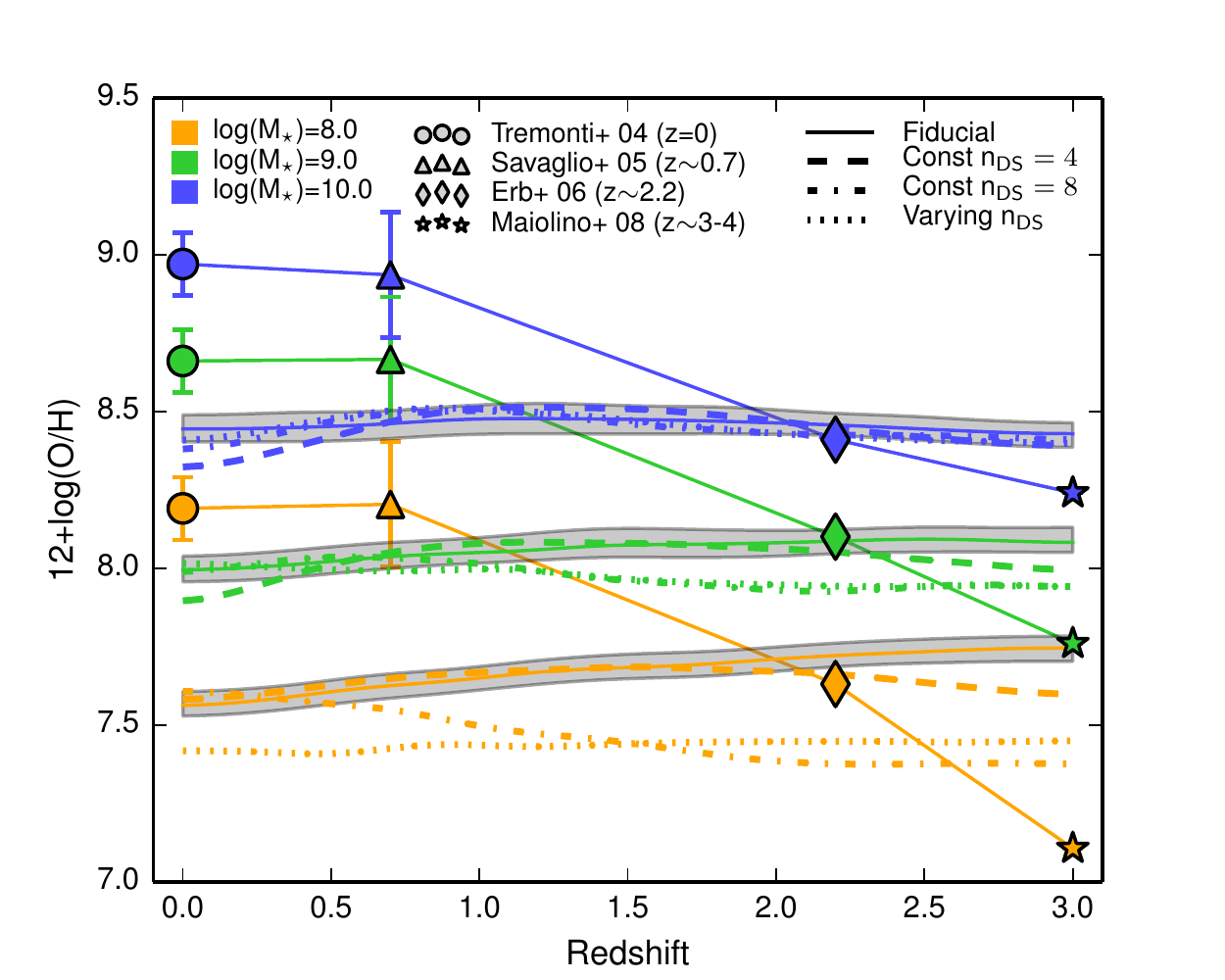}
\caption{Gas phase metallicities in the three direct suppression models for selected stellar masses as a function of redshift.  Metallicities are color coded according to the stellar mass bin they represent. The thin solid lines and shaded gray regions show the fiducial model's median and $\pm1\sigma$~region and the thick dashed, dotted, and dot-dashed lines show the direct suppression models.  Only galaxies with gas fractions greater than 0.2 are plotted and observations are shown as points whose shapes indicate which data set they represent.
\label{fig:thin_zgas}}
\end{figure}

The appropriate alterations to the star formation recipe are somewhat less clear than those to the stellar reheating recipe.  The Kennicutt law has no direct dependence on halo mass, only the surface density of the cold gas.  To simplify the recipe, we replaced the Kennicutt law with a constant star formation efficiency law (Eqn.~\ref{eq:constsfr}).  The default constant star formation efficiency model produces results similar to the fiducial model, mainly differing in cold gas fraction predictions: the constant efficiency star formation recipe produces cold gas fractions far lower than the fiducial model at low redshift.  To implement the direct star formation suppression, we multiply the constant star formation efficiency recipe by a factor, \fthin.  The inefficiency parameter \fthin\ can be made a function of galaxy mass and redshift and allows us to directly control the star formation rate.  This has the same effect as making the star formation timescale a function of mass and redshift, $\tau_{\rm{DS}}(M_{\rm{H}}, z)$.
\begin{equation}
\dot{M}_\star=f_{\rm{DS}}(M_{\rm{H}}, z)\frac{M_{\rm{cold}}}{\tau_{\rm{CE}}} = \frac{M_{\rm{cold}}}{\tau_{\rm{DS}}(M_{\rm{H}}, z)}
\label{eq:thin_sfr}
\end{equation}
The function \fthin\ effectively replaces the surface density threshold for star formation in our fiducial model.

The direct suppression factor \fthin\ should be unity above a certain halo mass, $M_{\rm{H,trans}}$, since high mass halos should remain unaffected, and halos with \mh $\lesssim 10^{11}$\ \msun\ should have low \fthin\ to prevent overproduction of stars.  To parameterize this, we set \fthin=0 at \mh$\leq 10^8$\ \msun, \fthin=1 at \mh $\geq M_{\rm{H,trans}}$, and a power law between with power \nthin.
\begin{align}
f_{\rm{DS}}&=\left\{
        \begin{array}{cl}
	0 & M_{\rm{H}}\leq 10^{8}M_\odot\\
	\left(\frac{{\rm{log}}(M_{\rm{H}})-8}{{\rm{log}}(M_{\rm{H,trans}})-8}\right)^{n_{\rm{DS}}} & 10^{8}M_\odot <M_{\rm{H}}<M_{\rm{H,trans}}\\
	1 & M_{\rm{H}}\geq M_{\rm{H,trans}}\\
        \label{eq:thinning}
	\end{array}
\right. %\\
\end{align}
We take $M_{\rm{H,trans}}=10^{12}$\ \msun.  Low mass halo properties are fairly insensitive to the choice of $M_{\rm{H,trans}}$.  This relation is shown for the three models we present in terms of the direct suppression timescale $\tau_{\rm{DS}}(M_{\rm{H}}, z)$~rather than \fthin\ itself in Fig.~\ref{fig:thinning}.  A low constant \nthin=4 follows the fiducial \fstar\ closely at low redshifts with only a slight decrease in normalization towards redshift $z=3$, as can be seen in the left panel of Fig.~\ref{fig:thin_fstar}.  A higher constant \nthin=8 also reproduces the $z=0$\ \fstarMh\ relation and does somewhat better than the fiducial model at redshifts $z=1$\ and 2, though it is still outside the 1-$\sigma$\ uncertainty at $z\sim 1$.  At $z=1$\ and $z=2$, \nthin=8 still overproduces stellar mass in halos of mass \mh$\cong10^{11}$\ \msun\ and at high redshifts, star formation is over-suppressed, producing a $z=3$\ \fstarMh\ below observations.  The \nthin=8 model does reproduce the sense of the \mh$\cong10^{10}$\ \msun\ \fstar (z) relation.  These same trends can be seen in the stellar mass functions in Fig.~\ref{fig:thin_smf}. Increasing \nthin\ beyond \nthin=8 would bring the redshift $z=1$\ and 2 \fstar\ relations closer to observations, but would make the discrepancy at higher redshifts worse. Moreover, none of the models with constant \nthin\ reproduce the observed cold gas fractions at $z=0$~(Fig.~\ref{fig:thin_cgas}) or correctly predict the observed slope in the sSFR-$M_\star$~relation at $z=0$~(Fig.~\ref{fig:thin_ssfr}).  The \nthin=8 model does produce a rising metallicity over time for the lowest mass galaxies, but the metallicity evolution is still too weak (Fig.~\ref{fig:thin_zgas}).

In the limit of very high \nthin, our model is reminiscent of the \citet{bouche:10}~model, but with the major difference that accretion itself is halted in the \citet{bouche:10}~model and only star formation is halted in ours. Our direct suppression model with \nthin~very high is approximately a step function as is the \citet{bouche:10} accretion floor, but with a transition mass of \mhalo=10$^{12}$\ \msun\ instead of 10$^{11}$\ \msun.  The step function fails in our model because we are only preventing star formation, not gas accretion, and as soon as a galaxy passes the threshold, it quickly forms enough stars to rejoin the fiducial \fstarMh.

In an attempt to bring the \fstarMh~relation at $z=1$\ and 2 into closer agreement with the observations, we try a model with constant \nthin=6 at $z>3$, then increase \nthin\ to \nthin=12 linearly between $z=3$\ and $z=0$.  This model matches \fstar\ at $z\sim 3$\ acceptably and matches well at $z=0$, but still over-predicts the $z=1$\ and $z=2$\ \fstar\ relations.  Our varying \nthin\ model over-predicts cold gas fractions for stellar masses $M_\star\lesssim10^9$~\msun\ at all redshifts and at stellar masses $M_\star\lesssim10^{10}$~\msun\ for $z\gtrsim2$.  It does produce a slightly negative sSFR slope at $z=0$\ but still does not match the observed normalization.

In terms of the star formation timescale, our models have a
  normal, constant star formation timescale above the transition halo
  mass $M_{\rm{H}} \cong 10^{12}$\ \msun\ and transition quickly to a
  very long star formation timescale below (see
  Fig.~\ref{fig:thinning}).  By adjusting \nthin, we control how
  quickly the star formation timescale increases below $M_{\rm{H}}
  \cong 10^{12}$\ \msun\ and thus how suppressed star formation is in
  low mass halos.  The way in which the model with varying
  \nthin\ failed suggests that monotonically increasing how steeply
  the star formation timescale rises below the transition mass is
  insufficient.  The transition to very long star formation timescales
  would most likely need to be extremely quick between $z=3$\ and
  $z=1$\ to match \fstar\ at redshifts $z=1$\ and 2, but must ease off
  towards $z=0$\ in order not to over-suppress star formation at
  $z=0$.  A successful direct suppression model would likely add at
least four new free parameters and the physical scenario that could
cause this sort of behavior is not obvious.  The failure of
  the direct suppression model highlights the resilience of the low
  mass galaxies' star formation histories against changes in the star
  formation efficiency, and suggests that the solution is unlikely to
  consist solely of adjustments to the star formation efficiency.

%----------------------------------------------------------------------------------------

\subsection{Parking lot: changing gas accretion rates}
\label{sec:discussion_park}

\begin{figure}
\centering
\includegraphics[width=\columnwidth]{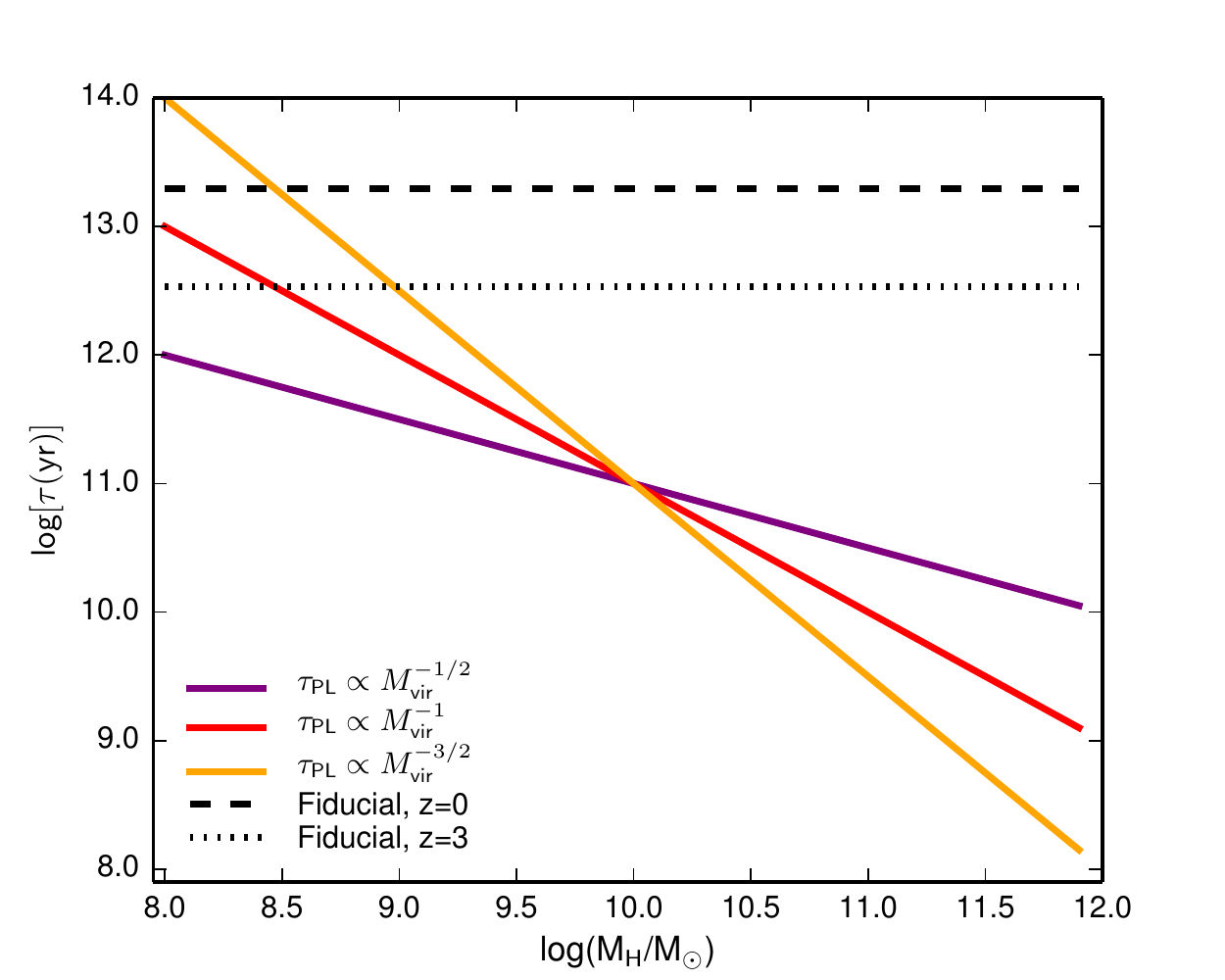}
\caption{Re-infall timescales as a function of halo mass for the three parking lot models as well as the fiducial model.  The fiducial model's infall timescale depends on redshift because it is a function of the dynamical time of the halo at the virial radius, which changes with redshift.
\label{fig:park_visualization}}
\end{figure}

\begin{figure*}
\centering
\includegraphics[width=\textwidth]{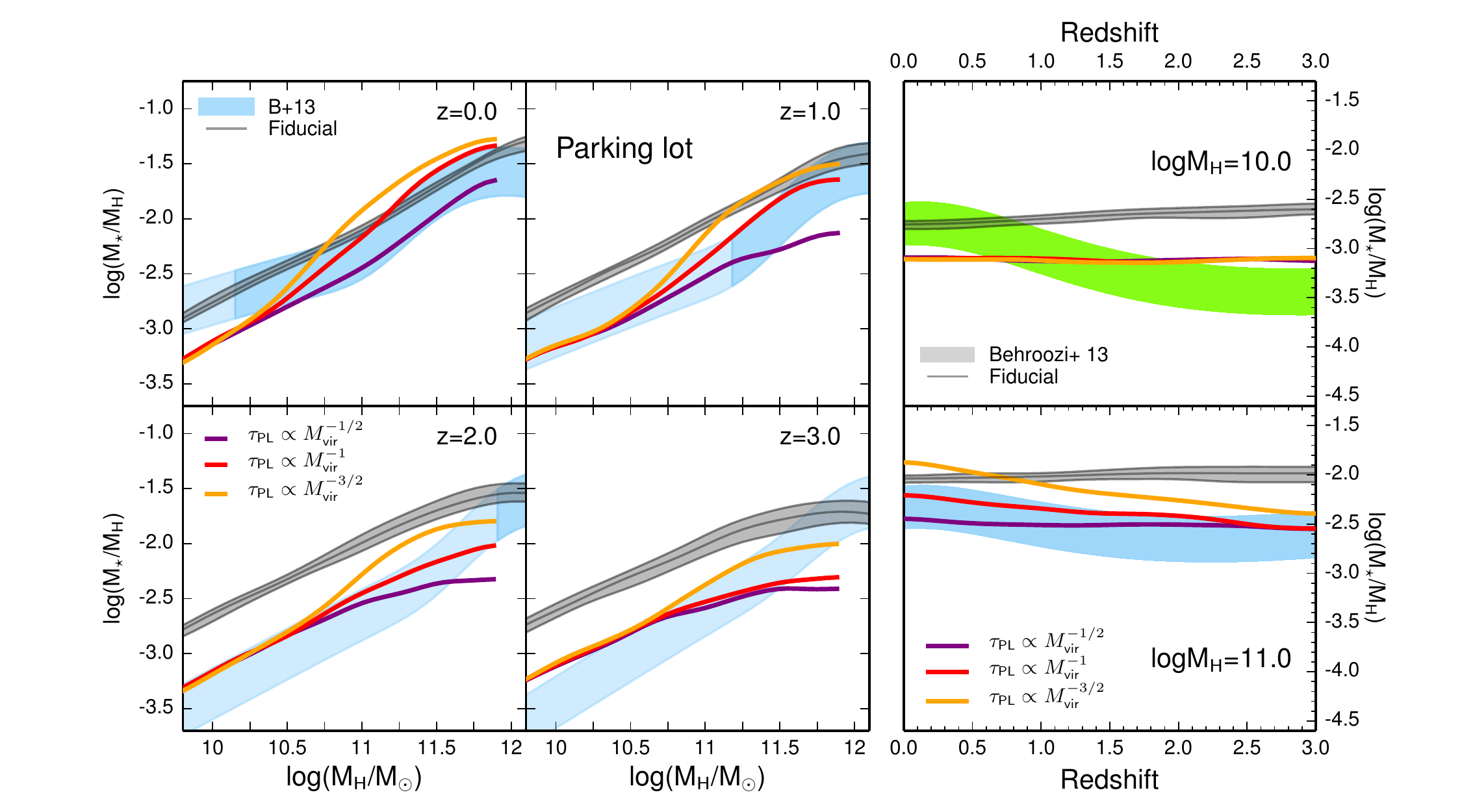}
\caption{Ratio of stellar mass to halo mass, \fstar, for the three parking lot models.  In all panels, the parking lot models are shown in yellow, red, and purple, and the median and $\pm1\sigma$\ region of the fiducial model are shown in gray.  Left panel: \fstarMh\ relation for four redshifts.  Right panel: \fstar(z) for halo mass $M_{\rm{H}}=10^{10}$\ \msun\ in the top panel and halo mass $M_{\rm{H}}=10^{11}$\ \msun\ in the lower panel.  Empirical constraints (shaded colored regions) are as described in Fig.~\ref{fig:fiducial_fstar}.
\label{fig:park_fstar}}
\end{figure*}

\begin{figure*}
\centering
\includegraphics[width=\textwidth]{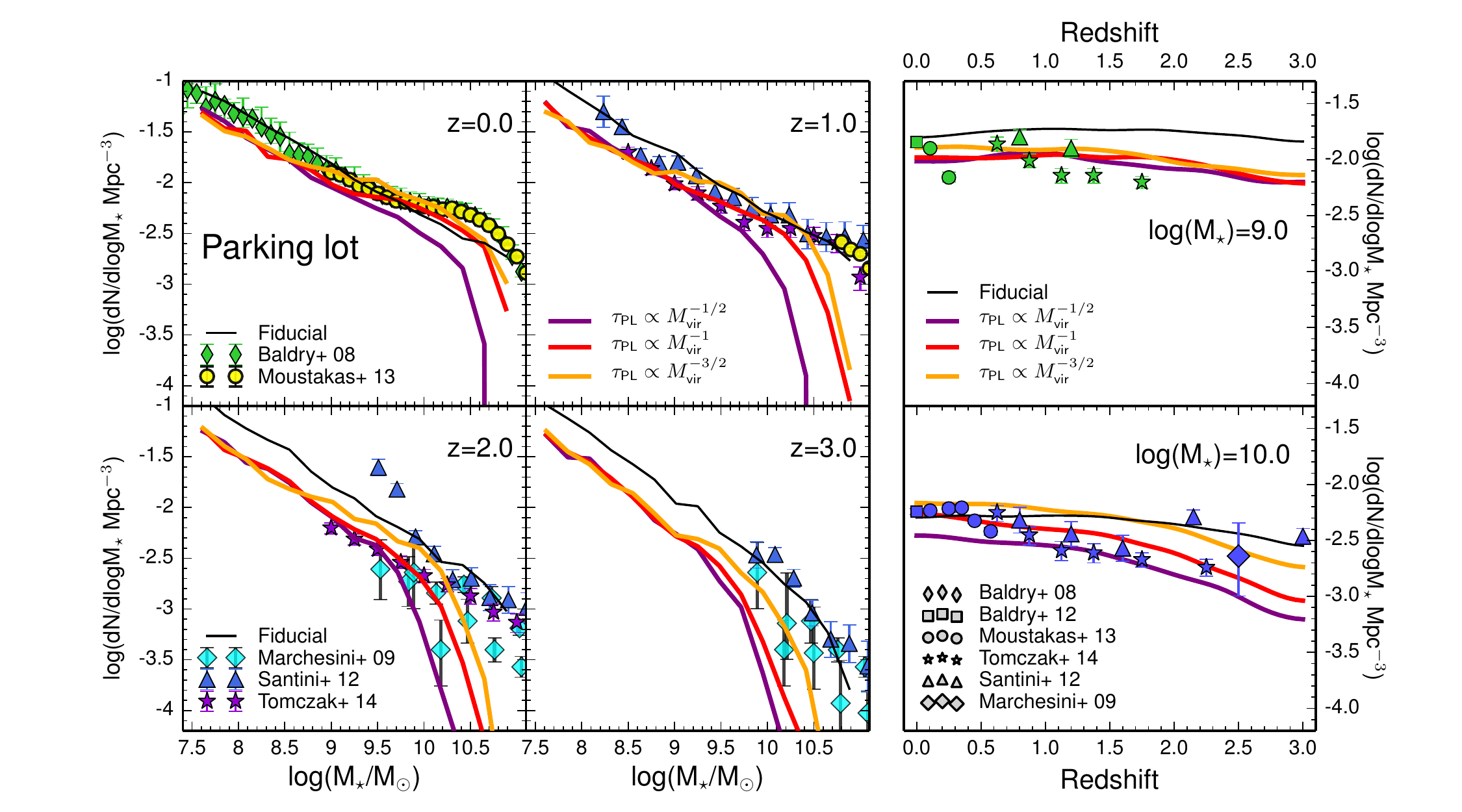}
\caption{Stellar mass functions for the parking lot models. In all panels, the parking lot models are shown in yellow, red, and purple, and the fiducial model is shown in black.  Data are as in Fig.~\ref{fig:fiducial_smf}.  Left panel: stellar mass functions for four redshifts. Right panel: number densities as a function of redshift for galaxies with \mstar$=10^9$\ \msun\ in the top panel and \mstar$=10^{10}$\ \msun\ in the lower panel.
\label{fig:park_smf}}
\end{figure*}

\begin{figure*}
\centering
\includegraphics[width=\textwidth]{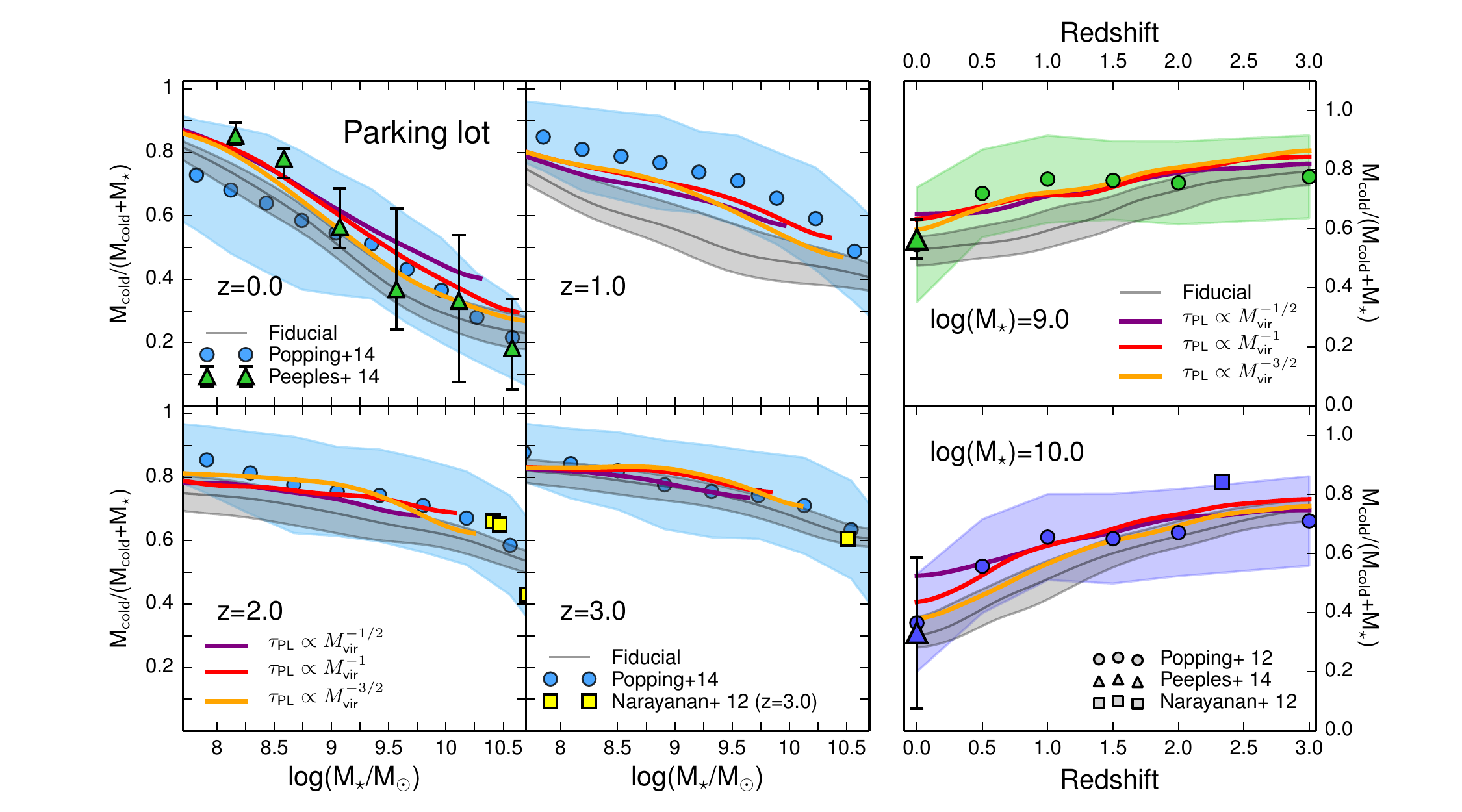}
\caption{Cold gas fractions for the parking lot models. In all panels, the parking lot models are shown in yellow, red, and purple, and the median and  $\pm1\sigma$~region of the fiducial model are shown in gray.  Data are as in Fig.~\ref{fig:fiducial_cgas}.  Left panel: gas fraction as a function of stellar mass for four redshifts.  Right panel: gas fraction as a function of redshift for galaxies with stellar mass \mstar$=10^9$\ \msun\ in the top panel and \mstar$=10^{10}$\ \msun\ in the lower panel. Only galaxies with nonzero gas fractions and bulge to total ratios B/T\textless0.4 are shown.  
\label{fig:park_cgas}}
\end{figure*}

\begin{figure*}
\centering
\includegraphics[width=\textwidth]{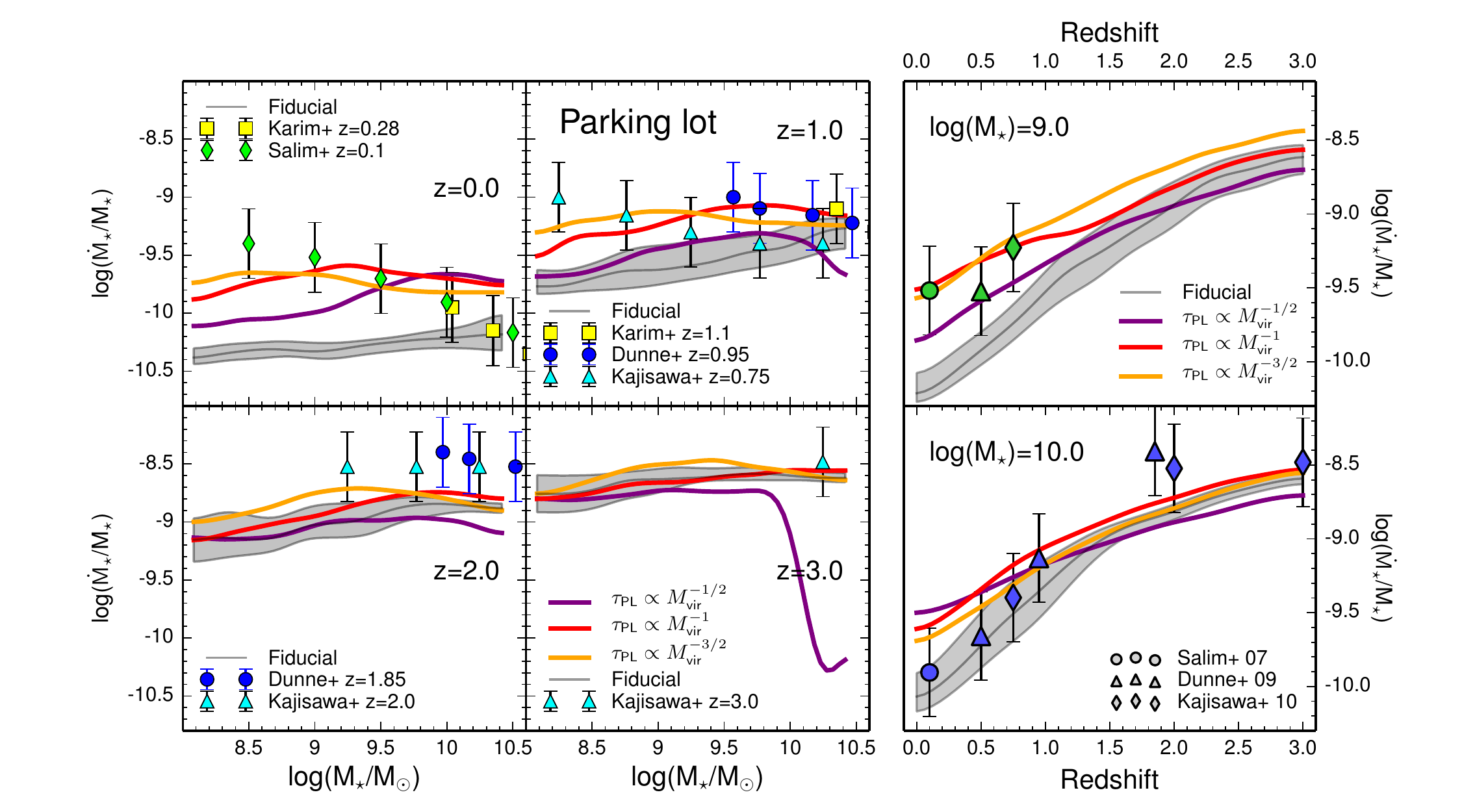}
\caption{Specific star formation rates for the three parking lot models.  In all panels, the parking lot models are shown in yellow, red, and purple, and the median and  $\pm1\sigma$~region of the fiducial model are shown in gray.  Data are as in Fig.~\ref{fig:fiducial_ssfr}.  Left panel: sSFR as a function of stellar mass for four redshifts.  Right panel: sSFR as a function of redshift for \mstar$=10^9$\ \msun\ in the top panel and \mstar$=10^{10}$\ \msun\ in the lower panel. 
\label{fig:park_ssfr}}
\end{figure*}

\begin{figure}
\centering
\includegraphics[width=\columnwidth]{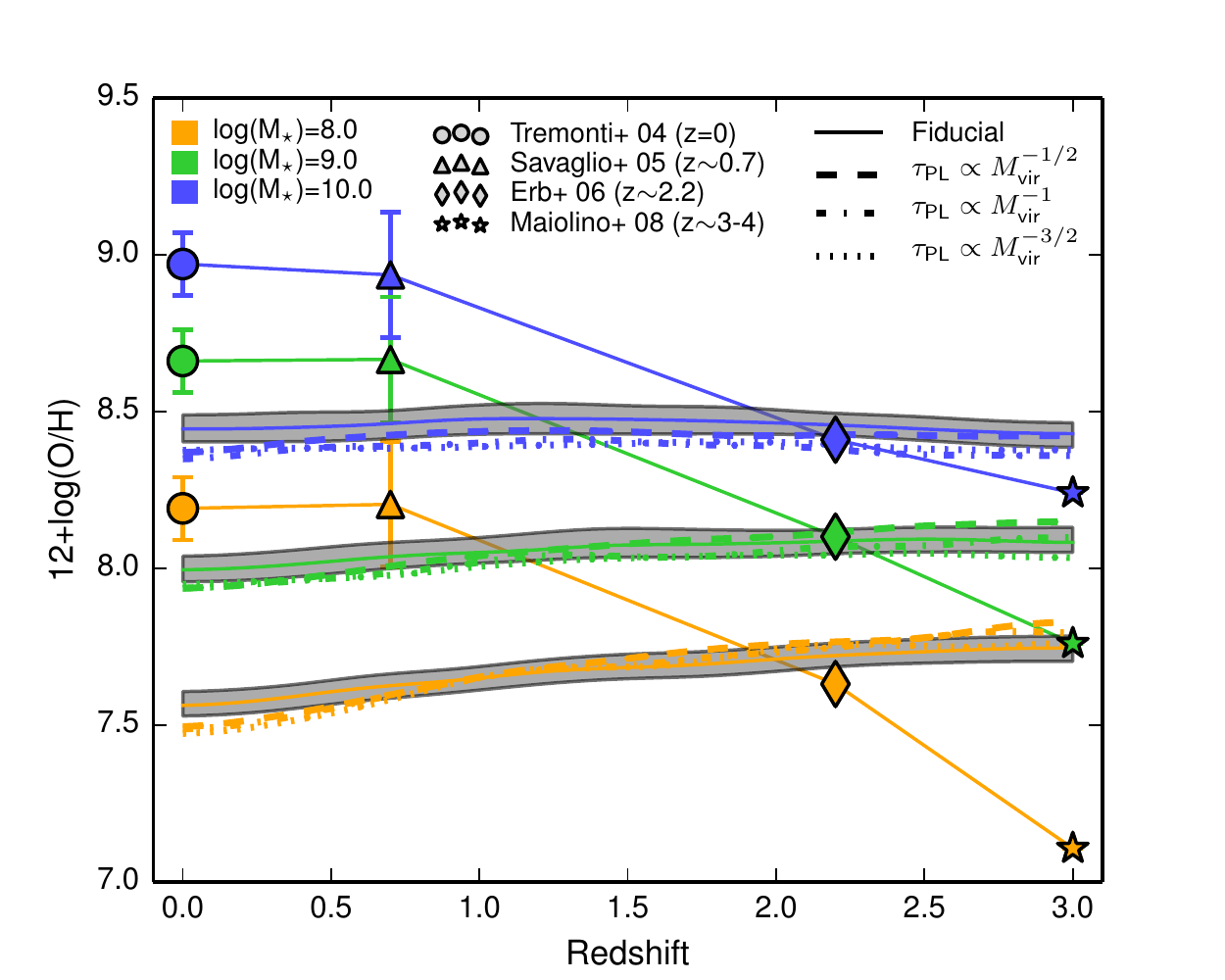}
\caption{Gas phase metallicities for the three parking lot models for selected stellar masses as a function of redshift.  Metallicities are color coded according to the stellar mass bin they represent.  The fiducial model is shown with thin solid lines and the three parking lot models are shown as dashed, dot-dashed, and dotted lines. Only galaxies with gas fractions greater than 0.2 are plotted and observations are shown as points whose shapes indicate which data set they represent, as in Fig.~\ref{fig:fiducial_zgas}.
\label{fig:park_zgas}}
\end{figure}

Every halo has three reservoirs of gas: cold ISM gas in
galaxies, hot halo gas (ICM), and ``ejected'' gas, which may be
associated with the circum-galactic medium (CGM) or IGM. Halos grow by accreting ``diffuse''
material that has never been in halos, as well as by subsuming
material from all of the progenitor halos. One must decide how to
combine these different reservoirs. In the Santa Cruz SAM, all galaxies keep their cold gas
reservoirs and the hot gas from halos that become
``satellites'' is assumed to be instantaneously subsumed into the hot
gas reservoir of the new halo. The ejected gas reservoir from the
largest progenitor halo becomes the ejected gas reservoir for the new
halo, and the ejected gas reservoirs from the other (minor)
progenitors are deposited into the \emph{hot gas reservoir} of the new central
halo. The ejected gas reservoirs also include IGM gas that was
prevented from accreting by the photo-ionizing background; for the
non-largest progenitors this is also subsumed into the new hot gas
reservoir of the larger halo.

We found that, within the usual set of assumptions of our fiducial model, simply changing the functional form of the re-infall timescale (Eqn.~\ref{eq:re-infall}) did not solve the dwarf galaxy problems we are trying to address here. We discuss reasons for this, and possible reasons for differences between our results and those of H13, in Appendix~\ref{sec:appendix_h13}. Briefly, we find that the significance of ``re-accreted'' gas to the total gas supply is quite sensitive to details of the bookkeeping for these different gas reservoirs when halos merge together.  The SAM used in the H13 model takes gas stripped from the ejected reservoir of an in-falling satellite and deposits it over time in the ejected reservoir of the central, whereas the fiducial Santa Cruz model instantaneously deposits all the gas from the ejected reservoir of the satellite into the hot gas reservoir of the new central.  This difference means that the H13 model's ejected reservoir handles a higher fraction of the galaxy's gas than the Santa Cruz model's and therefore changing the re-infall timescale in the H13 model has a larger effect than in the Santa Cruz model.

In the ``parking lot'' model, we
divert some of the gas that would normally be added directly to the
hot gas reservoir and instead store it along with the ejected gas.
This reservoir of ejected and diverted gas becomes our parking lot. We
then adopt various scalings for the timescale on which this parking
lot gas can accrete into the halo. With the addition of the diverted
gas, changing the rate of infall from this parking lot reservoir can
affect the evolution of galaxies' stellar masses at higher redshifts.

Before we choose how to alter the accretion timescale for the parking
lot gas, we must choose which gas is routed through the parking
lot. We found that when we diverted all the accreted gas to the
parking lot, our models produced an incorrect evolution similar to the
fiducial model with star formation happening too early in low mass
galaxies and too late in high mass galaxies. It may be the case that
if we made the parking lot accretion timescale a complex function of
halo mass and redshift, such models could be made to work, however,
this is beyond the scope of this paper. We found, though, that if we
divert only the hot and ejected gas reservoirs from the minor
progenitors following halo mergers, this has little effect on
accretion at high redshift (where it is dominated by accretion from
the IGM), but delays lower redshift accretion as required.  
Considering that numerical simulations
find that satellites' dark matter halos begin being stripped at
$5R_{\rm{vir}}$\ \citep{behroozi:13:5rvir}, it is perhaps not
unreasonable to think that the associated hot diffuse gas might also
be stripped and heated by the ejected reservoir.

The infall timescale must depend on halo mass in order to create a
differential between low mass and high mass halos.  We let the
timescale be proportional to the virial mass to a power:
\begin{align}
\label{eq:smoosh_recipe}
\dot{M}_{\rm{ReIn}} &= M_{\rm{PL}}/\tau_{\rm{PL}}\\
\tau_{\rm{PL}}&=\gamma_{\rm{PL}} \left(\frac{M_{0\rm{,PL}}}{M_{\rm{vir}}}\right)^{\alpha_{\rm{PL}}}
\end{align}
where we choose $M_{0\rm{,PL}}=10^{10}$\ \msun.
\citet{oppenheimer:10}~find that in their hydrodynamic simulations,
the gas recycling timescale (which in their case is the time between
gas ejection and re-infall into the ISM) scales as \alphapl=0.5
for momentum-driven winds or \alphapl=1.5 for energy-driven, constant
velocity winds.  Note that our timescale is that for reaccretion only, not the timescale for the full cycle of ejection and reaccretion as in \citet{oppenheimer:10}.

We test all three values of \alphapl\ with \gammapl=$10^{11}$\ yr and $M_{0\rm{,PL}}=10^{10}$\ \msun.  The reinfall timescales for these three models as a function of halo mass are shown in Fig.~\ref{fig:park_visualization}.  The results are insensitive to the exact values chosen for \gammapl\ and $M_{0\rm{,PL}}$.  Increasing $M_{0\rm{,PL}}$\ by two orders of magnitude somewhat decreases the normalization of \fstar\ for intermediate mass halos and changing \gammapl\ by an order of magnitude changes the normalization of \fstar\ by less than half a dex.  Other properties are essentially unaffected.  The
\alphapl=1 model is able to produce a gently rising value of
\fstar\ at $M_{\rm{H}} \geq 10^{11}$\ \msun, with a similar slope to the
\citet{behroozi:13} results, but \fstar\ at $M_{\rm{H}} \geq
10^{10}$\ \msun\ remains flat (Fig.~\ref{fig:park_fstar}). In addition, the parking lot models appear
to over-suppress star formation in halos with $M_{\rm{H}} \geq
10^{11.5}$\ \msun\ at $z=2$\ and $z=3$.  This problem would only be
exacerbated by turning AGN feedback back on. In order to improve the
\fstar\ results further, it appears that it would be necessary to
introduce a more complicated redshift and halo mass dependence for
\taupl. We note that in the results shown here, there is no
suppression of gas infall or evaporation of cold gas by the
photo-ionizing background included in our models. This is
important at the lowest halo masses considered ($M_{\rm{H}} \lesssim
10^{10}$).

Predicted cold gas fractions are slightly higher at high redshift (see Fig.~\ref{fig:park_cgas}), matching the \citet{popping:14gasfrac} well at most redshifts.  The $z=1$~gas fractions remain somewhat low but are within the $\pm1\sigma$~errors of the empirical model. The low mass slope of the sSFR
vs.~stellar mass relation is significantly improved, now lying within
the observational error bars except at the lowest masses for
$z\lesssim 1$\ (Fig.~\ref{fig:park_ssfr}). The ISM metallicity
evolution remains very similar to that in the fiducial model (see
Fig.~\ref{fig:park_zgas}).

%====================================================================
%====================================================================

\section{Conclusions}
\label{sec:conclusions}

\begin{figure}
\centering
\includegraphics[width=\columnwidth]{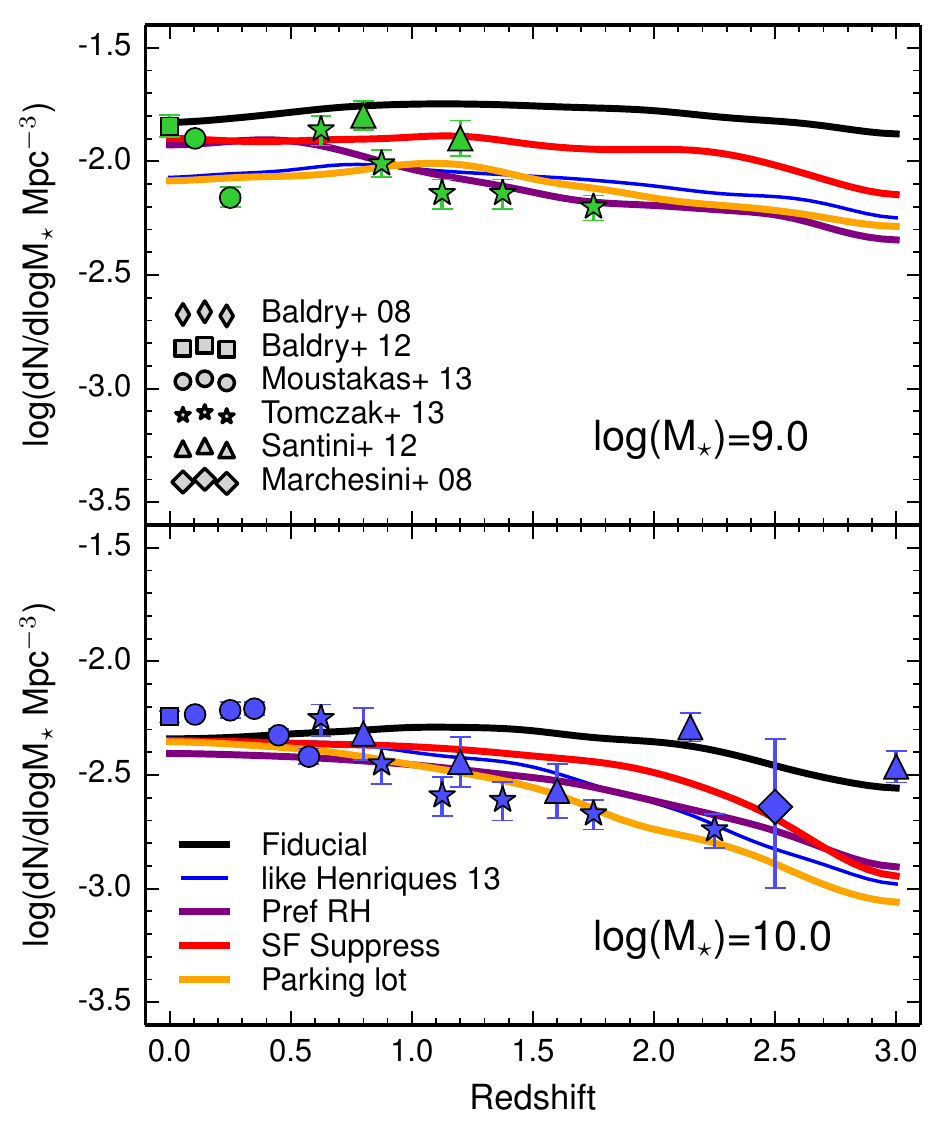}
\caption{Galaxy number densities as a function of redshift for galaxies with \mstar$=10^9$\ \msun\ in the top panel and \mstar$=10^{10}$\ \msun\ in the lower panel.  The fiducial model is shown in black, the ``H13-like'' version of our SAM described in Appendix~\ref{sec:appendix_h13} in blue, and the ``best'' versions of each modified scenario are shown with preferential reheating in purple, direct suppression with varying \nthin\ in red, and parking lot with $\tau_{\rm{PL}}\propto M_{\rm{vir}}^{-1}$ in yellow.  Observations are shown as points whose shapes indicate which data set they represent.  Data are as in Fig.~\ref{fig:fiducial_smf}.
\label{fig:summary}}
\end{figure}

The overarching theme of our study is that the interplay
between gas accretion, feedback, and star formation as commonly
implemented in \lcdm\ models of galaxy formation results in a
remarkable tendency to produce the ``upsizing'' behavior seen in the
fiducial model.  Models find that the \fstar\ and number density of
low mass galaxies is approximately constant or even \emph{decreases}
instead of following the increase with cosmic time implied by
observations. This is probably due to the failure of ``sub-grid''
recipes for star formation and stellar feedback to break the
characteristic self-similarity of halos' gas and dark matter accretion
histories.  It has been clear for some time that some modification
needs to be made to the sub-grid recipes in order to solve the cluster
of problems that constitutes the ``dwarf galaxy conundrum'' presented
here, assuming that the basic framework is correct. However,
it has remained unclear which set of physical recipes needs to be
modified or in what way.

In order to try to gain insight into this puzzle, we have considered a
broader set of complementary observables than have been presented in
most previous studies. In addition, we have considered three very
physically different classes of solution. Some previous works
\citep[e.g.][]{lu:14} have distinguished between ``ejective'' feedback
versus ``preventative'' feedback. Ejective feedback prevents
  star formation by ejecting cold gas and making it unavailable for
  forming stars, whereas preventative feedback prevents hot or
  in-falling gas from cooling and becoming available for star
  formation.  All of our models use ejective feedback, though the
  parking lot model could be interpreted as having both ejective and
  preventative feedback.  We show that the different classes of
solution, when tuned to match the qualitative behavior of \fstar, make
different predictions for other observables. For example, the direct
suppression model produced a larger change in galaxy cold gas
fractions at high redshift than other models, while the parking lot
model produced a greater change in the low mass slope of the
sSFR-\mstar\ relationship. Interestingly, \emph{none} of the scenarios
that we studied were able to reproduce the observed trend of strongly
increasing gas phase metallicities at fixed stellar mass with cosmic
time, suggesting that this problem may have a different origin and
solution.  Additionally, the fiducial implementation of squelching as
in S08 and S12 used in the altered recipes can decrease or reverse the
corrected ``sub-grid'' recipes' trend of increasing \fstar~with time.

Given the freedom that we allowed ourselves in parameterizing the
empirical recipes, it is perhaps not too surprising that we were able
to find solutions that qualitatively reproduced the increasing trend
of \fstar\ and comoving number density of low mass galaxies with
cosmic time for all three scenarios (see Fig.~\ref{fig:summary} for a
summary of all the scenarios, using the directly observable metric of
galaxy number density as a function of redshift). What is then
interesting is to try to assess how physically plausible the required
scalings are. The preferential reheating model gave perhaps the best
results overall, but requires a fairly extreme change in the slope of
the mass-loading factor, \arh, from \arh $\sim 4.5$\ at high redshift
to \arh$\sim 2$\ at low redshift. While there are physical reasons to
think that this scaling might have an effective redshift dependence,
as discussed above, it is unclear whether such a strong evolution in
the scaling can find a physical basis. The direct suppression model
requires an even more extreme scaling --- the star formation
efficiency or timescale must vary with halo mass as a power-law with a
slope of $\sim 8$\ (almost a step function). This would seem to be
already ruled out by direct observations of star formation
efficiencies in nearby galaxies, which do not vary by orders of
magnitude \citep[e.g.][]{bigiel:08}. We therefore disfavor the direct
suppression scenario as the primary solution to the dwarf galaxy
problem. The parking lot model was not quite as successful as the
preferential reheating model, but it did push the qualitative behavior
in the right direction, and the variations in the recipes are easily
within the uncertainties in our knowledge of the relevant physical
processes.  
Our parking lot model is very similar in spirit to the H13
  model, which by modifying the re-infall time for ejected gas was
  also quite successful at solving many of the problems we have
  highlighted here. However, we have pointed out that somewhat
  arbitrary choices in how ejected gas is handled in SAMs can have a
  large effect on the results.

Although here we have only considered solutions driven by one of the
three physical scenarios, it is entirely possible that more
than one of the kinds of variations we considered are important. In
particular, modified scalings for the mass-loading factor and the
re-accretion time of ejected gas are likely to be interconnected through the detailed microphysics of stellar-driven winds. 

It is also important to consider the possibility that some or all of
the observational data that have been used as constraints are not
correct. There are currently differences in galaxy stellar mass
function normalization that are comparable to the level of discrepancy
that we are discussing here, which is only factors of a few and not
orders of magnitude. Uncertainties on this level can arise from
field-to-field variance and systematic uncertainties in stellar mass
estimates. These will improve in the next few years as large areas are
surveyed to the depth necessary to probe these low mass objects out to
redshifts of $z\sim 1$--2. Similarly, direct estimates of cold gas
fractions in low mass galaxies at high redshift are currently
unfeasible, but this will change in the near future as the next
generation of radio telescopes comes online.  However, even with uncertainty in the observations we compare to, the fundamental disagreement in star formation histories remains.  Present-day low mass galaxies are bluer and more star-forming than models predict and these adjusted models are a step in the right direction.

In summary, in this paper we investigated three classes of empirical solution to
the ``dwarf galaxy conundrum,'' the mismatch of observed and predicted
star formation histories for galaxies forming in low mass dark matter
halos.  The three scenarios involve 1)
changing the halo mass dependence of the mass outflow rate of
stellar-driven winds as a function of redshift, 2) changing the star
formation efficiency as a function of halo mass and redshift, and 3)
trapping accreting gas in a ``parking lot'' reservoir with a halo
mass-dependent infall timescale.  We compared the predictions of the
three scenarios to observational estimates of the ratio of stellar mass to halo mass, SMF, sSFR,
metallicity, and cold gas fractions from $z\sim 0$--3 and we find that:

\begin{itemize}

\item All three scenarios are able to qualitatively reproduce the
  rising behavior of \fstar\ and the comoving number density of
  low mass galaxies when we allow the parameterizations to be arbitrary
  functions of both halo mass and redshift, provided we do not include photo-ionization squelching.  The three
  adjusted scenarios make different predictions for
  other observables such as cold gas fractions and sSFRs, which may
  help to discriminate between them.

\item In the preferential reheating model, we altered the way in
  which the mass-loading of stellar-driven winds scaled with mass and
  redshift. 
 This required a fairly dramatic change in the mass-loading factor's power law dependence on circular velocity as a function of time.  Our model starts from \arh=4.5 at $z>2$\ and transitions to \arh=2 at $z<1$, compared with a constant value of \arh=1--2 for conventional momentum- or energy-driven winds. 
Although it is expected that wind
  scalings may deviate from the simple energy- or momentum-driven
  case, it is not clear whether physical processes can lead to such a
  large slope or effective redshift evolution.

\item Direct suppression of star formation via explicit manipulation
  of the star formation timescale requires an aggressive suppression
  factor and a complicated redshift dependence. We expect that such
  strong variation in the star formation efficiency will be ruled out
  by direct observations. The direct suppression scenario is less
  tractable than might be expected because of the strongly
  self-regulated nature of star formation in the present paradigm. We
  therefore disfavor this class of solution relative to the other two.

\item In the parking lot model, gas is held temporarily in a
  reservoir outside the galaxy and allowed to accrete on a specified
  timescale.  This required fairly minor alteration to the standard scalings
  assumed in SAMs, especially relative to the very large uncertainties
  in our current understanding and parameterization of this process. Gas that has been
  heated either by gravitational interaction with other halos or by a
  global or local radiation field may have longer accretion times
  than expected in the standard picture of cosmological accretion. The
  SAMs predict, in agreement with results from recent numerical
  simulations, that this ``pre-heated'' gas may comprise a very
  significant component of the accretion. 

\item The predicted evolution of low mass galaxies in
  SAMs is quite sensitive to details of the bookkeeping for the hot
  and ejected gas reservoirs following halo mergers, as well as (at
  the lowest masses) to the modeling of accretion suppression and
  photo-evaporation by an ionizing background. This may explain why
  the proposed solution of H13 is not effective when implemented in
  some other SAM codes, including the fiducial Santa Cruz code. 

\end{itemize}

%===================================
\section*{Acknowledgments}
\begin{small}

We thank Sandra Faber, Bruno Henriques, Michaela Hirschmann, and Yu Lu for enlightening discussions. We also thank Peter Behroozi, Molly Peeples, Yu Lu, Desika Narayanan, and Gergo Popping for sharing tabulated data in electronic form and the anonymous referee for his or her helpful feedback.  This work is by the CANDELS Multi-Cycle Treasury program HST-GO-12060 from the NASA/ESA HST, which is operated by the Association of Universities for Research in Astronomy, Inc., under NASA contract NAS5-26555.

\end{small}

%====================================================================
%====================================================================

\bibliographystyle{apj}
\bibliography{article}

\begin{thebibliography}{63}
\expandafter\ifx\csname natexlab\endcsname\relax\def\natexlab#1{#1}\fi

\bibitem[{{Baldry} {et~al.}(2012){Baldry}, {Driver}, {Loveday}, {Taylor},
  {Kelvin}, {Liske}, {Norberg}, {Robotham}, {Brough}, {Hopkins}, {Bamford},
  {Peacock}, {Bland-Hawthorn}, {Conselice}, {Croom}, {Jones}, {Parkinson},
  {Popescu}, {Prescott}, {Sharp}, \& {Tuffs}}]{baldry:12}
{Baldry}, I.~K., {et~al.} 2012, \mnras, 421, 621

\bibitem[{{Baldry} {et~al.}(2008){Baldry}, {Glazebrook}, \&
  {Driver}}]{baldry:08}
{Baldry}, I.~K., {Glazebrook}, K., \& {Driver}, S.~P. 2008, \mnras, 388, 945

\bibitem[{{Behroozi} {et~al.}(2010){Behroozi}, {Conroy}, \&
  {Wechsler}}]{behroozi:10}
{Behroozi}, P.~S., {Conroy}, C., \& {Wechsler}, R.~H. 2010, \apj, 717, 379

\bibitem[{{Behroozi} {et~al.}(2013){Behroozi}, {Wechsler}, \&
  {Conroy}}]{behroozi:13}
{Behroozi}, P.~S., {Wechsler}, R.~H., \& {Conroy}, C. 2013, \apj, 770, 57

\bibitem[{{Behroozi} {et~al.}(2014){Behroozi}, {Wechsler}, {Lu}, {Hahn},
  {Busha}, {Klypin}, \& {Primack}}]{behroozi:13:5rvir}
{Behroozi}, P.~S., {Wechsler}, R.~H., {Lu}, Y., {Hahn}, O., {Busha}, M.~T.,
  {Klypin}, A., \& {Primack}, J.~R. 2014, \apj, 787, 156

\bibitem[{{Bigiel} {et~al.}(2008){Bigiel}, {Leroy}, {Walter}, {Brinks}, {de
  Blok}, {Madore}, \& {Thornley}}]{bigiel:08}
{Bigiel}, F., {Leroy}, A., {Walter}, F., {Brinks}, E., {de Blok}, W.~J.~G.,
  {Madore}, B., \& {Thornley}, M.~D. 2008, \aj, 136, 2846

\bibitem[{{Bouch{\'e}} {et~al.}(2010){Bouch{\'e}}, {Dekel}, {Genzel}, {Genel},
  {Cresci}, {F{\"o}rster Schreiber}, {Shapiro}, {Davies}, \&
  {Tacconi}}]{bouche:10}
{Bouch{\'e}}, N., {et~al.} 2010, \apj, 718, 1001

\bibitem[{{Bower} {et~al.}(2006){Bower}, {Benson}, {Malbon}, {Helly}, {Frenk},
  {Baugh}, {Cole}, \& {Lacey}}]{bower:06}
{Bower}, R.~G., {Benson}, A.~J., {Malbon}, R., {Helly}, J.~C., {Frenk}, C.~S.,
  {Baugh}, C.~M., {Cole}, S., \& {Lacey}, C.~G. 2006, \mnras, 370, 645

\bibitem[{{Boylan-Kolchin} {et~al.}(2008){Boylan-Kolchin}, {Ma}, \&
  {Quataert}}]{boylan-kolchin:08}
{Boylan-Kolchin}, M., {Ma}, C.-P., \& {Quataert}, E. 2008, \mnras, 383, 93

\bibitem[{{Conroy} \& {Wechsler}(2009)}]{conroy:09}
{Conroy}, C., \& {Wechsler}, R.~H. 2009, \apj, 696, 620

\bibitem[{{Dav{\'e}} {et~al.}(2012){Dav{\'e}}, {Finlator}, \&
  {Oppenheimer}}]{dave_fo:11}
{Dav{\'e}}, R., {Finlator}, K., \& {Oppenheimer}, B.~D. 2012, \mnras, 421, 98

\bibitem[{{Dav{\'e}} {et~al.}(2011){Dav{\'e}}, {Oppenheimer}, \&
  {Finlator}}]{dave_of:11}
{Dav{\'e}}, R., {Oppenheimer}, B.~D., \& {Finlator}, K. 2011, \mnras, 415, 11

\bibitem[{{Diemand} {et~al.}(2007){Diemand}, {Kuhlen}, \& {Madau}}]{diemand:07}
{Diemand}, J., {Kuhlen}, M., \& {Madau}, P. 2007, \apj, 657, 262

\bibitem[{{Dunne} {et~al.}(2009){Dunne}, {Ivison}, {Maddox}, {Cirasuolo},
  {Mortier}, {Foucaud}, {Ibar}, {Almaini}, {Simpson}, \& {McLure}}]{dunne:09}
{Dunne}, L., {et~al.} 2009, \mnras, 394, 3

\bibitem[{{Efstathiou}(1992)}]{efstathiou:92}
{Efstathiou}, G. 1992, \mnras, 256, 43P

\bibitem[{{Erb} {et~al.}(2006){Erb}, {Shapley}, {Pettini}, {Steidel}, {Reddy},
  \& {Adelberger}}]{erb:06}
{Erb}, D.~K., {Shapley}, A.~E., {Pettini}, M., {Steidel}, C.~C., {Reddy},
  N.~A., \& {Adelberger}, K.~L. 2006, \apj, 644, 813

\bibitem[{{Fontanot} {et~al.}(2009){Fontanot}, {De Lucia}, {Monaco},
  {Somerville}, \& {Santini}}]{fontanot:09}
{Fontanot}, F., {De Lucia}, G., {Monaco}, P., {Somerville}, R.~S., \&
  {Santini}, P. 2009, \mnras, 397, 1776

\bibitem[{{Gallazzi} {et~al.}(2005){Gallazzi}, {Charlot}, {Brinchmann},
  {White}, \& {Tremonti}}]{gallazzi:05}
{Gallazzi}, A., {Charlot}, S., {Brinchmann}, J., {White}, S.~D.~M., \&
  {Tremonti}, C.~A. 2005, \mnras, 362, 41

\bibitem[{{Gnedin}(2000)}]{gnedin:00}
{Gnedin}, N.~Y. 2000, \apj, 542, 535

\bibitem[{{Guo} {et~al.}(2011){Guo}, {White}, {Boylan-Kolchin}, {De Lucia},
  {Kauffmann}, {Lemson}, {Li}, {Springel}, \& {Weinmann}}]{guo:11}
{Guo}, Q., {et~al.} 2011, \mnras, 413, 101

\bibitem[{{Haas} {et~al.}(2013){Haas}, {Schaye}, {Booth}, {Dalla Vecchia},
  {Springel}, {Theuns}, \& {Wiersma}}]{haas:13}
{Haas}, M.~R., {Schaye}, J., {Booth}, C.~M., {Dalla Vecchia}, C., {Springel},
  V., {Theuns}, T., \& {Wiersma}, R.~P.~C. 2013, \mnras, 435, 2931

\bibitem[{{Hanasz} {et~al.}(2013){Hanasz}, {Lesch}, {Naab}, {Gawryszczak},
  {Kowalik}, \& {W{\'o}lta{\'n}ski}}]{hanasz:13}
{Hanasz}, M., {Lesch}, H., {Naab}, T., {Gawryszczak}, A., {Kowalik}, K., \&
  {W{\'o}lta{\'n}ski}, D. 2013, \apjl, 777, L38

\bibitem[{{Henriques} {et~al.}(2013){Henriques}, {White}, {Thomas}, {Angulo},
  {Guo}, {Lemson}, \& {Springel}}]{henriques:13}
{Henriques}, B.~M.~B., {White}, S.~D.~M., {Thomas}, P.~A., {Angulo}, R.~E.,
  {Guo}, Q., {Lemson}, G., \& {Springel}, V. 2013, \mnras, 431, 3373

\bibitem[{{Hopkins} {et~al.}(2012){Hopkins}, {Quataert}, \&
  {Murray}}]{hopkins:12}
{Hopkins}, P.~F., {Quataert}, E., \& {Murray}, N. 2012, \mnras, 421, 3522

\bibitem[{{Kajisawa} {et~al.}(2010){Kajisawa}, {Ichikawa}, {Yamada},
  {Uchimoto}, {Yoshikawa}, {Akiyama}, \& {Onodera}}]{kajisawa:10}
{Kajisawa}, M., {Ichikawa}, T., {Yamada}, T., {Uchimoto}, Y.~K., {Yoshikawa},
  T., {Akiyama}, M., \& {Onodera}, M. 2010, \apj, 723, 129

\bibitem[{{Karim} {et~al.}(2011){Karim}, {Schinnerer},
  {Mart{\'{\i}}nez-Sansigre}, {Sargent}, {van der Wel}, {Rix}, {Ilbert},
  {Smol{\v c}i{\'c}}, {Carilli}, {Pannella}, {Koekemoer}, {Bell}, \&
  {Salvato}}]{karim:11}
{Karim}, A., {et~al.} 2011, \apj, 730, 61

\bibitem[{{Kennicutt}(1998)}]{kennicutt:98}
{Kennicutt}, Jr., R.~C. 1998, \apj, 498, 541

\bibitem[{{Kewley} \& {Ellison}(2008)}]{kewley:08}
{Kewley}, L.~J., \& {Ellison}, S.~L. 2008, \apj, 681, 1183

\bibitem[{{Komatsu} {et~al.}(2009){Komatsu}, {Dunkley}, {Nolta}, {Bennett},
  {Gold}, {Hinshaw}, {Jarosik}, {Larson}, {Limon}, {Page}, {Spergel},
  {Halpern}, {Hill}, {Kogut}, {Meyer}, {Tucker}, {Weiland}, {Wollack}, \&
  {Wright}}]{wmap5}
{Komatsu}, E., {et~al.} 2009, \apjs, 180, 330

\bibitem[{{Kravtsov} {et~al.}(2004){Kravtsov}, {Gnedin}, \&
  {Klypin}}]{kravtsov:04}
{Kravtsov}, A.~V., {Gnedin}, O.~Y., \& {Klypin}, A.~A. 2004, \apj, 609, 482

\bibitem[{{Krumholz} \& {Dekel}(2012)}]{krumholz:12}
{Krumholz}, M.~R., \& {Dekel}, A. 2012, \apj, 753, 16

\bibitem[{{Lu} {et~al.}(2015){Lu}, {Mo}, \& {Wechsler}}]{lu:14}
{Lu}, Y., {Mo}, H.~J., \& {Wechsler}, R.~H. 2015, \mnras, 446, 1907

\bibitem[{{Lu} {et~al.}(2014){Lu}, {Wechsler}, {Somerville}, {Croton},
  {Porter}, {Primack}, {Behroozi}, {Ferguson}, {Koo}, {Guo}, {Safarzadeh},
  {Finlator}, {Castellano}, {White}, {Sommariva}, \& {Moody}}]{lu_candels:14}
{Lu}, Y., {et~al.} 2014, \apj, 795, 123

\bibitem[{{Maiolino} {et~al.}(2008){Maiolino}, {Nagao}, {Grazian}, {Cocchia},
  {Marconi}, {Mannucci}, {Cimatti}, {Pipino}, {Ballero}, {Calura}, {Chiappini},
  {Fontana}, {Granato}, {Matteucci}, {Pastorini}, {Pentericci}, {Risaliti},
  {Salvati}, \& {Silva}}]{maiolino:08}
{Maiolino}, R., {et~al.} 2008, \aap, 488, 463

\bibitem[{{Marchesini} {et~al.}(2009){Marchesini}, {van Dokkum}, {F{\"o}rster
  Schreiber}, {Franx}, {Labb{\'e}}, \& {Wuyts}}]{marchesini:09}
{Marchesini}, D., {van Dokkum}, P.~G., {F{\"o}rster Schreiber}, N.~M., {Franx},
  M., {Labb{\'e}}, I., \& {Wuyts}, S. 2009, \apj, 701, 1765

\bibitem[{{Mo} {et~al.}(1998){Mo}, {Mao}, \& {White}}]{mo_mao_white:98}
{Mo}, H.~J., {Mao}, S., \& {White}, S.~D.~M. 1998, \mnras, 295, 319

\bibitem[{{Moster} {et~al.}(2013){Moster}, {Naab}, \& {White}}]{moster:13}
{Moster}, B.~P., {Naab}, T., \& {White}, S.~D.~M. 2013, \mnras, 428, 3121

\bibitem[{{Moustakas} {et~al.}(2013){Moustakas}, {Coil}, {Aird}, {Blanton},
  {Cool}, {Eisenstein}, {Mendez}, {Wong}, {Zhu}, \& {Arnouts}}]{moustakas:13}
{Moustakas}, J., {et~al.} 2013, \apj, 767, 50

\bibitem[{{Narayanan} {et~al.}(2012){Narayanan}, {Bothwell}, \&
  {Dav{\'e}}}]{narayanan:12}
{Narayanan}, D., {Bothwell}, M., \& {Dav{\'e}}, R. 2012, \mnras, 426, 1178

\bibitem[{{Okamoto} {et~al.}(2008){Okamoto}, {Gao}, \& {Theuns}}]{okamoto:08}
{Okamoto}, T., {Gao}, L., \& {Theuns}, T. 2008, \mnras, 390, 920

\bibitem[{{Oppenheimer} {et~al.}(2010){Oppenheimer}, {Dav{\'e}}, {Kere{\v s}},
  {Fardal}, {Katz}, {Kollmeier}, \& {Weinberg}}]{oppenheimer:10}
{Oppenheimer}, B.~D., {Dav{\'e}}, R., {Kere{\v s}}, D., {Fardal}, M., {Katz},
  N., {Kollmeier}, J.~A., \& {Weinberg}, D.~H. 2010, \mnras, 406, 2325

\bibitem[{{Peeples} {et~al.}(2014){Peeples}, {Werk}, {Tumlinson},
  {Oppenheimer}, {Prochaska}, {Katz}, \& {Weinberg}}]{peeples:14}
{Peeples}, M.~S., {Werk}, J.~K., {Tumlinson}, J., {Oppenheimer}, B.~D.,
  {Prochaska}, J.~X., {Katz}, N., \& {Weinberg}, D.~H. 2014, \apj, 786, 54

\bibitem[{{Popping} {et~al.}(2014{\natexlab{a}}){Popping}, {Behroozi}, \&
  {Peeples}}]{popping:14gasfrac}
{Popping}, G., {Behroozi}, P.~S., \& {Peeples}, M.~S. 2014{\natexlab{a}}, ArXiv
  e-prints

\bibitem[{{Popping} {et~al.}(2014{\natexlab{b}}){Popping}, {Somerville}, \&
  {Trager}}]{popping:14}
{Popping}, G., {Somerville}, R.~S., \& {Trager}, S.~C. 2014{\natexlab{b}},
  \mnras, 442, 2398

\bibitem[{{Primack}(2003)}]{primack:03}
{Primack}, J.~R. 2003, Nuclear Physics B Proceedings Supplements, 124, 3

\bibitem[{{Quinn} {et~al.}(1996){Quinn}, {Katz}, \& {Efstathiou}}]{quinn:96}
{Quinn}, T., {Katz}, N., \& {Efstathiou}, G. 1996, \mnras, 278, L49

\bibitem[{{Salim} {et~al.}(2007){Salim}, {Rich}, {Charlot}, {Brinchmann},
  {Johnson}, {Schiminovich}, {Seibert}, {Mallery}, {Heckman}, {Forster},
  {Friedman}, {Martin}, {Morrissey}, {Neff}, {Small}, {Wyder}, {Bianchi},
  {Donas}, {Lee}, {Madore}, {Milliard}, {Szalay}, {Welsh}, \& {Yi}}]{salim:07}
{Salim}, S., {et~al.} 2007, \apjs, 173, 267

\bibitem[{{Santini} {et~al.}(2012){Santini}, {Fontana}, {Grazian}, {Salimbeni},
  {Fontanot}, {Paris}, {Boutsia}, {Castellano}, {Fiore}, {Gallozzi},
  {Giallongo}, {Koekemoer}, {Menci}, {Pentericci}, \&
  {Somerville}}]{santini:12}
{Santini}, P., {et~al.} 2012, \aap, 538, A33

\bibitem[{{Savaglio} {et~al.}(2005){Savaglio}, {Glazebrook}, {Le Borgne},
  {Juneau}, {Abraham}, {Chen}, {Crampton}, {McCarthy}, {Carlberg}, {Marzke},
  {Roth}, {J{\o}rgensen}, \& {Murowinski}}]{savaglio:05}
{Savaglio}, S., {et~al.} 2005, \apj, 635, 260

\bibitem[{{Somerville}(2002)}]{somerville:02}
{Somerville}, R.~S. 2002, \apjl, 572, L23

\bibitem[{{Somerville} {et~al.}(2008{\natexlab{a}}){Somerville}, {Barden},
  {Rix}, {Bell}, {Beckwith}, {Borch}, {Caldwell}, {H{\"a}u{\ss}ler}, {Heymans},
  {Jahnke}, {Jogee}, {McIntosh}, {Meisenheimer}, {Peng}, {S{\'a}nchez},
  {Wisotzki}, \& {Wolf}}]{somerville_barden:08}
{Somerville}, R.~S., {et~al.} 2008{\natexlab{a}}, \apj, 672, 776

\bibitem[{{Somerville} {et~al.}(2012){Somerville}, {Gilmore}, {Primack}, \&
  {Dom{\'{\i}}nguez}}]{somerville:12}
{Somerville}, R.~S., {Gilmore}, R.~C., {Primack}, J.~R., \& {Dom{\'{\i}}nguez},
  A. 2012, \mnras, 423, 1992

\bibitem[{{Somerville} {et~al.}(2008{\natexlab{b}}){Somerville}, {Hopkins},
  {Cox}, {Robertson}, \& {Hernquist}}]{somerville:08}
{Somerville}, R.~S., {Hopkins}, P.~F., {Cox}, T.~J., {Robertson}, B.~E., \&
  {Hernquist}, L. 2008{\natexlab{b}}, \mnras, 391, 481

\bibitem[{{Somerville} \& {Kolatt}(1999)}]{somerville_kolatt:99}
{Somerville}, R.~S., \& {Kolatt}, T.~S. 1999, \mnras, 305, 1

\bibitem[{{Somerville} \& {Primack}(1999)}]{somerville:99}
{Somerville}, R.~S., \& {Primack}, J.~R. 1999, \mnras, 310, 1087

\bibitem[{{Springel} {et~al.}(2005){Springel}, {White}, {Jenkins}, {Frenk},
  {Yoshida}, {Gao}, {Navarro}, {Thacker}, {Croton}, {Helly}, {Peacock}, {Cole},
  {Thomas}, {Couchman}, {Evrard}, {Colberg}, \& {Pearce}}]{springel:05}
{Springel}, V., {et~al.} 2005, \nat, 435, 629

\bibitem[{{Stadel} {et~al.}(2009){Stadel}, {Potter}, {Moore}, {Diemand},
  {Madau}, {Zemp}, {Kuhlen}, \& {Quilis}}]{stadel:09}
{Stadel}, J., {Potter}, D., {Moore}, B., {Diemand}, J., {Madau}, P., {Zemp},
  M., {Kuhlen}, M., \& {Quilis}, V. 2009, \mnras, 398, L21

\bibitem[{{Tomczak} {et~al.}(2014){Tomczak}, {Quadri}, {Tran}, {Labb{\'e}},
  {Straatman}, {Papovich}, {Glazebrook}, {Allen}, {Brammer}, {Kacprzak},
  {Kawinwanichakij}, {Kelson}, {McCarthy}, {Mehrtens}, {Monson}, {Persson},
  {Spitler}, {Tilvi}, \& {van Dokkum}}]{tomczak:14}
{Tomczak}, A.~R., {et~al.} 2014, \apj, 783, 85

\bibitem[{{Torrey} {et~al.}(2013){Torrey}, {Cox}, {Kewley}, \&
  {Hernquist}}]{torrey:13}
{Torrey}, P., {Cox}, T.~J., {Kewley}, L., \& {Hernquist}, L. 2013, in
  Astronomical Society of the Pacific Conference Series, Vol. 477, Galaxy
  Mergers in an Evolving Universe, ed. W.-H. {Sun}, C.~K. {Xu}, N.~Z.
  {Scoville}, \& D.~B. {Sanders}, 237

\bibitem[{{Tremonti} {et~al.}(2004){Tremonti}, {Heckman}, {Kauffmann},
  {Brinchmann}, {Charlot}, {White}, {Seibert}, {Peng}, {Schlegel}, {Uomoto},
  {Fukugita}, \& {Brinkmann}}]{tremonti:04}
{Tremonti}, C.~A., {et~al.} 2004, \apj, 613, 898

\bibitem[{{Vogelsberger} {et~al.}(2014){Vogelsberger}, {Genel}, {Springel},
  {Torrey}, {Sijacki}, {Xu}, {Snyder}, {Nelson}, \&
  {Hernquist}}]{vogelsberger:14}
{Vogelsberger}, M., {et~al.} 2014, \mnras, 444, 1518

\bibitem[{{Weinmann} {et~al.}(2012){Weinmann}, {Pasquali}, {Oppenheimer},
  {Finlator}, {Mendel}, {Crain}, \& {Macci{\`o}}}]{weinmann:12}
{Weinmann}, S.~M., {Pasquali}, A., {Oppenheimer}, B.~D., {Finlator}, K.,
  {Mendel}, J.~T., {Crain}, R.~A., \& {Macci{\`o}}, A.~V. 2012, \mnras, 426,
  2797

\bibitem[{{White} \& {Frenk}(1991)}]{white:91}
{White}, S.~D.~M., \& {Frenk}, C.~S. 1991, \apj, 379, 52

\end{thebibliography}

\appendix
\section{Exploration of parameter space}
\label{sec:param_appendix}
\begin{figure*}
\centering
\includegraphics[width=\textwidth]{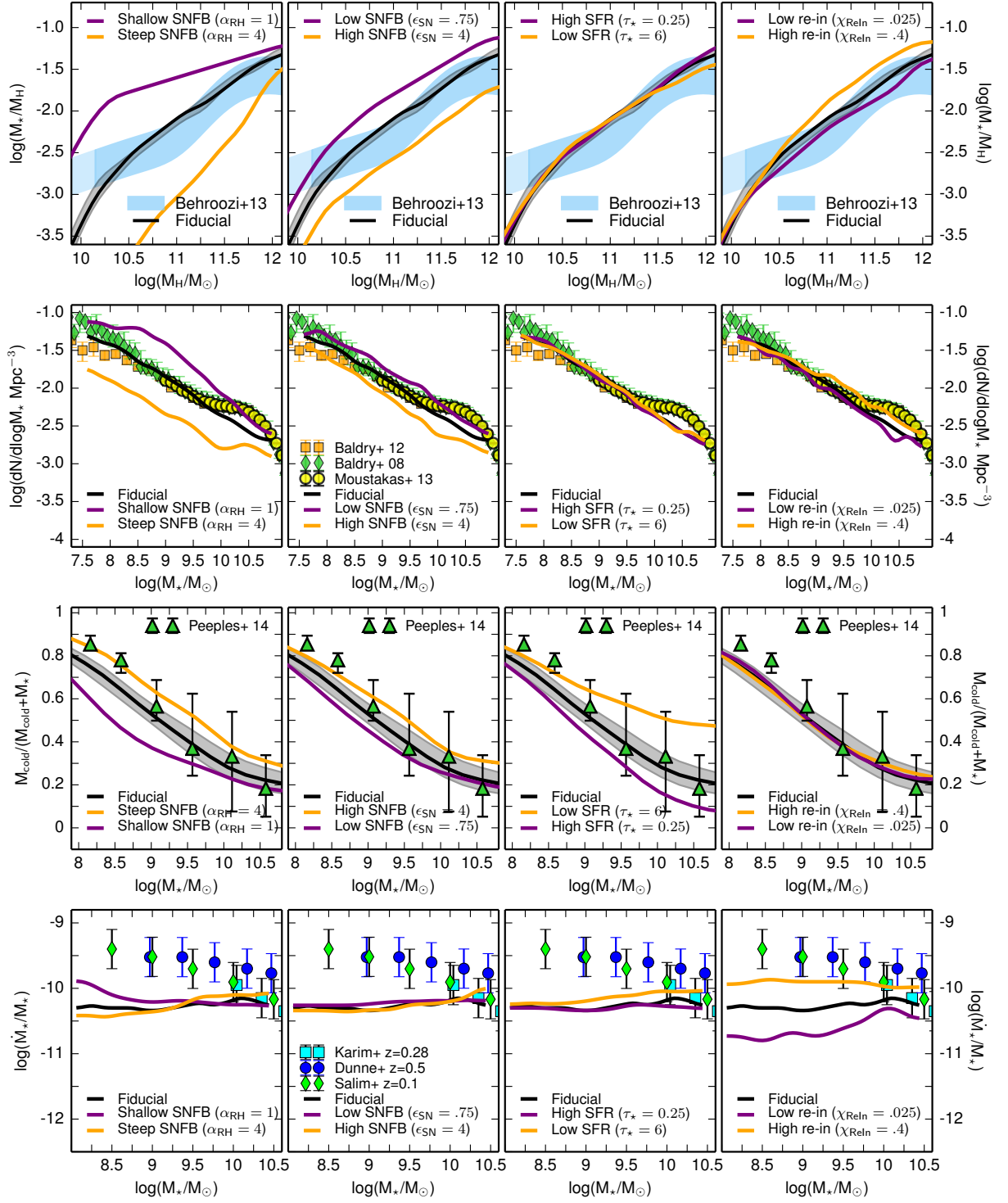}
\caption{\fstar, SMF, cold gas fraction, and sSFR results at redshift
  $z=0$\ for high and low values of \arh\ in the left column, \esn\ in the second column, \tst\ in the third column, and \chirein\ in the right-most column.   The fiducial model is
  shown in black, the low parameter value is shown in purple, and the
  high parameter value is shown in yellow.  Data are as noted in
  legends.
\label{fig:all_grid}}
\end{figure*}

\begin{figure*}
\centering
\includegraphics[width=\textwidth]{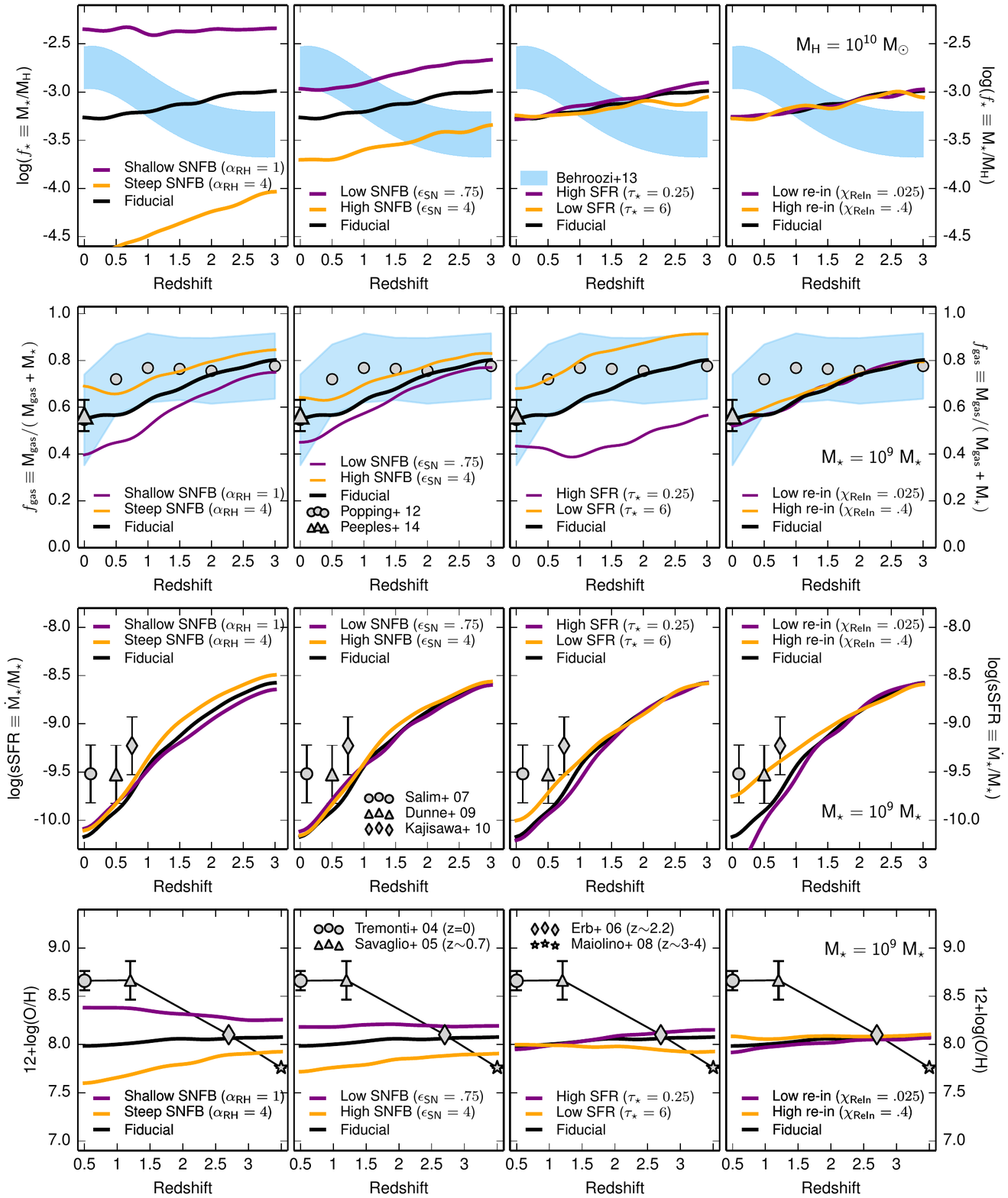}
\caption{\fstar\ as a function of redshift for halo mass
  $M_{\rm{H}}=10^{10}$~\msun\ and cold gas fractions, sSFR, and ISM
  metallicity as a function of redshift for stellar mass
  $M_\star=10^{9}$~\msun\ galaxies.  We show models with low, fiducial, and
  high values of \arh\ in the left column, \esn\ in the second column, \tst\ in the third column, and \chirein\ in the right-most column.  Low parameter value models are shown
  in purple, the fiducial model is shown in black, and high parameter
  value models are shown in yellow.  Data are as noted in legends.
\label{fig:vs_z_grid}}
\end{figure*}

In order to inform our approach with the empirical models, we ran
simulations with all but one of the model's parameters held at the
fiducial value.  We chose the set of parameters we examined to be those
most likely to affect star formation: the wind parameters, \arh, \esn,
and $V_{\rm{eject}}$; the star formation parameters, \tst,
$\chi_{\rm{gas}}$, and $\Sigma_{\rm{crit}}$; and the re-infall
parameter \chirein.  Of these, only \esn, \arh, \tst\, and \chirein\ had
significant impact on low mass galaxy properties.  We find that no
fixed value of these parameters can alter the undesirable fiducial
trends with redshift.  In Fig.~\ref{fig:all_grid}, we show the redshift $z=0$\ \fstarMh, SMF, cold gas fraction, and sSFR for the fiducial model and variations in the four interesting parameters.  Each column shows the fiducial model and a high and low value of a different parameter.  Fig.~\ref{fig:vs_z_grid} shows \fstar(z) for \mh=$10^{10}$\ \msun\ halos and the cold gas fraction, sSFR, and ISM metallicity as a function of redshift for \mstar=$10^9$\ \msun\ galaxies for the same parameter variations.

%----------------------------------------------------------------------------------------

\subsection{Uninteresting parameters}
\label{sec:appendix_uninteresting}
\label{sec:appendix_veject}
\label{sec:appendix_chire-infall}
\label{sec:appendix_chigas}
\label{sec:appendix_sigmacrit}

These parameters have little leverage on the overall properties of low
mass galaxies.  The changes described for each parameter are the only
noticeable effects on the galaxy population.  Varying $\chi_{\rm{gas}}$\ does not affect galaxy population properties at all.

% . . . . . . . . . . . . . . . . . . . . . . . . . . . . . . . . . . . . . . . . . . . . . . . . . . . . . . . . . . . . . . . . . . . . . . . . . . . . . . . . . . . . .
\begin{itemize}

\item[$V_{\rm{eject}}$]: The value of $V_{\rm{eject}}$~dictates the
  halo mass of the transition in the fate of gas that is reheated by
  stellar-driven winds as detailed in Sec.~\ref{sec:snfb}. For galaxies with $V_{\rm
    circ}<<V_{\rm{eject}}$, reheated gas is deposited entirely in the ejected
  reservoir, while for larger circular velocities, it is deposited entirely in
  the hot gas reservoir.  Changing $V_{\rm{eject}}$~primarily affects
  intermediate mass halos with masses between 10$^{11.5}$-10$^{12.5}$
  \msun.  These halos see a decrease in star formation when
  $V_{\rm{eject}}$~is increased because there is an additional delay in
  the cooling of reheated gas for these halos, since it is first ejected from the halo before it rejoins the hot halo and can cool.

\item[$\Sigma_{\rm{crit}}$]: The value of $\Sigma_{\rm{crit}}$~sets
  the critical surface density for star formation: only cold gas in
  parts of the gas disk with surface density higher than
  $\Sigma_{\rm{crit}}$~is considered available for star formation (see
  Sec.~\ref{sec:star_formation}).  Higher $\Sigma_{\rm{crit}}$\ leads
  to lower star formation efficiency and higher cold gas fractions,
  because less gas is available for star formation and more inert cold
  gas remains in the disk.

\end{itemize}

%-----------------------------------------------------------------------------------

\subsection{Interesting parameters}
\label{sec:appendix_interesting}
\label{sec:appendix_epssn}
\label{sec:appendix_alpharh}
\label{sec:appendix_taustar}

\begin{itemize}
\item[\arh]: This parameter controls the slope of the mass-loading factor's
  dependence on galaxy circular velocity (Eqn.~\ref{eq:snfb}).
  Increasing~\arh\ steepens the low mass end of the \fstarMh\ relation.  As \arh\ increases, winds in low mass halos are
strengthened relative to high mass halos.  Thus, when low mass halos form stars, gas
  is more efficiently driven out of the disk and  suppresses star
  formation.  Increasing~\arh\ increases the cold gas fraction
  slightly because star formation is made less efficient.  It also
  decreases metallicity because metals are ejected by the winds.

\item[\esn]: This is the normalization parameter in the mass-loading
  factor for stellar-driven winds (Eqn.~\ref{eq:snfb}).  As
  \esn\ increases, winds become more efficient at removing gas from the
  cold disk.  This is a mass-independent effect, so the shape of the
  \fstarMh\ relation doesn't change but the overall star formation
  efficiency decreases.  Galaxies form a few stars, eject a large
  amount of gas, then have to wait for the reheated gas to cool again
  before forming any more stars.  This effect is also
  reflected in the cold gas fractions, which decrease as reheating becomes less efficient because more gas is converted to stars.

\item[\tst]: The normalization of the star
  formation law (Eqn.~\ref{eq:kennicutt})
  is inversely proportional to \tst.  Interestingly, changing
  \tst\ does not have much effect on star formation rates or the
  overall mass of stars formed in galaxies by redshift $z=0$.
  Increasing \tst\ delays star formation, but by redshift
  $z\lesssim6$, the \fstarMh\ relation is nearly unaffected by
  factor of a few changes in \tst.  The main effect of increasing
  \tst\ is to increase the amount of cold gas in the disk, particularly
  at high redshifts.  This effect is caused by the interplay between
  the star formation efficiency and the cold gas mass.  As star
  formation becomes less efficient, cold gas mass builds up and a
  lower efficiency is still able to produce the same overall star
  formation rate.  

\item[\chirein]: The value of \chirein\ controls the rate at which gas
  can fall from the ejected reservoir back into the hot halo
  (Eqn.~\ref{eq:re-infall}). Choosing an extremely high value for this
  parameter would mean that gas lost to the ejected reservoir would be
  immediately returned to the hot gas halo, which is functionally
  equivalent to assuming that the reheated gas is deposited in the hot
  halo.  Conversely, turning \chirein\ to 0 would mean that gas ejected
  to the ejected reservoir would be lost forever.  If a halo is
  massive ($V_{\rm circ} \gtrsim V_{\rm{eject}}$), little or no gas
  will be ejected from the halo and therefore the value of
  \chirein\ will not affect it.  For low and intermediate mass halos,
  the higher \chirein, the more gas the galaxy has to work with at
  redshifts z$\lesssim1.5$.  At high redshifts, the re-infall timescale
  is long enough with respect to the age of the universe that galaxies
  do not get a significant amount of gas from re-infall and
  \chirein\ doesn't affect the galaxies' overall properties.  At lower
  redshift, as \chirein\ is increased, more gas can get back in and
  star formation is increased somewhat, raising \fstar, the SMF, and
  the sSFR. It is interesting to note that \esn\ and \chirein\ have a
  degenerate effect on \fstar, but a different relative impact on the
  sSFR: \chirein\ has much more leverage on sSFR for a given change in
  \fstar\ than \esn. This is because of the more direct and more
  immediate coupling between the stellar winds and the star formation
  efficiency.
\end{itemize}

\section{Comparison with the results of Henriques et al. (2013)}
\label{sec:appendix_h13}
\begin{figure*}
\centering
\includegraphics[width=\textwidth]{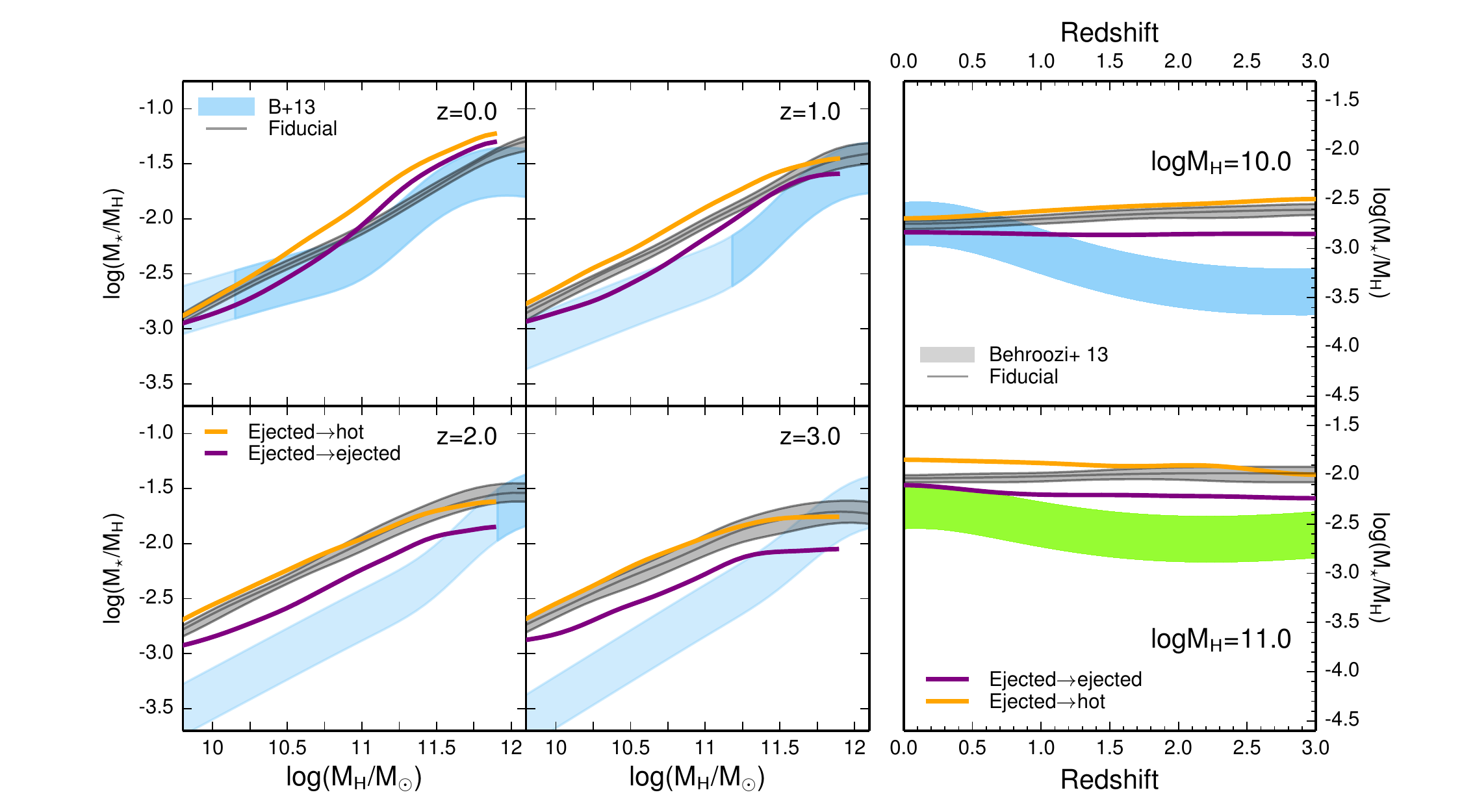}
\caption{Ratio of stellar mass to halo mass, \fstar, for our implementation of the H13 model with in-falling satellites' gas handled in two ways.  In all panels, the fiducial model's median and $\pm1\sigma$\ region are shown in gray and our H13-like models are shown in yellow for the default handling of satellites' ejected reservoirs and purple for gas in satellites' ejected reservoirs being deposited in the central's ejected reservoir.  Left panel: \fstarMh\ relation for four redshifts.  Right panel: \fstar(z) for halo mass $M_{\rm{H}}=10^{10}$\ \msun\ in the top panel and halo mass $M_{\rm{H}}=10^{11}$\ \msun\ in the lower panel.  Empirical constraints (shaded colored regions) are as described in Fig.~\ref{fig:fiducial_fstar}.
\label{fig:h13_fstar}}
\end{figure*}

\begin{figure*}
\centering
\includegraphics[width=\textwidth]{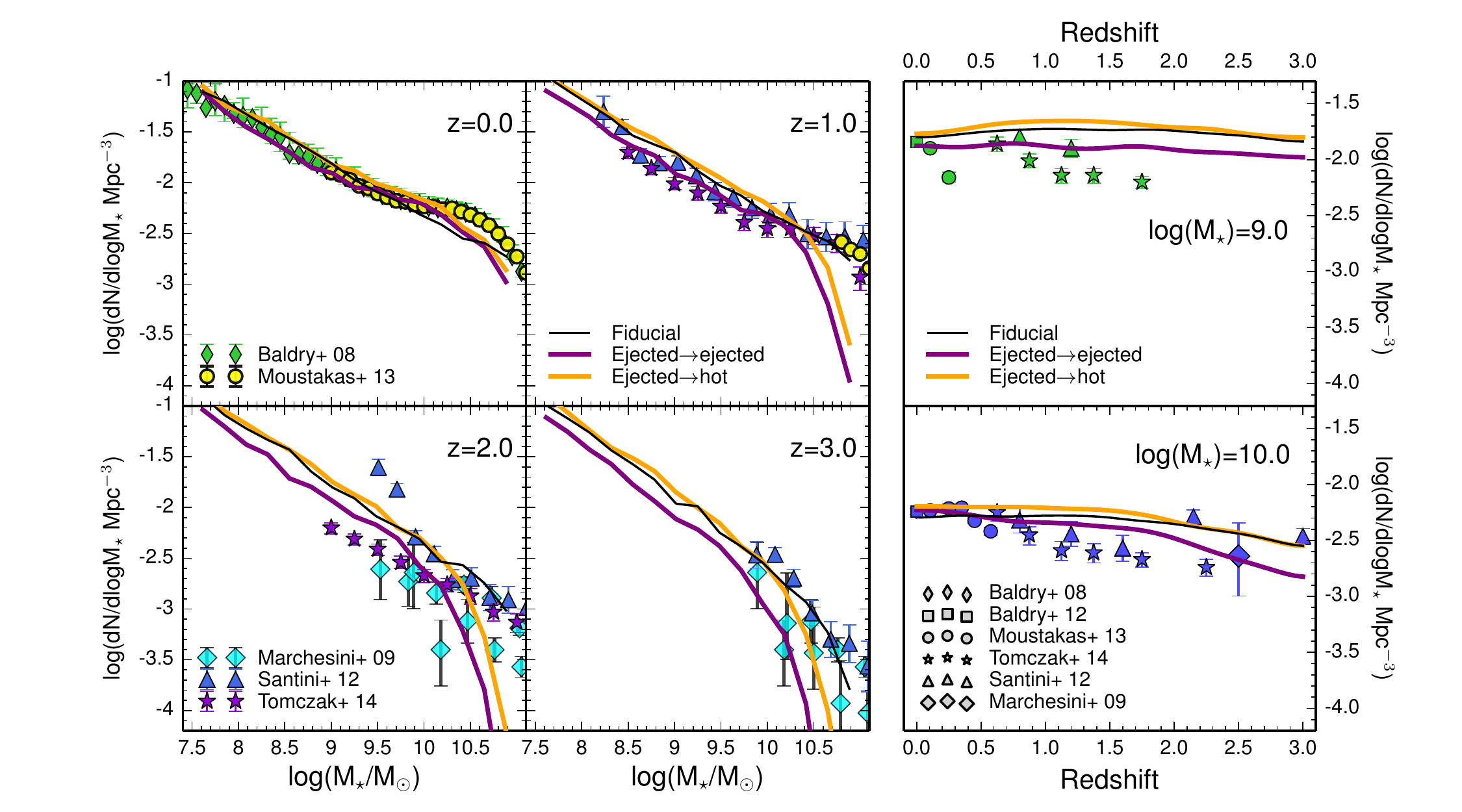}
\caption{Stellar mass functions for our implementation of the H13 model with satellite gas handled two different ways.  In all panels, the fiducial model is shown in black and our H13-like models are shown in yellow for the default handling of satellites' ejected reservoirs and purple for gas in satellites' ejected reservoirs being deposited in the central's ejected reservoir.  Left panel: stellar mass function for four redshifts.  Right panel: galaxy number density as a function of redshift for galaxies with \mstar$=10^9$\ \msun\ in the top panel and \mstar$=10^{10}$\ \msun\ in the lower panel.  Data are as described in Fig.~\ref{fig:fiducial_smf}.
\label{fig:h13_smf}}
\end{figure*}

\begin{figure*}
\centering
\includegraphics[width=\textwidth]{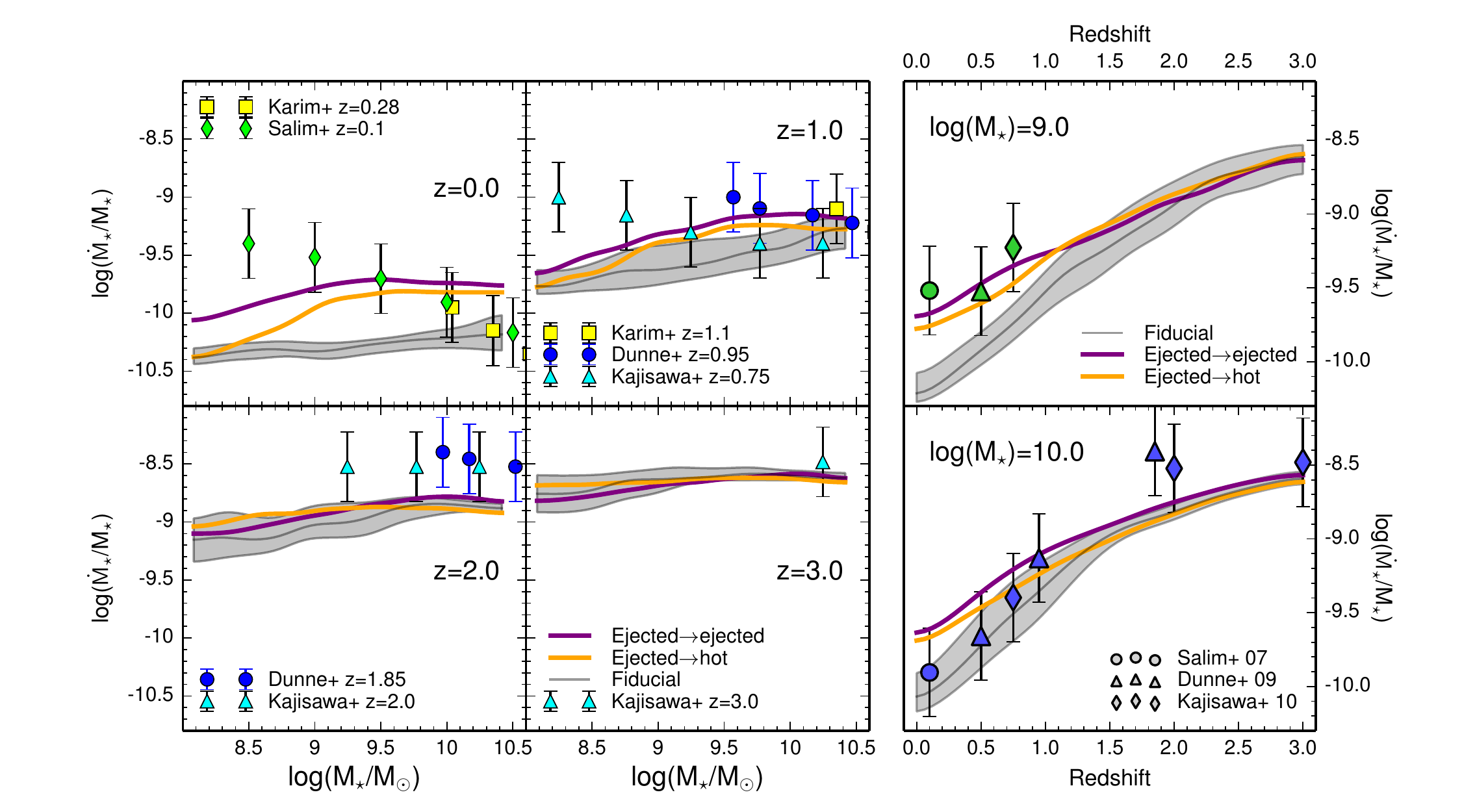}
\caption{Specific star formation rates for our implementation of the H13 model with satellite gas treated two different ways. In all panels, the fiducial model's median and $\pm1\sigma$\ region are shown in gray and our H13-like models are shown in yellow for the default handling of satellites' ejected reservoirs and purple for gas in satellites' ejected reservoirs being deposited in the central's ejected reservoir.  Left panel: sSFR as a function of stellar mass for four redshifts.  Right panel: sSFR as a function of redshift for \mstar$=10^9$\ \msun\ in the top panel and \mstar$=10^{10}$\ \msun\ in the lower panel.  Data are as described in Fig.~\ref{fig:fiducial_ssfr}.
\label{fig:h13_ssfr}}
\end{figure*}

In a recent paper, \citet[][H13]{henriques:13} also addressed the
problems models have with reproducing low mass galaxy properties. They made use of
Monte Carlo Markov Chains (MCMC) coupled with the \citet{guo:11}
semi-analytic model, and found that no single set of parameters could
simultaneously reproduce the abundances of low mass galaxies at all
redshifts. They found, however, that changing the halo mass and time
dependence of the timescale for the re-infall of ejected gas
significantly improved the agreement between their model and
the B- and K-band luminosity function from $z\sim 0$--3 as well as the stellar mass function. Specifically, they proposed the revised scaling:
\begin{equation}
\dot{M}_{\rm{ReIn}} = \left(\frac{M_{\rm{eject}}}  {t_{\rm{ReIn}}}\right)
\label{eq:h13_rein}
\end{equation}
with 
\begin{equation}
t_{\rm{ReIn}} = -\gamma_{\rm{ReIn}} \left( \frac{10^{10} M_\odot}{M_{\rm{H}}}\right)
\label{eq:h13_timescale}
\end{equation}
where the constant $\gamma_{\rm ReIn}$\ has dimensions of time. 
The re-infall timescale is now an explicit function of halo mass but not of time or redshift, while previously it was a function of redshift but not explicitly of halo mass. 
The previous form, Eqn.~\ref{eq:re-infall}, depends on the halo dynamical time, which is independent of halo mass but depends on cosmic time; halos that form at high redshift are denser and have a smaller dynamical time for a given mass. 
Low mass halos now take longer to re-accrete their ejected gas (see Fig.~\ref{fig:park_visualization}). As pointed out
by H13, this is in qualitative agreement with the wind return scalings
found in some numerical hydrodynamic simulations
\citep[e.g.][]{oppenheimer:10}.

We implemented the revised re-accretion timescale functional form
above in our fiducial SAM by simply replacing Eqn.~\ref{eq:re-infall} with Eqns.~\ref{eq:h13_rein} and \ref{eq:h13_timescale}, but found that this did not improve our
predictions for \fstar\ or the comoving number density of low mass
galaxies; instead, it had very little effect on these quantities. The
results are shown in Fig.~\ref{fig:h13_fstar} and
Fig.~\ref{fig:h13_smf}, labeled `Ejected $\rightarrow$ hot' (the
reason for this label will be explained presently).
In order to understand why our model behaves differently, we conducted
several experiments. We found that we could get behavior similar to
that reported by H13 by changing the bookkeeping for the ejected gas
reservoirs of non-largest progenitors following halo mergers. In our
fiducial model, as published in S08, S12 and elsewhere, when halos
merge the gas in the ejected reservoir of the largest progenitor halo
becomes the ejected reservoir of the new halo and the ejected
reservoirs of all other halos are deposited in the \emph{hot gas}
reservoir of the new halo. In addition, all the gas in the hot
reservoirs of the non-largest progenitors is assumed to be
instantaneously stripped and added to the hot gas reservoir of the new
host halo, where it is only allowed to accrete onto the central galaxy
from then on. It turns out that the hot and ejected gas in these
non-largest progenitor halos is a significant component of the total
accretion budget, particularly at late times.

In the H13 models, both the hot and ejected gas from non-largest
progenitors remains bound to the halos even after they become
satellites in the new halo. These hot and ejected reservoirs are then
stripped from the satellites on timescales dictated by tidal and ram
pressure stripping. The stripped gas from the satellites' hot
reservoir is added to the central's hot reservoir, and the stripped
gas from the ejected reservoir is added to the central's ejected
reservoir (B. Henriques 2014, private communication; see also Guo et
al. 2011). When we made the alternate assumption that the ejected
reservoirs are added to the new host's ejected reservoir when halos
become satellites, as well as adopting the revised reaccretion
timescale, we find that the comoving number density of galaxies with
stellar masses $M_\star \sim 10^9$--$10^{10}$\ \msun\ decreases relative to
our fiducial model by about 0.25 dex at $z\sim 1$--2, consistent with
the findings of H13 (see Fig.~\ref{fig:h13_smf}, `Ejected
$\rightarrow$ ejected'). In this model, we find that \fstar\ is
roughly flat from $z\sim 3$\ to 0 at a halo mass of $M_H \sim 10^{11}$,
and is still decreasing (rather than increasing) at lower halo mass
$M_H \sim 10^{10}$\ \msun. This is also consistent with the results
shown by H13, in which the SMF in their modified model is changed only
over a limited range in stellar mass, and steepens again to match the
unmodified slope at masses $M_\star \lesssim 10^9$\ \msun.

H13 do not show the sSFR as a function of stellar mass, but they do
show that the peak in sSFR in the stellar mass range $10^9$\ \msun$ < M_\star<
10^{9.5}$\ \msun\ shifts to higher sSFR, in better agreement with
observations. This is consistent with the results from our
``H13-like'' model; however, we find that the sSFR decreases again at
$M_\star \lesssim 10^9$\ \msun, in conflict with observations of nearby
galaxies (see Fig.~\ref{fig:h13_ssfr}). H13 do not show predictions
for cold gas fractions nor the evolution of the mass-metallicity
relation in their model. We find that the cold gas fractions at $z\sim
0.5$--1 and $M_\star \gtrsim 10^8$\ \msun~are about 20\% higher in our
H13-like model, and that the mass-metallicity evolution is not
significantly different from the fiducial model.

\end{document}